\documentclass[letterpaper,11pt]{article}
\pdfoutput=1 

\usepackage[T1]{fontenc} 
\usepackage[margin=1.2in,letterpaper]{geometry}
\usepackage{latexsym}
\usepackage{amsfonts}
\usepackage[usenames,dvipsnames]{xcolor}
\usepackage{graphicx,amssymb,amsmath,epsfig}
\usepackage[english]{babel}
\usepackage{graphicx}
\usepackage{dcolumn}
\usepackage{bm}
\usepackage{slashed}
\usepackage{tabularx} 
\usepackage{physics}
\usepackage{url}
\usepackage{amssymb}
\usepackage{setspace}
\usepackage{amsmath}  
\usepackage{graphicx} 
\usepackage[nottoc]{tocbibind}
\usepackage{tikz-cd}
\usepackage{bm}\usepackage{pdfpages}
\usepackage{hyperref}
\usepackage{xcolor}
\usepackage{ulem}

\usepackage{lipsum}%
\usepackage[most]{tcolorbox}

\definecolor{mpurple}{rgb}{0.44, 0.16, 0.39}
\definecolor{myblue}{rgb}{0,0,0.8}

\usepackage[english]{babel} 
\usepackage{lscape}
\usepackage{float}
\usepackage{listings}
\usepackage{xcolor}
\usepackage{amsfonts}
\usepackage{comment}
\usepackage{mathtools}
\definecolor{codegreen}{rgb}{0,0.6,0}
\definecolor{codegray}{rgb}{0.5,0.5,0.5}
\definecolor{codepurple}{rgb}{0.58,0,0.82}
\definecolor{backcolour}{rgb}{0.95,0.95,0.92}
\lstdefinestyle{mystyle}{
	backgroundcolor=\color{backcolour},   
	commentstyle=\color{codegreen},
	keywordstyle=\color{magenta},
	numberstyle=\tiny\color{codegray},
	stringstyle=\color{codepurple},
	basicstyle=\ttfamily\footnotesize,
	breakatwhitespace=false,         
	breaklines=true,                 
	captionpos=b,                    
	keepspaces=true,                 
	showspaces=false,                
	showstringspaces=false,
	showtabs=false,                  
	tabsize=2
}
\numberwithin{equation}{section}
\usepackage{enumitem}
\usepackage{enumitem,amssymb}
\newlist{todolist}{itemize}{2}
\setlist[todolist]{label=$\square$}
\usepackage{pifont}

\usepackage{subfigure}

\usepackage{cite}

\newcommand{\be}{\begin{equation}}
	\newcommand{\ee}{\end{equation}}

\newcommand{\br}{\begin{eqnarray}}
	\newcommand{\er}{\end{eqnarray}}

\newcommand{\skdv}{\text{sKdV}}
\newcommand{\smkdv}{\text{smKdV}}

\newcommand{\lie}{\mathcal{G}}

\newcommand{\cD}{\mathcal{D}}
\newcommand{\cE}{{\mathcal{E}}^{(1)}}
\newcommand{\de}{\delta}

\newcommand{\bpsi}{\bar{\psi}}

\newcommand{\g}{\gamma}
\newcommand{\G}{\Gamma}

\newcommand{\kp}{\kappa}
\newcommand{\la}{\lambda}

\newcommand{\bnu}{\bar{\nu}}
\newcommand{\bg}{\bar{\gamma}}
\newcommand{\om}{\omega}
\newcommand{\Om}{\Omega}

\newcommand{\pa}{\partial}

\newcommand{\s}{\sigma}
\newcommand{\ta}{\tau}

\newcommand{\cS}{{\cal S}}
\newcommand{\bp}{\bar{\psi}}
\newcommand{\bchi}{\bar{\chi}}

\renewcommand{\v}{v}

\begin{document}
	
	\vspace*{1cm}
	\noindent
	
	\vskip 1 cm
	\begin{center}
		{\Large\bf New negative grade solitonic sector for supersymmetric KdV and mKdV hierarchies}
	\end{center}
	\normalsize
	\vskip 1cm

	\begin{center}
		{Y. F. Adans}\footnote{\href{mailto:ysla.franca@unesp.br}{ysla.franca@unesp.br}}$^{a}$,
		{A. R. Aguirre}\footnote{\href{mailto:alexis.roaaguirre@unifei.edu.br}{alexis.roaaguirre@unifei.edu.br}}$^{,b}$,
		{J. F. Gomes}\footnote{\href{mailto:francisco.gomes@unesp.br}{francisco.gomes@unesp.br}}$^{,a}$, 
		{G. V. Lobo}\footnote{\href{mailto:gabriel.lobo@unesp.br}{gabriel.lobo@unesp.br}}$^{,a}$, 
		and 
		{A. H. Zimerman}\footnote{\href{mailto:a.zimerman@unesp.br}{a.zimerman@unesp.br}}$^{,a}$
		\\[.5cm]
		
		\par \vskip .1in \noindent
		$^{a}$\emph{Institute of Theoretical Physics - IFT/UNESP,
			Rua Dr. Bento Teobaldo Ferraz 271, 01140-070, São Paulo, SP, Brazil}\\[0.3cm]
		
		$^{b}$\emph{Institute of Physics and Chemistry - IFQ/UNIFEI, Federal University of Itajubá,\\ Av. BPS 1303, 37500-903, Itajub\'a, MG, Brazil.}\\[0.3cm]
		
		\vskip 2cm
	\end{center}

	\begin{abstract}
		A systematic construction for supersymmetric negative graded (non-local) flows for mKdV and KdV based on $sl(2,1)$ with a principal gradation is proposed in this paper. We show that smKdV and sKdV can be mapped onto each other through a gauge super Miura transformation, together with an additional condition for the negative flows, which ensure the supersymmetry of the negative sKdV flow. In addition, we classify both smKdV and sKdV flows with respect to the vacuum (boundary) solutions. These are classified according to zero or non-zero vacuum. Each vacuum solution is used to derive both soliton solutions and the corresponding Heisenberg subalgebra for the smKdV hierarchy. We present the new solutions corresponding to non-zero bosonic and fermionic vacuum by constructing the deformed vertex operators. Finally, the gauge Miura transformation is employed to obtain the sKdV solutions, which exhibit a rich degeneracy due to both multiple gauge super Miura transformations and multiple vacuum possibilities.
	\end{abstract}
	
	\newpage
	
	\section{Introduction}
	\label{sec:intro}
	
	Integrable field theories are very peculiar models  admitting  infinite number of conservation laws. These are, in turn   responsible for  the stability of soliton solutions.
	The Korteweg–de Vries (KdV) \cite{miura_korteweg-vries_1968, gardner_method_1967, gardner_kortewegdevries_1974} and  modified KdV (mKdV) \cite{gomes_negative_2009, aratyn_integrable_2003} equations are for instance,    typical examples  of such a class of models  and have acted as  prototypes  for many new developments in the subject. More recently integrable hierarchies   appears  in many areas of theoretical physics,  as in 2D CFTs \cite{zamolodchikov_infinite_1985, sasaki_virasoro_nodate, eguchi_deformations_1989, bazhanov_integrable_1996} and string theory \cite{ douglas_strings_1990,  frohlich_intersection_1992}. In fact, apart from  single equations, say mKdV (or KdV)  a series of other (higher/lower grade) evolution equations of motion  (flows)  can be systematically constructed given rise to 
	an {\it Integrable Hierarchy} \cite{lax_integrals_1968, gelfand_asymptotic_1975, miwa_solitons_2000}.  It has been realized that these class of  field theoretical models can be described and classified in terms of an   affine algebraic structure \cite{drinfeld_lie_1985,leznov_two-dimensional_1983,olive_local_1985,olive_affine_1993,olive_solitons_1993,babelon_dressing_1992,babelon_affine_1993,babelon_introduction_2003,aratyn_kac-moody_1991,de_groot_generalized_1992,hollowood_tau-functions_1993,luis_miramontes_tau-functions_1999,ferreira_tau-functions_1997} within the  framework  of zero curvature representation.

	The zero curvature representation  appears   as  a very powerful  method in constructing systematically  flow equations . 
	It has the virtue of being gauge invariant and this property has  been explored in order to construct  soliton solutions  from a vacuum solution and moreover, to generate Backlund  and Miura transformations for generalized $A_r-$ mKdV hierachy \cite{de_carvalho_ferreira_generalized_2021, de_carvalho_ferreira_gauge_2021}. The key ingredient in the construction involves the construction of an object  of 
	algebraic origin, the Lax operator, universal   to all flows of the hierarchy. It  depends upon   the  decomposition of the affine Lie algebra ${\lie} $ according to a grading operator denoted by ${Q}$ 
	The Lax operator is further specified by a second decomposition according to a choice of  a constant grade one  generator $E^{(1)}$.  The construction of   (multi) soliton solutions consists in   mapping  the zero curvature  for a specific vacuum configuration  into a non-trivial solution by gauge  transformation. 
	It therefore  follows  that  an hierarchy is  defined  and classified by the following  Lie algebraic ingredients, namely, i) the affine Lie algebra ${\lie} $, ii) the grading operator $Q$, iii) the constant  grade one generator $E^{(1)}$ and iv) a vacuum orbit\footnote{See for instance \cite{gomes_negative_2009} for a review.}.

	The ${\mathcal{N}}=1$ supersymmetric version of KdV and mKdV equations (sKdV and smKdV) was originally introduced in \cite{mathieu_supersymmetric_1988}, and generalized to  odd sKdV hierarchy in \cite{kersten_higher_1988}. Following the algebraic approach of constructing integrable hierarchies based on superalgebras \cite{delduc_supersymmetric_1998}, the sKdV equation has been studied both under a construction based upon the $osp(2|2)$ \cite{inami_lie_1991, inami_n_1991} or  $sl(2,1)$ affine algebra \cite{madsen_non-local_2001, aratyn_supersymmetry_2004}. In addition, the smKdV and the supersymmetric sinh-Gordon (sshG) equations have also been largely explored \cite{yamanaka_super_1988, di_vecchia_classical_1977, chaichian_method_1978, chodos_simple_1980, fioravanti_hidden_2000}, and it turns  out that both equations belongs to the same integrable hierarchy when negative odd flows are taken into account \cite{aratyn_supersymmetry_2004}. Soliton solutions within the zero vacuum orbit were also explored for both smKdV and sshG equations using the dressing method \cite{gomes_soliton_2006}, as well as integrable defects and Bäcklund transformation for such systems, by using both the Lagrangian and Gauge-Bäcklund transformations formalisms \cite{aguirre_n1_2015, aguirre_type-ii_2015, aguirre_defects_2018, xue_backlund-darboux_2014, zhou_darboux_2014}.
	
	More recently, novels results on negative even flows for the bosonic mKdV hierarchy required non-zero vacuum orbit in order to  construct soliton solutions \cite{gomes_negative_2009, gomes_nonvanishing_2012}.  
	It follows that the structure of the vacuum and  the commutativity of the flows are directly connected.  Two sub-hierarchies  can be considered, namely, mKdV-I (positive and negative odd flows with a zero vacuum) and  mKdV-II (positive odd and negative even flows with non-zero vacuum) \cite{adans_negative_2023}.
	
	Similarly, the same spliting in terms of different vacuum  structures can be observed  within the KdV hierarchy. This can be accomplished  from the Miura transformation  realized as a gauge transformation \cite{de_carvalho_ferreira_gauge_2021}.  In fact it has  been shown that the KdV hierarchy also splits in two sub-hierarchies, both having the same flows, positive and negative odd, but with different vacuum orbits.
	Consistency requires   additional conditions upon the temporal derivative of the KdV fields  for negative flows, the so-called temporal Miura relations.

	
	
	This paper aims to fill some gaps concerning the supersymmetric versions of  mKdV and KdV equations, namely:
	\begin{itemize}
		\item[$\diamond$] To explore the negative even flows of smKdV hierarchy and determine whether  they are supersymmetric. As expected, in order to do that, we analyse the hierarchy in terms of its orbit vacuum solution.
		
		\item[$\diamond$] To obtain the super Miura transformation as a gauge transformation, and use it to study the negative flows of sKdV hierarchy, as well as its implications. 
		
		\item[$\diamond$] To construct novel soliton solutions for the smKdV and sKdV equations by combining both the new non-zero dressing orbits and super Miura transformation.
	\end{itemize}
	
	This paper is organized as follows. In section \ref{sec.flows}, we review the algebraic formulation for the Lax pair of smKdV and sKdV hierachies, and exhibit some examples of flows, including a novel negative even  non-local flow for smKdV, and a negative odd for sKdV.  In section \ref{sec:Miura}, smKdV and sKdV hierarchies are connected by a gauge transformation leading to four types of super Miura transformations 
	together with additional conditions for the time derivatives for the negative flows. In section \ref{sec.dressing}, we introduce the dressing method for $sl(2,1)$ in order to prove, in section \ref{heisenberg.commuting.flows}, the existence of different sub-hierarchies, as well as to evaluate the supersymmetry of each flow. Finally, we calculate the smKdV solution considering different vacuum orbits in section \ref{mkdv.solutions}. In   \ref{kdv.solutions} we  discuss  the super Miura transformation to evaluate sKdV solutions. 

	\section{Lax pair for the smKdV and sKdV flows}
	\label{sec.flows}
	
	In this section, we shall employ the algebraic formalism to construct  integrable smKdV and sKdV hierarchies. The general structure for a given flow $t_{N}$, with $N\in \mathbb{Z}$, is encoded in the zero curvature condition ({ZCC}), 
	\begin{equation}
		\comm{\pa_x + A_x}{\; \pa_{t_N} + A_{t_N}} = 0
		\label{zcc}
	\end{equation}
	for the universal spatial Lax potential, $\mathcal{L}_x =\pa_x + A_x$, and the temporal Lax potential, $\mathcal{L}_N = \pa_{t_N} + A_{t_N}$ which is specified for each integer $N$. Both smKdV and sKdV hierarchies are based upon the affine $sl(2,1)$ superalgebra endowed with a principal grading operator $Q$ (see Appendix \ref{algebra.sl(2,1)}). For the smKdV spatial gauge potential, we have
	\begin{equation} \label{ax.smkdv}
		A_x^{\smkdv} = \cE + A_0 + A_{\frac{1}{2}} 
		= \mqty(
		\sqrt{\la} + v & 1 & - \bpsi \\
		\la & \sqrt{\la} - v & \sqrt{\la} \; \bpsi \\
		\sqrt{\la} \; \bpsi & - \bpsi& 2 \sqrt{\la} 				 
		)
	\end{equation}
	where $\cE = K_1^{(1)} + K_2^{(1)} \in \lie_1$ is a constant grade one  element, $A_0 = v \; M_1^{(0)} \in \lie_0$ contains the {bosonic field} $v = v(x,t_N)$, and $A_{\frac{1}{2}} = \bpsi \; G_2^{(\frac{1}{2})} \in \lie_{\frac{1}{2}}$  the {fermionic field} $\bpsi = \bpsi(x,t_N)$. On the other hand, the spatial gauge potential for sKdV hierarchy differs from smKdV  in the algebraic elements associated to the fields,
	\begin{equation} \label{ax.skdv}
		A_x^{\skdv} = \cE + A_{-1} + A_{-\frac{1}{2}} 
		= \mqty(
		\sqrt{\la} & 1 & 0 \\
		\la + J & \sqrt{\la} & \bchi / \sqrt{\la} \\
		0 & - \bchi / \sqrt{\la}   & 2 \sqrt{\la} 				 
		)
	\end{equation}
	where $A_{-1} = \frac{1}{2} \; J \left( K_1^{(-1)} - M_2^{(-1)} \right) \in \lie_{-1}$ and  $A_{-\frac{1}{2}} = \frac{1}{2} \; \bchi \left(F_2^{(-\frac{1}{2})} + G_1^{(-\frac{1}{2})} \right) \in \lie_{-\frac{1}{2}}$  contain the {bosonic } $J=J(x,t_N)$ and the  {fermionic }  $  \bchi = \bchi(x,t_N)$ fields respectively. Now, regarding the temporal gauge potential, it is possible to propose a different ansatz for each integer flow:
	\begin{itemize}
		\item For the smKdV positive sub-hierarchy 
		\begin{align}
			A_{t_N}^{\smkdv} = D^{(N)}_{N} + D^{(N-\frac{1}{2})}_{N} + \cdots + D^{(\frac{1}{2})}_{N} + D^{(0)}_{N},
			\qquad
			\qquad
			(D^{(a)}_{N} \in \lie_a),
			\label{gauge.t.smkdv.pos}
		\end{align}
		\item For the sKdV positive sub-hierarchy 
		\begin{equation}
			A_{t_N}^{\skdv} = \cD^{(N)}_{N} + \cD^{(N-\frac{1}{2})}_{N} + \cD^{(N-1)}_{N} + \cdots + \cD^{(0)}_{N} + \cD^{(-\frac{1}{2})}_{N} + \cD^{(-1)}_{N}, 
			\qquad
			(\cD^{(a)}_{N} \in \lie_a).
			\label{gauge.t.skdv.pos}
		\end{equation}
	\end{itemize}
	The elements $D^{(a)}_N$ and $\cD^{(a)}_{N}$ can be determined recursively using grade by grade decomposition of ZCC (for details see Appendix \ref{ap.gradezcc}). This leads to a constraint upon the highest grade elements for the associated  positive sub-hierarchies, i.e.,
	\begin{equation}
		\comm{ \cE }{ D^{(N)}_N } =   \comm{ \cE }{ \cD^{(N)}_N } =0.
	\end{equation}
	Consequently both $D^{(N)}_N$  and ${ \cD^{(N)}_N }$ belong to the kernel of $\cE$ (see more details in \eqref{kernel.super.E1}), implying that $N = 2m + 1$, with $m \in \mathbb{Z}$. 
	In both positive sub-hierarchies {only  odd flows are allowed.} 
	
	Considering now the temporal gauge potentials for the negative flows, we propose the following structure:
	\begin{itemize}
		\item For the smKdV negative sub-hierarchy 
		\begin{equation}
			A^{\smkdv}_{t_{-N}} =  D^{(-N)}_{-N} + D^{(-N+\frac{1}{2})}_{-N} + \cdots +D^{(-\frac{1}{2})}_{-N},
			\qquad
			\qquad
			(D^{(a)}_{-N} \in \lie_a).
			\label{gauge.t.smkdv.neg}
		\end{equation}
		\item For the  sKdV negative sub-hierarchy 
		\begin{equation}
			A^{\skdv}_{t_{-N}} =  \cD^{(-N-2)}_{-N} + \cD^{(-N-\frac{3}{2})}_{-N} + \cD^{(-N-1)}_{-N} + \cdots + D^{(-1)}_{-N} + D^{(-\frac{1}{2})}_{-N},
			\qquad
			(\cD^{(a)}_{-N} \in \lie_a),
			\label{gauge.t.skdv.neg}
		\end{equation}
	\end{itemize}
	where, analogously to the positive case, we can decompose the ZCC and determine each component $D^{(a)}_{-N}$ or    $\cD^{(a)}_{-N}$. From the smKdV lowest grade, we find the non-local equation for  $ D^{(-N)}_{-N}$, 
	\begin{equation}
		\pa_x D^{(-N)}_{-N} + \comm{A_0}{D^{(-N)}_{-N}} = 0.
		\label{lowest.grade.smkdv.neg}
	\end{equation}
	Now, unlike the smKdV negative case \eqref{lowest.grade.smkdv.neg}, we do obtain a constraint for the negative sub-hierarchy of sKdV,
	\begin{equation}
		\comm{A_{-1}}{\cD^{(-N-2)}_{-N}} = 0, 
		\label{lowest.grade.skdv.neg.a}
	\end{equation}
	implying that $\cD^{(-N-2)}_{-N}$ is proportional to $A_{-1} = \frac{1}{2} \; J ( K_1^{(-1)} - M_2^{(-1)} )$, and restricting the negative flows to $N = 2m+1$. We can therefore conclude that {\it the negative sub-hierarchy of sKdV admits only odd flows}, while there are no restrictions on the negative smKdV sub-hierarchy\footnote{In Sect. 3  we shall demonstrate how equations can be mapped from one hierarchy to another through super Miura transformations. The fact that there are fewer temporal flows in sKdV side leads to a coalescence of   smKdV flows into   sKdV flows. }. 
	
	Let us now consider the  smKdV hierarchy and  present  the  first few flows:
	\begin{itemize}
		\item $N=3$
		\begin{subequations}
			\begin{align}
				4 \pa_{t_3} v &= \pa_x^3 v - 6 v^2 \pa_x v - 3 \bar{\psi} \pa_x \left( v \pa_x \bar{\psi} \right),
				\label{smkdv.t3.b}
				\\[0.2cm]
				4 \pa_{t_3} \bar{\psi} &= \pa_x^3 \bar{\psi} - 3 v \pa_x \left( v \bar{\psi}\right),
				\label{smkdv.t3.f}
			\end{align}	\label{smkdv.t3}
		\end{subequations}
		leads to the super mKdV equation, which names the whole hierarchy.
		\item $N=5$
		\begin{subequations}\label{smkdv.t5}
			\begin{align}
				16 \pa_{t_5} v &= \pa_x \left(\pa_x^{4}v+6v^5-10v\left(\pa_x v\right)^2-10v^2 \pa_{x}^2 v- 5v\bp \pa_x^3\bp \right. \nonumber \\
				& -5\pa_{x}^2v\bp \pa_x \bp - 5 \pa_x v \bp \pa_x^2 \bp+20v^3\bp \pa_x \bp \Bigl),
				\label{smkdv.t5.b}
				\\[0.2cm]
				16 \pa_{t_5} \bar{\psi} & =  \pa_x^5 \bp - 5 v^2 \pa_x^3 \bp- 15 v \pa_x v \pa_x^2 \bp - 10 (\pa_x v)^2 \pa_x \bp + 10 v^4 \pa_x \bp \nonumber \\
				&-10 \pa_x v \pa_x^2 v \bp - 5 v \pa_x^3v \bp - 15 v \pa_x^2v \pa_x \bp+ 20 v^3 \pa_x v \bp   ,
				\label{smkdv.t5.f}
			\end{align}	
		\end{subequations}
		\item  $N=-1$
		\begin{subequations}\label{smkdv.tm1}
			\begin{align}
				\pa_x \pa_{t_{-1}} \phi &= 2\sinh{2 \phi} - 2\bpsi \psi \sinh{\phi},
				\\[0.2cm]
				\pa_{t_{-1}} \bpsi &= 2 \psi \cosh{\phi},
				\\[0.2cm]
				\pa_x \psi &= 2 \bp \cosh\phi,
			\end{align}		
		\end{subequations}
		leads to the super sinh-Gordon equation. Here, we have introduced a convenient reparametrization $v(x,t_N) = \partial_x \phi(x,t_N)$ (or  $ \phi (x,t_N)= \pa_x^{-1} v(x,t_N)$).
		\item  $N=-2$
		\begin{subequations}
			\begin{align}
				\pa_{t_{-2}}\pa_x \phi &= -2\left(a_{-}+a_{+}\right)+2 \bp\left(\Om_{-}+\Om_{+}\right),
				\\[0.2cm]
				\pa_{t_{-2}} \bp &= -2 \left(\Omega_{+}-\Omega_{-}\right),
			\end{align}
			\label{smkdv.tm2}
		\end{subequations}
		with
		\begin{subequations}
			\begin{align}
				\psi_{\pm} &= \pa_x^{-1}\left( e^{\pm \phi} \bp\right),
				\label{psi.pm}
				\\[0.2cm]
				a_{\pm} &= e^{\pm 2\phi} \pa_{x}^{-1} \left[e^{\mp 2\phi} \left( 1 + \psi_{\mp} \pa_x \psi_{\pm} \right)\right],
				\label{a.pm}
			\end{align}
		\end{subequations}
		and
		\begin{equation} \label{om.pm}
			\Om_{\pm} = \frac{e^{\pm \phi}}{2} \pa_x^{-1} \left[e^{\mp 2\phi} \psi_{\pm}-\psi_{\mp} \mp \left( \pa_x \psi_{\mp} \right) \left(1 +\psi_{-}\psi_{+} \mp 2a_{\pm} \right)\right].
		\end{equation}
		Here we have used the anti-derivative operator, defined as $\pa_x^{-1} f = \int^x f(y) \dd{y}$.
	\end{itemize} 
	Taking the limit $\bp \to 0$, we recover the mKdV bosonic hierarchy, which is based on the $sl(2)$ affine algebra \cite{adans_negative_2023}. Now, if we consider the ${N=\frac{1}{2}}$ case, we obtain
	\begin{subequations}\label{susy.transf.mkdv}
		\begin{align}
			\de v\equiv \pa_{t_{\frac{1}{2}}} v &=\xi \; \pa_x\bp,
			\\[0.2cm]
			\de \bp \equiv  \pa_{t_{\frac{1}{2}}} \bp &=   -\xi \; v,
		\end{align}	
	\end{subequations}
	\noindent
	the supersymmetry transformation  relating the bosonic and fermionic fields of smKdV. Here $\xi$ is a Grassmann constant parameter.
	
	For the sKdV hierarchy the first temporal non-trivial flows leads to  the following equations:
	\begin{itemize}
		\item  $N=3$
		\begin{subequations}\label{skdv.t3}
			\begin{align}
				4 \partial_{t_3} J &= \partial_x^3 J-6 J \partial_x J-3 \bar{\chi} \partial_x^2 \bar{\chi},
				\label{skdv.t3.b}
				\\[0.2cm]
				4 \partial_{t_3} \bar{\chi} &= \partial_x^3 \bar{\chi}-3 \partial_x\left(J \bar{\chi}\right),
				\label{skdv.t3.f}
			\end{align}
		\end{subequations}
		which is  the supersymmetric  KdV equation and  name the entire hierarchy.
		\item $N=5$  leads to the supersymmetric Sawada-Kotera equation, 
		\begin{subequations}\label{skdv.t5}
			\begin{align}
				16 \partial_{t_5} J &= \pa_x^5 J - 10 \pa_x J \pa_x^2 J - 10 J \pa_x^3 J + 30 J^2 \pa_x J \nonumber \\
				&+ 26 J \bar{\chi} \pa_x^2\bar{\chi} + 32 \pa_x J \bar{\chi} \pa_x \bar{\chi} - 8 \bar{\chi} \pa_x^4 \bar{\chi} - 4 \pa_x \bar{\chi} \pa_x^3 \bar{\chi}
				\label{skdv.t5.b}
				\\[0.2cm]
				16 \partial_{t_5} \bar{\chi} &= \pa_x^5 \bar{\chi} - 5 \pa_x^3 J \bar{\chi} - 10 \pa_x^2 J \pa_x \bar{\chi} - 10 \pa_x J \pa_x^2 \bar{\chi} \nonumber\\
				& -5J \pa_x^3 \bar{\chi} + 20 J \pa_x J \bar{\chi} + 10 J^2 \pa_x \bar{\chi}
				\label{skdv.t5.f}
			\end{align}		
		\end{subequations}
		\item $N=-1$
		\begin{subequations}
			\begin{align} 
				\pa_{t_{-1}}&\pa_x^3 \eta - 2 \; \pa_x^2 \eta \left(\pa_{t_{-1}}\eta + \g \right) - 4 \; \pa_{t_{-1}}\pa_x\eta \; \pa_x \eta - \pa_x \left(\pa_{t_{-1}} \bg \; \pa_x^2 \bg \right) \nonumber \\[2mm]
				&- \pa_x \eta \; \pa_x \bg \; \pa_{t_{-1}}	\bg +\pa_x^2 \bg \left( \bnu_{-} - \bnu_{+} + 2 \bg \right) + 2 \; \pa_{t_{-1}} \pa_x^2 \bg \; \pa_x \bg \,\, =\,\, 0,
				\label{skdv.tm1.b}		
				\\[4mm]
				&  \pa_{t_{-1}}\left(\pa_x \eta \; \pa_x \bg\right)+ \pa_x^2 \left( \bnu_{+} + \bnu_{-} \right) + \pa_x \left( \pa_{t_{-1}}\pa_x^2 \bg + 2 \pa_x \bg + 2 \; \pa_x \bg \; \g \right) \nonumber \\[2mm]
				&- \pa_x \eta (\bnu_{-}-\bnu_{+}+2\bg)  \,\, =\,\, 0,
				\label{skdv.tm1.f}		
			\end{align} \label{skdv.tm1}
		\end{subequations}
		here we have used the following reparametrization,
		\begin{subequations}
			\begin{align} \label{skdv.tm1.aux}
				J(x,t_N) &= \pa_x \eta(x,t_N), \quad \quad \quad \bchi(x,t_N) = \pa_x \bg(x,t_N), \quad \quad \quad \pa_x \g = \pa_x \bg \pa_{t_{-1}} \bg,
				\\[2mm]
				\pa_x \bnu_{+} &= \pa_{t_{-1}} \eta \pa_x \bg, \qquad \qquad \quad	\pa_x \bnu_{-} = \pa_x \eta \pa_{t_{-1}} \bg.
			\end{align}
		\end{subequations}
		\item  $N=\frac{1}{2}$
		\begin{subequations}\label{susy.transf.kdv}
			\begin{align}
				\de J =\pa_{t_{\frac{1}{2}}} J &= -\xi \; \pa_x \bchi,
				\\[0.2cm]
				\de \bchi =	\pa_{t_{\frac{1}{2}}} \bchi &= \; \xi J,
			\end{align}		
		\end{subequations}
		provides the supersymmetry transformation for sKdV model.
	\end{itemize}
	
	The sKdV $(-1)$ equations \eqref{skdv.tm1} were  derived in \cite{adans_skdv_2024}. Notice that these equations present a very remarkable feature compared to negative flows of smKdV  \eqref{smkdv.tm1} and \eqref{smkdv.tm2}, since they do not depend on exponential terms and are mainly composed of mixtures of temporal and spatial derivatives. 
	
	Regarding supersymmetry property, it was already proven in \cite{aratyn_supersymmetry_2004} that the positive flows of sKdV were supersymmetric, as well as the positive and odd negative flows for the smKdV hierarchy.
	In this work, we extend this verification to the novels negative flows of sKdV and smKdV exhibited here. 
	It is possible to check by direct calculations that the equation of motion for  smKdV $(-2)$ is indeed supersymmetric (see Appendix \ref{ap.susy.tm2}). 
	For the sKdV $(-1)$, the supersymmetry proof is accomplished  by making use of the  super Miura transformation (SMT) in section \ref{sec:Miura}.
	
	The supersymmetric mKdV and KdV  hierarchies are related through via super Miura transformation (SMT), and it is known that each positive flow of the sKdV hierarchy is mapped (SMT)  into a positive flow of smKdV \cite{inami_lie_1991}. Now,  with the presentation of these novel negative flows for sKdV and even negative flows for smKdV, a discussion opens up on how to map the negative flows from one hierarchy into the other,  since  that there are more negative flows in smKdV sector than in sKdV.  This indicates a coalescence similar to those we have indicated  in  earlier work  for the pure bosonic case \cite{adans_negative_2023}. In the next section, we present the super Miura transformation as a gauge transformation and demonstrate how it provides a map between the sKdV and smKdV hierarchies.

				\section{
					Gauge super Miura transformation} \label{sec:Miura}

				In this section, we discuss the  connection between the supersymmetric versions of mKdV and KdV hierarchies. For the bosonic case, the mKdV and KdV equations can be related through the well-known {Miura transformation}  \cite{miura_korteweg-vries_1968, fordy_factorization_1980, guil_homogeneous_1991}. The extension to the complete hierarchy was shown to be connected to the $sl(2)$  affine algebra  and considered in   \cite{gomes_miura_2016}.  This construction was later  generalized to the  $sl(r+1)$ case in \cite{de_carvalho_ferreira_gauge_2021} and more recently,  extended to negative flows in \cite{adans_negative_2023}. 
				
				For the supersymmetric case, a connection between smKdV and sKdV equations was established  by  the super Miura transformation  \cite{mathieu_supersymmetric_1988}.  More recently, we have recovered and extended these results for the complete hierarchies in \cite{adans_skdv_2024}. The key ingredient is to construct  a gauge transformation $\mathcal{S}$ which maps the gauge potentials given by \eqref{ax.smkdv} and \eqref{ax.skdv} as
				\begin{equation}
					A_x^{\skdv} = \cS A_x^{\smkdv} \cS^{-1} + \cS \pa_x \cS^{-1}.
					\label{gauge.smiura.x}
				\end{equation}
				Following  the algebraic approach proposed in \cite{de_carvalho_ferreira_gauge_2021}, two different graded ansatz can be considered for $\cS$:
				\begin{equation}
					\cS_1 = s^{(0)} + s^{(-\frac{1}{2})} + s^{(-1)} 
					\qquad \text{or} \qquad
					\cS_2 = s^{(-1)} + s^{(-3/2)} + s^{(-2)}
				\end{equation}        
				where $	s^{(n)}$ are given by the following graded elements
				\small{\begin{equation}
						s^{(2m)}=  \la ^{m} \mqty(
						a_{11} & 0 & 0 \\
						0 & a_{22} & 0 \\
						0 & 0 & a_{33}
						)
						, \quad 
						s^{(2m+1)}= \la^{m} \mqty(
						\sqrt{\la}\; b_{11} &  a_{12} & 0 \\
						\la\; a_{21} & \sqrt{\la}\; b_{22} & 0 \\
						0 & 0 & \sqrt{\la}\; b_{33}
						)
					\end{equation}
					\begin{equation}
						s^{(2m+\frac{1}{2})}= \la ^{m}\mqty(
						0 & 0 & a_{13} \\
						0 & 0 & \sqrt{\la}\; a_{23} \\
						\sqrt{\la}\; a_{31} &  a_{32} & 0
						), \quad 
						s^{(2m+\frac{3}{2})}= \la ^{m}\mqty(
						0 & 0 & \sqrt{\la} a_{13} \\
						0 & 0 & \la a_{23} \\
						\la\;a_{31} & \sqrt{\la}\; a_{32} & 0
						)
				\end{equation}}
				\normalsize leading to four different solutions, namely, 
				\begin{equation}
					\cS_{1,\pm} = \mqty(
					1 & 0 & 0 \\
					v & 1 & - \bp \\
					\pm \bp & 0 & \pm 1
					), \qquad \cS_{2,\pm} = 
					\mqty(
					0 & \frac{1}{\la} & 0 \\
					1 & - \frac{v}{\la} & \frac{\bpsi}{\sqrt{\la}} \\
					0 & \mp \frac{\bpsi}{\la}  & \frac{\pm 1}{\sqrt{\la}}
					) .  
				\end{equation}
				The sKdV and smKdV fields are related by
				\begin{equation} \label{smiura.x.s.1}
					\begin{split}
						J^{(1,\pm)} &= v^2 - \pa_x v + \bpsi \pa_x \bpsi ,
						\\
						\bar{\chi}^{(1,\pm)} &= \mp v \bpsi \pm \pa_x \bpsi,
					\end{split}
				\end{equation}
				for $\cS_{1,\pm}$ and by 
				\begin{equation} \label{smiura.x.s.2}
					\begin{split}
						J^{(2,\pm)} &= v^2 + \pa_x v + \bpsi \pa_x \bpsi  ,
						\\
						\bar{\chi}^{(2,\pm)}&= \mp v \bpsi \mp \pa_x \bpsi.
					\end{split}
				\end{equation}
				for $\cS_{2,\pm}$. The existence of several different gauge transformations is nothing more than a manifestation of the symmetry of super mKdV equation under the parity transformation $v \to -v$ and $\bp \to -\bp$. Also, we might consider the inversion matrix $L_{-}$ acting upon the fermionic subspace, i.e.,	
				\begin{equation}
					L_{-} =  \mqty(
					1 & 0  & 0  \\
					0& 1 &0   \\
					0&  0 & - 1
					)
				\end{equation}
				so we relate
				\begin{equation*}
					\cS_{i,-} = L_{-} \; \cS_{i,+}.
					\qquad
				\end{equation*}
				Thus, we can proceed with our analysis assuming that $\cS = \cS_{1,+}$ without loss of generality. This is indeed highly convenient, as it allows the gauge-Miura $\cS_{+}$  to be expressed in exponential form 
				\begin{equation} \label{smiura.exp}
					\cS \equiv	\cS_{+} = e^{ \frac{v}{2} \left(K_1^{(-1)}-M_2^{(-1)}\right) - \frac{\bp}{2} \left(F_2^{(-\frac{1}{2})}+G_1^{(-\frac{1}{2})}\right)  }. 
				\end{equation}
				It therefore follows that  it is possible to show that each positive  \textit{smKdV} $N$ flow can be  mapped  into a one-to-one correspondence with  \textit{sKdV} $N$ flow:
				\begin{equation} 
					\begin{tikzcd}[row sep=0.2em, column sep=2em]
						t_{N}^{\smkdv} \arrow[r, "\cS"] & t_{N}^{\skdv}.
					\end{tikzcd}
					\label{corresp.pos.fer}
				\end{equation}
				In order to   extend  this analysis to the negative sector, consider (\ref{smiura.exp}) and a generic negative \textit{odd}  flow  $ A_{t_{-2n+1}}^{\smkdv}$ under the gauge transformation induced by $\cS$, yielding the following graded structure
				\begin{equation}
					\begin{split}
						A_{t_{-2n+1}}^{\skdv} & \equiv \cS \left( D^{(-2n+1)}_{-2n+1}+ D^{(-2n+\frac{3}{2})}_{-2n+1} + \cdots + D^{(-\frac{1}{2})}_{-2n+1} \right) \cS^{-1} + \cS \pa_{t_{-2n+1}}\cS^{-1}  \\
						&= {\cD}^{(-2n-1)}_{-2n+1} + {\cD}^{(-2n-\frac{1}{2})}_{-2n+1} +  {\cD}^{(-2n)}_{-2n+1} +\cdots + {\cD}^{(-1)}_{-2n+1} + \cD^{(-\frac{1}{2})}_{-2n+1} \label{kdv1} .
					\end{split}
				\end{equation}
				Considering the subsequent negative \textit{even}  flow  $ A_{t_{-2n}}^{\smkdv}$, the lower grade operator is  proportional to   $D^{(-2n)}_{-2n} \sim M_1^{(-2n)}$,  leaving the  algebraic structure invariant, i.e.,
				\be
				\begin{split}
					\widetilde{A}_{t_{-2n+1}}^{\skdv} & \equiv  \cS \left( D^{(-2n)}_{-2n}+ D^{(-2n+\frac{1}{2})}_{-2n}+\cdots +D^{(-\frac{1}{2})}_{-2n} \right) \cS^{-1} + \cS \pa_{t_{-2n}}\cS^{-1}  \\
					&= \widetilde{\cal D}^{(-2n-1)}_{-2n+1} + \widetilde{\cal D}^{(-2n-\frac{1}{2})}_{-2n+1} +  \widetilde{\cal D}^{(-2n)}_{-2n+1} +\cdots + \widetilde{\cal D}^{(-1)}_{-2n+1}+\widetilde{\cal D}^{\left(-\frac{1}{2}\right)}_{-2n+1}.
				\end{split}
				\label{kdv2}
				\ee
				Since  the potentials $A_x^{\skdv}$ and $ A_x^{\smkdv} $ are universal within the hierarchies, 
				the zero curvature  condition for \eqref{kdv1} and \eqref{kdv2} must yield the same
				operator, i.e., 
				\be \label{equal_op}
				{A}_{t_{-2n+1}}^{\skdv}=\widetilde{A}_{t_{-2n+1}}^{\skdv}  ,
				\ee
				and the two  gauge potentials provide the  same  evolution equations. Therefore,  a negative even and its subsequent odd flow collapse  into 
				the same negative odd sKdV flow, which is consistent with the fact that there is no \textit{even} negative flow within the sKdV hierarchy, i.e.,
				\be\label{corresp.neg.sup}
				\begin{tikzcd}[row sep=0.1em, column sep=3em]
					t_{-N}^{\smkdv} \arrow[dr, "\cS"] & \\
					&  t_{-N}^{\skdv} \\ 
					t_{-N-1}^{\smkdv} \arrow[ur, swap, "\cS"] 
				\end{tikzcd}
				\ee
				for $N = 2n - 1$, $n \in \mathbb{Z}^{+}$.
				Let us consider the explicit example of  $A_{t_{-1}}^{\smkdv}$ and $A_{t_{-2}}^{\smkdv}$  given by \eqref{smkdv.tm1.lax} and \eqref{smkdv.tm2.lax} in Appendix \ref{ap.laxpairs}. 
				The mapping  
				\begin{equation}
					A_{t_{-1}}^{\skdv} = \cS A_{t_{-1}}^{\smkdv} \cS^{-1} + \cS \pa_x \cS^{-1},
					\label{gauge.smiura.tm1}
				\end{equation}
				implies  the following relations among  the \textit{smKdV} and 
				the  \textit{sKdV} fields, 
				\begin{subequations} \label{smiura.tm1}
					\begin{align}
						\eta_{t_{-1}}&=2\left(e^{-2\phi}+e^{-\phi} \psi\bp\right) , \label{smiura.tm1.b}\\
						\bg_{t_{-1}}&=2e^{-\phi} \psi, \label{smiura.tm1.f}
					\end{align}
				\end{subequations}
				where the smKdV fields in the r.h.s are solutions of $t_{-1}$ eqns. (\ref{smkdv.tm1}). However, if the map follows $t_{-2}^{\smkdv}$ into  $t_{-1}^{\skdv}$, i.e.,
				\begin{equation}
					A_{t_{-1}}^{\skdv} = \cS A_{t_{-2}}^{\smkdv} \cS^{-1} + \cS \pa_x \cS^{-1},
					\label{gauge.smiura.tm2}
				\end{equation}
				this relation requires 
				\begin{subequations} \label{smiura.tm2}
					\begin{align}
						\eta_{t_{-1}}&=4\left(a_{-}+ \Om_{-}\bp+\frac{1}{2} \psi_{+} \psi_{-}\right),\label{smiura.tm2.b}\\
						\bg_{t_{-1}}&=4 \Om_{-},\label{smiura.tm2.f}
					\end{align}
				\end{subequations}
				where $\psi_{\pm}$, $a_{\pm}$ and $\Om_{\pm}$ are given by \eqref{psi.pm}, \eqref{a.pm} and \eqref{om.pm} and the fields satisfy (\ref{smkdv.tm2}).
				Notice that such set of relations involving KdV fields,  $(\eta,\bg)$ define a distinct set of  {solutions} for equation \eqref{skdv.tm1}. This allows us to determine a larger class of solutions for the negative flows within the {sKdV} hierarchy and will be  quite  useful  in determining the Heisenberg sKdV subalgebra as follows in Appendix \ref{ap.vacuumkdv}. Moreover, one of the most remarkable applications of such relations is to allow us to directly verify supersymmetry of  \eqref{skdv.tm1}.  It is  straightforward  to show that combining the reparametrization $J=\pa_x \eta$ and $\bchi= \pa_x \bg$ together with relations \eqref{susy.transf.kdv}, we have
				\begin{subequations} \label{smiura.susy}
					\begin{align}
						\delta \bchi &= \xi J \qquad \quad \!\!\rightarrow \quad \delta \bg = \xi \eta + f(t),\\
						\delta J &= -\xi \pa_x \bchi \quad \rightarrow \quad \delta \eta = -\xi \pa_x \bg + g(t).
					\end{align}
				\end{subequations}
				For the positive sKdV equations, $f(t)$ and $g(t)$ has no relevance, since  there  is always  at least one derivative with respect  to $x$ acting upon $\eta$ or  $\bg$. Nevertheless, for sKdV$(-1)$ and other negative flow equations, there will  always be an additional  explicit relations involving temporal derivatives like  \eqref{smiura.tm1} or \eqref{smiura.tm2} to take care of those terms. Consider for instance sKdV$(-1)$ as gauge transformed from the sshG and use \eqref{smiura.susy} together with \eqref{smiura.tm1}, it is possible to determine $f(t)$ and $g(t)$ to obtain
				\begin{subequations} \label{smiura.susy.tm1}
					\begin{align}
						\delta \bg &= \xi \eta - 2 \xi t,\\
						\delta \eta &= -\xi \pa_x \bg.
					\end{align}
				\end{subequations}
				Furthermore, in order to compute the supersymmetric transformations of the auxiliary fields $\g$ and $\nu_\pm$ \eqref{skdv.tm1.aux} sKdV$(-1)$, we will need to use \eqref{smiura.x.s.1} and \eqref{smiura.tm1} to get the expressions
				\begin{subequations}
					\begin{align} \label{auxv.1}
						\g &= 2 e^{-\phi} \bp\psi,\\
						\Bar{\nu_-} -\Bar{\nu_{+}}+2\bg&= 2\bp- 2(\pa_x\phi) e^{-\phi} \psi - 2e^{-2\phi} \bp ,
					\end{align}
				\end{subequations}
				and finally obtain the variations
				\begin{subequations} \label{smiura.susy.tm1.aux}
					\begin{align}
						\delta \g &= \xi \left(\nu_- -\nu_{+}+2\bg\right),\\
						\delta \left(\Bar{\nu_-} -\Bar{\nu_{+}}+2\bg\right)&= -\xi \pa_x \g.
					\end{align}
				\end{subequations}
				It is clear that this pair acts as supersymmetric variables among themselves. Finally, after a straightforward but long calculation, by using \eqref{smiura.susy.tm1} and \eqref{smiura.susy.tm1.aux}, it is possible to show that the pair of equations \eqref{skdv.tm1} is supersymmetric invariant.
				On the other hand, if we had used \eqref{smiura.tm2}, the procedure will follow the same steps, with a minor adjustment in the auxiliary variables, which will now be given by
				\begin{subequations}
					\begin{align} \label{auxv.2}
						\g &= 4 \bp \Om_{-}-2 \psi_{+}\psi_{-}\\
						\Bar{\nu_-} -\Bar{\nu_{+}}+2\bg&= 2\bp+ 2 e^{\phi}\psi_{-}+ 2 e^{-\phi}\psi_{+}+ 2\bp \psi_{-}\psi_{+} - 4\bp a_{-} - 4\pa_x \phi \Om_{-}.
					\end{align}
				\end{subequations}
				This however,  will not introduce any changes to the supersymmetry relations \eqref{smiura.susy.tm1} and \eqref{smiura.susy.tm1.aux}, leading to the same result, i.e., the sKdV($-1$) flow is a supersymmetric. In this way, the importance of the gauge-Miura mapping in determining sKdV properties in terms of smKdV becomes evident once again.
				
				Despite the fact that the temporal Miura is an essential ingredient and  highly non-trivial relation, 
				it is possible to obtain a general formula for any flow using both the zero curvature equation for sKdV and the super Miura transformation. Let us consider an arbitrary negative sKdV flow. The ZCC decomposition in appendix \ref{ap.gradezcc} give us eqns. \eqref{10g} and \eqref{10f} that set both  $\cD_{-N}^{\left(-\frac{1}{2}\right)}$  and $\cD_{-N}^{(-1)}$in the \textit{Kernel} of $\cE$. Next, one can use the equations \eqref{10e} and \eqref{10d} to obtain the following form for the lowest elements $\cD_{-N}$ for any $N$, namely \footnote{Here, we have chosen the integration constant appearing in the $K_2$ term to be 1. We also recall that $\pa_x \g = \pa_x \bg \pa_{t_{-N}} \bg$}
				\begin{equation} \label{mt1.k}
					\cD_{-N}^{(-\frac{1}{2})}[\bg,\eta] = \frac{\pa_{t_{-N}} \bg(x,t)}{2} F_2^{\left(-\frac{1}{2}\right)}
				\end{equation}
				and
				\begin{equation} \label{mt2.k}
					\cD_{-N}^{(-1)}[\bg,\eta] = \frac{1}{2}\left(\g +\pa_{t_{-N}} \eta(x,t) \right)K_1^{(-1)}+\frac{1}{2}\left(\g +2 \right)K_2^{(-1)}.
				\end{equation}
				Additionally, the super Miura transformation \eqref{smiura.exp} give us the following results for a generic mapping upon smKdV($-M$) flow to  sKdV($-N$) flow for the minus one-half graded  ($\cD_{-N}^{(-\frac{1}{2})}[\bg,\eta]$) element
				\begin{equation}
					\cD_{-N}^{(-\frac{1}{2})}[\bg,\eta]= D_{-M}^{(-\frac{1}{2})}[\bp,\phi]+\frac{\pa_{t_{-M}}\bp}{2} \left(F_2^{(-\frac{1}{2})}+G_1^{(-\frac{1}{2})}\right).
				\end{equation}
				Using \eqref{2d},  the equation of motion for $\bp$ is determined 
				\begin{equation}
					\pa_{t_{-M}}\bp =-2 b_{-M}^{(-\frac{1}{2})}
				\end{equation}
				where
				\begin{equation}
					D_{-M}^{(-\frac{1}{2})}[\bp,\phi] = a_{-M}^{(-\frac{1}{2})} F_2^{(-\frac{1}{2})}+b_{-M}^{(-\frac{1}{2})} G_1^{(-\frac{1}{2})}.
				\end{equation}
				Combining these two equations, it is easy to see that 
				\begin{equation}\label{mt1.m}
					\cD_{-N}^{(-\frac{1}{2})}[\bg,\eta]= \left( a_{-M}^{(-\frac{1}{2})} -b_{-M}^{(-\frac{1}{2})}\right) F_2^{(-\frac{1}{2})}
				\end{equation}
				and due  to the Lax pair uniqueness, one can compare \eqref{mt1.k} and \eqref{mt1.m} to obtain the first temporal super Miura
				\begin{equation} \label{gen.mt.1}
					\pa_{t_{-N}} \bg(x,t)= 2\left( a_{-M}^{(-\frac{1}{2})} -b_{-M}^{(-\frac{1}{2})}\right). 
				\end{equation}
				The other one is obtained in the same way, by using the results obtained from the super Miura transformation for the grade minus one element, $\cD_{-N}^{(-1)}[\bg,\eta] $ :
				\begin{equation} \label{mt2.m}
					\cD_{-N}^{(-1)}[\bg,\eta] = \left(a_{-M}^{(-1)} +c_{-M}^{(-1)} \right)K_1^{(-1)}+\frac{1}{2}\left(1+b_{-M}^{(-1)} +\bp a_{-M}^{(-\frac{1}{2})} -\bp b_{-M}^{(-\frac{1}{2})} \right)K_2^{(-1)}.
				\end{equation}
				where 
				\begin{equation}
					D_{-M}^{(-1)}[\bp,\phi] = a_{-M}^{(-1)}K_1^{(-1)}+(1+b_{-M}^{(-1)})K_2^{(-1)}+c_{-M}^{(-1)}M_2^{(-1)}
				\end{equation}
				and when compared with  \eqref{mt2.k}  yields
				\begin{equation} \label{gen.mt.2}
					\pa_{t_{-N}} \eta(x,t) = 2(a_{-M}^{(-1)} +c_{-M}^{(-1)}) +\pa_{t_{-N}} \bg(x,t) \bp.
				\end{equation}
				It therefore follows that,  knowing  the elements $D_{-M}^{(-1)}[\bp,\phi]$ and $D_{-M}^{(-\frac{1}{2})}[\bp,\phi]$ of the smKdV hierarchy 
				we are able to determine the temporal Miura relations (consider the notation $f(x,t_{N}) \equiv f_{N}$ for simplicity):
				\be \label{mtgen2}
				\pa_{t_{-2n+1}} \Bar{\g} = 
				\begin{cases} 
					2\left(a_{-2n+1}^{(-\frac{1}{2})} -b_{-2n+1}^{(-\frac{1}{2})}\right) \left[ \phi_{-2n+1},\bp_{-2n+1} \right]  & \mbox{for odd smKdV flows,}  \\  
					2\left(a_{-2n}^{(-\frac{1}{2})} -b_{-2n}^{(-\frac{1}{2})}\right)\left[ \phi_{-2n},\bp_{-2n} \right]  & \mbox{for even smKdV flows.}
				\end{cases}
				\ee
				\\
				\be \label{mtgen1}
				\pa_{t_{-2n+1}} \eta = 
				\begin{cases} 
					2\left(a_{-2n+1}^{(-1)} +c_{-2n+1}^{(-1)}\right)\left[ \phi_{-2n+1},\bp_{-2n+1} \right] + \pa_{t_{-2n+1}} \bg \; \bp_{-2n+1} & \mbox{for odd smKdV}  \\[-0.2cm] \mbox{} & \mbox{flows,}  \\[0.2cm]
					2\left(a_{-2n}^{(-1)} +c_{-2n}^{(-1)}\right) \left[ \phi_{-2n},\bp_{-2n} \right] + \pa_{t_{-2n+1}} \bg \; \bp_{-2n} & \mbox{for even smKdV} \\[-0.2cm] \mbox{} & \mbox{flows.} 
				\end{cases}
				\ee
				As examples, eqns. \eqref{smiura.tm1.b}, \eqref{smiura.tm1.f},  \eqref{smiura.tm2.b} and \eqref{smiura.tm2.f}, are direct applications of this relation.
				Now we have established all the structure concerning the equations of motion for both the smKdV and sKdV hierarchy as well as the ingredients to map the equations of motion of one into another. 
				In next section, we will consider the remaining ingredient, namely, the solution of the field equations.
				
				
				{\color{black}
					\section{The  dressing method and tau functions for $sl(2,1)$}
					\label{sec.dressing}
					
					We shall now discuss the construction of   soliton solutions for  smKdV and sKdV hierarchies. In order to do that, we will introduce the \textit{dressing method}  and classify each hierarchy  according to its vacuum orbit.
					\noindent
					Let us start with a brief compilation of the most important ideas of  this method. The main point of the {dressing} method is to start with a simple solution  ({vacuum solution or vacuum orbit}) to obtain the non-trivial  (multi)-soliton solutions for the entire hierarchy. Let us define 
					\begin{equation}
						{A}^{\text{vac}}_{\mu}=A_{\mu}\big|_{v=v_{\text{vac}}}
					\end{equation}
					where $\mu=x,t_N$, and $v_{\text{vac}}$ is the vacuum solution solution for the zero curvature condition \eqref{zcc}. We use the  fact that ZCC is invariant under a gauge transformation  $\Theta_{\pm}$ to map the vacuum into  some non-trivial  soliton solution, i.e.,
						\begin{equation} \label{gauge.zcc}
							A_\mu= \Theta_{\pm}(\partial_\mu + {A}^{\text{vac}}_\mu)\Theta_{\pm}^{-1}.
						\end{equation}
						The plus-minus label upon $\Theta$ indicates that it can be decomposed either as a exponential of positive or negative graded elements of the algebra,i.e.,
						\begin{equation} \label{expansion.gauge}
							\Theta_+= \prod_{n\geq0} \exp\left(\theta_{n/2}\right), \qquad 
							\Theta_-=\prod_{n>0} \exp\left(\theta_{-n/2}\right),
						\end{equation}
						where $\theta_a \in \lie_a$. Another important property is the fact that the Lax potentials can be written as a pure gauge potential
						\begin{equation} \label{pure.gauge}
							{A}^{\text{vac}}_\mu=T_0\partial_\mu {T_0}^{-1}, \quad A_\mu=T\partial_\mu {T}^{-1}
						\end{equation}
						where
						\begin{equation} \label{pure.gauge.decomp}
							T_0=e^{-xA_x^{\text{vac}}}\,e^{-tA_t^{\text{vac}}} \qquad \text{and} \qquad T=\Theta_{\pm} T_0.
						\end{equation}
						Combining \eqref{gauge.zcc}, \eqref{pure.gauge} and \eqref{pure.gauge.decomp}, we can show that the following condition holds,
						\begin{equation} \label{gauge.fun.rel}
							{\Theta_-^{-1}}\Theta_+=T_0\,g\,T_0^{-1}. 
						\end{equation}
						where $g$ is a constant group element.
						Let us  introduce highest weight representation (based upon the Kac-Moody algebra $\lie$) such that
						\begin{equation} 
							\theta_i|\lambda_j\rangle=0, \quad \text{with} \quad i>0.
						\end{equation}
						Consider the  matrix elements $\tau_{ij}(\boldsymbol{v})$
						\begin{equation}
							\tau_{ij}(\boldsymbol{v},\s) \equiv \langle \lambda_i| \lie_\s{\Theta_-^{-1}}\Theta_+|\lambda_j\rangle,
						\end{equation}
						where $\lie_\s$ is a properly chosen algebra element of grade $\s$. On the other hand, using \eqref{gauge.fun.rel}  we can also write the following equivalence
						\begin{equation}\label{eq4.9}
							\tau_{ij}(\boldsymbol{v},\s)= \langle \lambda_i| \lie_\s T_0\,g\,T_0^{-1}|\lambda_j\rangle .
						\end{equation}
						Now, let us consider the constant element $g$ in the following form,  
						\begin{equation} \label{vertex.def}
							g= \prod_{i=1}^{n} \exp(V(k_i)) \quad \text{such}  \quad \comm{A_{\mu}^{\text{vac}}}{ V(k_i)}=\om_\mu (k_i) V(k_i),
						\end{equation}
						where $V(k_i)$ is the so-called \textit{vertex operator} \cite{babelon_dressing_1992, babelon_affine_1993, hollowood_tau-functions_1993}. It there follows that expression \eqref{gauge.fun.rel} became rather simple
						\begin{equation}
							T_0\,g\,T_0^{-1}=  \prod_{i=1}^{n}\exp(  \rho_i(t_N,x) V(k_i) ),
						\end{equation}
						with
						\begin{equation}
							\rho_i(t_N,x) \equiv \exp(-\om_{t_N}(k_i) t_N-\om_x(k_i) x).
						\end{equation}
						Substituting the above expressions in (\ref{eq4.9}),  we find
						\begin{equation}
							\tau_{ij}(\boldsymbol{v},\s)=\langle  \lambda_i| \lie_{\s}\prod_{i=1}^{n} \exp(\rho_i(t_N,x) V(k_i))|\lambda_j\rangle
						\end{equation}
						and assuming that  $V(k_i)$ is a nilpotent  operator, we have
						\begin{equation}
							\tau_{ij}(\boldsymbol{v},\s) =\langle \lambda_i|\prod_{i=1}^{n}\left(\lie_{\s}+ \rho_i \lie_{\s}V(k_i)+\cdots+\frac{1}{m!}\rho_i^m \lie_{\s} V^m(k_i)\right)|\lambda_j\rangle
						\end{equation}
						where $m$ depends on the vertex operator and $n$ is chosen to match the number of solitons.  For instance, for models based on the $sl(2,1)$ superalgebra, $m=2$, and we are mainly interested in one-soliton solution, i.e. $n=1$, so we can simplify the expression above as 
						\begin{equation} \label{tau.vertex}
							\tau_{ij}(\boldsymbol{v},\s)=\langle \lambda_i|\lie_\s \Theta_-^{-1}\Theta_+|\lambda_j\rangle=\langle \lambda_i|\lie_\s|\lambda_j\rangle +\langle \lambda_i| \lie_\s V(k_i)|\lambda_j\rangle \rho(t,x).
						\end{equation}
						Now, let us do some practical calculations concerning the $sl(2,1)$ smKdV case. First, we establish the fundamental weight system as follow 
						\begin{equation}
							\begin{split}
								\hat{\kappa} \ket{\lambda_i} &= \ket{\lambda_i}, \\
								M_1^{(0)} \ket{\lambda_i} &= \delta_{i, 1} \ket{\lambda_i}, \\
								\lie^{(n / 2)} \ket{\lambda_i} &= 0,
							\end{split}
							\qquad
							\qquad
							\begin{split}
								\left\langle\lambda_i\right| \hat{\kappa}  &=\left\langle\lambda_i\right|, \\
								\left\langle\lambda_i\right| M_1^{(0)}  &=\left\langle\lambda_i\right| \delta_{i, 1}, \\
								\left\langle\lambda_i\right|\lie^{(-\frac{n}{2})}  &=0,
							\end{split}
							\qquad
							\qquad
							\begin{split}
								(n = 1,2,3 \dots)
							\end{split}
						\end{equation}
						and consider the spatial Lax for super mKdV hierarchy \footnote{We have added a central extension contribution $\nu_x \kappa$ in order to complete the model, but the $\nu$ field has no physical relevance.}:
						\begin{equation} \label{ax.mkdv}
							A_x^{\text{smKdV}} = \cE+ A_0(\phi, \nu)+A_{\frac{1}{2}}(\bar{\psi})=K_1^{(1)}+K_2^{(1)}+\pa_x \phi_1 \; M_1^{(0)}+\pa_x \nu \kappa+ \bar{\psi} G_2^{(\frac{1}{2})}.
						\end{equation}
						Consider the most general vacuum projection $(\nu, \phi,\bar{\psi})=(0,v_0x,\bar{\psi}_0 )$ 
						\begin{equation} \label{ax.vac.mkdv}
							A_{x,\text{vac}}^{\text{smKdV}} = K_1^{(1)}+K_2^{(1)}+v_0\; M_1^{(0)}+ \bar{\psi_0} G_2^{(\frac{1}{2})}.
						\end{equation}
						Since the field appears in grade $0$ and $\frac{1}{2}$ terms, we propose $\tau$ functions with projections on these grades. The first obvious choice for the grade zero term is 
						\begin{subequations}
							\begin{align}
								\tau_{00}(\boldsymbol{v},0)\equiv\tau_0&=\langle \label{tau.0} \lambda_0|\Theta_-^{-1}\Theta_+|\lambda_0\rangle=\langle \lambda_0|e^{\theta_{0}}|\lambda_0\rangle,\\ \label{tau.1}
								\tau_{11}(\boldsymbol{v},0)\equiv\tau_1&= \langle \lambda_1|\Theta_-^{-1}\Theta_+|\lambda_1\rangle=\langle \lambda_1|e^{\theta_{0}}|\lambda_1\rangle.
							\end{align}
						\end{subequations}
						In turn, since the matrix element must have zero grade in general, and it must project on grade $\frac{1}{2}$ in order to extract some information of the fermionic $\bp$ field, we will propose the following projections,
						\begin{equation}
							\begin{split}
								\langle \lambda_i|\lie_{\frac{1}{2}}\Theta_-^{-1}\Theta_+|\lambda_i\rangle&=\langle \lambda_i|\lie_{\frac{1}{2}} \left(1 -\theta_{-\frac{1}{2}} +\mathcal{O}^2(\theta_{-\frac{1}{2}})\right)e^{\theta_{0}}|\lambda_i\rangle \\
								&= -\langle \lambda_i|\lie_{\frac{1}{2}} \theta_{-\frac{1}{2}}e^{\theta_{0}}|\lambda_i\rangle
							\end{split}
						\end{equation}
						and define
						\begin{subequations}
							\begin{align} \label{tau.2}
								\tau_{00}\left(\boldsymbol{v},\frac{1}{2}\right)\equiv \tau_2&= \langle \lambda_0|\lie_{\frac{1}{2}}\Theta_-^{-1}\Theta_+|\lambda_0\rangle=-\langle \lambda_0|\lie_{\frac{1}{2}} \theta_{-\frac{1}{2}}e^{\theta_{0}}|\lambda_0\rangle \quad \text{and} \quad\\ \label{tau.3}
								\tau_{11}\left(\boldsymbol{v},\frac{1}{2}\right)\equiv	\tau_3&= \langle \lambda_1|\lie_{\frac{1}{2}}\Theta_-^{-1}\Theta_+|\lambda_1\rangle=-\langle \lambda_1|\lie_{\frac{1}{2}} \theta_{-\frac{1}{2}}e^{\theta_{0}}|\lambda_1\rangle.
							\end{align}
						\end{subequations}
						Now, we use eqs. \eqref{gauge.zcc}, \eqref{expansion.gauge} for $\Theta_{+}$, and \eqref{ax.vac.mkdv}, in order to establish the relation between $\theta_0$, $\theta_{-\frac{1}{2}}$, and the fields of the model. Then, from the grade zero term, we have 
						\begin{equation}
							(\pa_x\phi) M_1^{(0)}+ (\pa_x \nu) \hat{\kappa} =  -\pa_x \theta_0+ e^{\theta_0} \left(v_0 M_1^{(0)}\right)e^{-\theta_0}   
						\end{equation}
						with solution given by
						\begin{equation}
							e^{-\theta_0}=e^{(\phi-v_0x) M_1^{(0)}+\nu \hat{\kappa}},
						\end{equation}
						and then
						\begin{eqnarray} \label{tau.0.f}
							\tau_0&=& \langle \lambda_0|\Theta_-^{-1}\Theta_+|\lambda_0\rangle=e^{-\nu}\\
							\tau_1&=& \langle \lambda_1|\Theta_-^{-1}\Theta_+|\lambda_1\rangle=e^{-\nu-\phi+v_0x},   \label{tau.1.f}
						\end{eqnarray}
						from where we find  the bosonic field  written as
						\begin{equation}  \label{tau.boson}
							\phi(x,t) = v_0 x + \ln \left( \frac{\tau_{0}(x,t)}{\tau_{1}(x,t)} 
							\right).
						\end{equation}
						Of course, we could continue the calculation of the following higher grades terms in order to completely determine $\Theta_{+}$. However, since we are mainly interested in determining the field components, we now apply the same procedure using eqs. \eqref{gauge.zcc}, \eqref{expansion.gauge}, \eqref{ax.mkdv} and \eqref{ax.vac.mkdv} with $\Theta_{-}$. The simplest projection is given on the grade $\frac{1}{2}$ term, 
						\begin{equation}
							A_{\frac{1}{2}}(\bp)=[ \theta_{-\frac{1}{2}},\cE]+A_{\frac{1}{2},\text{vac}}(\bp_0)
						\end{equation}
						from which we get
						\begin{equation}
							\theta_{-\frac{1}{2}}=f_{1,-\frac{1}{2}}F_2^{(-\frac{1}{2})}+\frac{1}{2}(\bp-\bp_0)G_1^{(-\frac{1}{2})}.\label{eq4.28}
						\end{equation}
						The others projections are slightly more complicated, given by
						\begin{eqnarray}
							A_{0} &=&[ \theta_{-\frac{1}{2}},A_{\frac{1}{2},\text{vac}}]+\frac{1}{2}[\theta_{-\frac{1}{2}},[ \theta_{-\frac{1}{2}},\cE]]+[ \theta_{-1},\cE]+A_{0,\text{vac}}
							\\[0.2cm]
							\pa_x\theta_{-\frac{1}{2}}&=& [ \theta_{-\frac{1}{2}},A_{0,\text{vac}}]+[ \theta_{-\frac{3}{2}},\cE]+[\theta_{-\frac{1}{2}},[ \theta_{-1},\cE]]\nonumber\\			&&+\frac{1}{2}[\theta_{-\frac{1}{2}},[ \theta_{-\frac{1}{2}},A_{\frac{1}{2},\text{vac}}]]+\frac{1}{3!}[\theta_{-\frac{1}{2}},[ \theta_{-\frac{1}{2}},[ \theta_{-\frac{1}{2}},\cE]]]+[ \theta_{-1},A_{\frac{1}{2},\text{vac}}]
							\\[0.2cm]
							\pa_x \theta_{-1}&=&[ \theta_{-2},\cE]+[ \theta_{-1},A_{0,\text{vac}}]+[ \theta_{-\frac{3}{2}},A_{\frac{1}{2},\text{vac}}]+\frac{1}{2}[\pa_x \theta_{-\frac{1}{2}},\theta_{-\frac{1}{2}}]+\frac{1}{2}[\theta_{-1},[\theta_{-1},\cE]]\nonumber\\
							&&+[\theta_{-\frac{1}{2}},[\theta_{-\frac{3}{2}},\cE]]+\frac{1}{2}[\theta_{-\frac{1}{2}},[\theta_{-\frac{1}{2}},A_{0,\text{vac}}]]+[\theta_{-\frac{1}{2}},[\theta_{-1},A_{\frac{1}{2},\text{vac}}]]\nonumber\\
							&&+\frac{1}{2}[\theta_{-\frac{1}{2}}, [\theta_{-\frac{1}{2}},[\theta_{-1},\cE]]]+\frac{1}{3!}[\theta_{-\frac{1}{2}}, [\theta_{-\frac{1}{2}},[\theta_{-\frac{1}{2}},A_{\frac{1}{2},\text{vac}}]]]\nonumber\\&&+\frac{1}{4!}[\theta_{-\frac{1}{2}}, [\theta_{-\frac{1}{2}},[\theta_{-\frac{1}{2}},[\theta_{-\frac{1}{2}},\cE]]]].
						\end{eqnarray}
						Thus, by proposing a general form for the other $\theta$ components appearing in the equations, namely
						\begin{equation}
							\begin{split}
								\theta_{-1}&= f_{1,-1} K_1^{(-1)}+f_{2,-1} K_2^{(-1)}+f_{3,-1} M_2^{(-1)}\\
								\theta_{-\frac{3}{2}}&=f_{1,-\frac{3}{2}} F_1^{(-\frac{3}{2})}+f_{2,-\frac{3}{2}} G_2^{(-\frac{3}{2})}\\
								\theta_{-2}&= f_{1,-2} h_1 M_2^{(-2)}
							\end{split}
						\end{equation}
						we find the following set of equations
						\begin{subequations}
							\begin{align}
								\pa_x\phi &\,=f_{1,-\frac{1}{2}}\bp+2 f_{1,-\frac{1}{2}}\bp _0+2f_{3,-1}+v_0 \label{g01}\\[0.2cm]
								\pa_x \nu &\,= -f_{1,-1}+f_{2,-1}-f_{3,-1}-\frac{1}{2}f_{1,-\frac{1}{2}}\bp-f_{1,-\frac{1}{2}}\bp_0+\frac{1}{2}\bp \bp_0 \label{g02}\\[0.2cm]
								\pa_x f_{1,-\frac{1}{2}}&\,= \bp f_{3,-1} + \frac{ v_0}{2}(\bp-\bp_0) -\frac{1}{2}f_{1,-\frac{1}{2}}\bp_0 \bp \label{g121}\\[0.2cm]
								\pa_x \bp &\, = 2 f_{1,-\frac{1}{2}} v_0+4f_{2,-\frac{3}{2}}+4f_{1,-\frac{1}{2}} f_{3,-1} -2(f_{1,-1}+f_{2,-1})\bp_0. \label{g122}
							\end{align}
						\end{subequations}
						Since, we are mainly interested in determine $\theta_{-\frac{1}{2}}$,  let us consider equation \eqref{g01} and \eqref{g121}, to find the following  differential equation (where we have renamed $f_{1,-\frac{1}{2}}\equiv \frac{\Xi}{2}$ for simplicity) 
						\begin{equation} \label{eqx}
							(\pa_x\phi) \bp \,=\pa_x \Xi - \bp_0 v_0 +\frac{3\Xi}{2}\bp_0 \bp
						\end{equation}
						which completely determine $\Xi$ in terms of the $(\phi, \bar\psi)$ fields for a given vacuum configuration (see Appendix \ref{dif.dress}). 
						Once we have determined $ \theta_{-\frac{1}{2}}$, we find ourselves in  position to write the functional form of $\bp$ in terms of $\tau$ functions. From eqs. \eqref{tau.2} and \eqref{tau.3}, we have
						\begin{subequations}
							\begin{align}
								\tau_2 &=-\langle \lambda_0|\lie_{\frac{1}{2}} \theta_{-\frac{1}{2}}|\lambda_0\rangle e^{-\nu} = - \langle \lambda_0|\lie_{\frac{1}{2}} \theta_{-\frac{1}{2}}|\lambda_0\rangle \tau_0 \\
								\tau_3 &=-\langle \lambda_1|\lie_{\frac{1}{2}} \theta_{-\frac{1}{2}}|\lambda_1\rangle e^{-\nu-\phi+v_0x}=-\langle \lambda_1|\lie_{\frac{1}{2}} \theta_{-\frac{1}{2}}|\lambda_1\rangle \tau_1 ,
							\end{align}
						\end{subequations}
						and by choosing $\lie_{\frac{1}{2}}= F_1^{(\frac{1}{2})}$, from eq.  \eqref{eq4.28}, we find that the $\tau$ functions are given by
						\begin{eqnarray}
							\tau_2 &=&  \frac{1}{2}\left(\Xi - (\bp-\bp_0) \right) \tau_0\label{tau.2.f}\\
							\tau_3 &=&  \frac{1}{2}\left(\Xi + (\bp-\bp_0) \right) \tau_1
							\label{tau.3.f}  
						\end{eqnarray} 
						and, the fermionic field  given in the following form
						\begin{equation} \label{tau.fermion}
							\bp =\bp_0 +\frac{\tau_3}{\tau_1} - \frac{\tau_2}{\tau_0}.
						\end{equation}
						Having established the relationship between the fields of the model and the $\tau$  functions, we can now compute the $\tau$  functions using equation \eqref{tau.vertex}. The next natural step is to determine the vertex operators as described in equation \eqref{vertex.def}. However, the vertex operator is highly dependent on the vacuum configuration accepted by the model. Therefore, we will first focus on understanding the possible vacuum states for each flow.		
						
						
						\section{Heisenberg subalgebras and the commutativity of flows}
						\label{heisenberg.commuting.flows}
						
						Let us now discuss the different  vacuum configurations  and its implications in the commutativity of the flows.  Consider the zero curvature in the vacuum configuration, 
						\begin{equation} \label{zc.vac}
							\comm{A_{\text{$x$,vac}}}{A_{\text{$t_{N}$,vac}}}=0,
						\end{equation}
						for both smKdV and sKdV hierarchies. Firstly, it is important to remark that  the higher graded component  (positive flows) and the lowest graded component (negative flows) $A_{t_{N}}$ for the smKdV case, is always non vanishing (including the vacuum configuration). Indeed, one can use \eqref{highgrade.smkdv}  and \eqref{lowgrade.smkdv} to show that for each flow the higher graded element is always determined and non-zero, namely
						\begin{itemize}
							\item Positive odd $N=2m+1$:
							\begin{equation} \label{vacua.proj.high.a}
								D^{(2m+1)}= \mathcal{E}^{(2m+1)}
							\end{equation}
							\item Negative even $N=-2m$:
							\begin{equation} \label{vacua.proj.high.b}
								D^{(-2m)}= M_1^{(2m)}
							\end{equation}
							
							\item Negative odd $N=-2m+1$: 
							\begin{equation}\label{vacua.proj.high.c}
								D^{(-2m+1)}=  \cosh\phi K_1^{(-2m+1)} +K_2^{(-2m+1)}.
							\end{equation}
						\end{itemize}
						Starting from the simplest case, i.e,  a {positive} flow projected in a general constant vacuum $\left(v_0, \bp_0\right)$  configuration, we find the following relation from eqn. \eqref{zc.vac}, 
						\begin{equation} \label{zc.vac.mkdvp}
							\comm{A_{\text{$x$,vac}}^{\text{mKdV}}}{A_{\text{$t_{N}$,vac}}^{\text{mKdV}}}=\comm{\cE+v_0 M_1^{(0)}+\bp_0 G_2^{\frac{1}{2}}}{ D^{(N)}_{N,\text{vac}} + D^{(N-\frac{1}{2})}_{N,\text{vac}} + \cdots + D^{(0)}_{N,\text{vac}}}=0.
						\end{equation}
						Therefore, we must find the kernel $\mathcal{K}_\Om$ 
						of the operator $\Om \equiv \cE+v_0 M_1^{(0)}+\bp_0 G_2^{\frac{1}{2}} $, and analyse whether   it coincides with the vacuum projection for the positive flows. Such kernel is given by:
						\begin{equation}
							\mathcal{K}_\Om = \left\{\Omega_{2m+1}, \Gamma_{2m+1}\right\} \quad \text{such} \quad \comm{X}{\cE+v_0 M_1^{(0)}+\bp_0 G_2^{\frac{1}{2}}}=0, \; \quad X \in \mathcal{K}_\Om ,
						\end{equation}
						with 
						\begin{eqnarray} \label{o1}
							\Omega_{2m+1} &=& v_0 M_1^{(2m)}+\bp_0 G_2^{(2m+\frac{1}{2})}+ K_1^{(2m+1)}+K_2^{(2m+1)},\\
							\Gamma_{2m+1} &=& K_2^{(2m+1)}+\frac{\bp_0}{v_0}F_2^{(2m+\frac{3}{2})}. \label{g1}
						\end{eqnarray}
						It is straightforward to show that the vacuum projection of the  smKdV  Lax operators given in the appendix \ref{ap.laxpairs} are linear combinations of the $	\mathcal{K}_\Om$ elements, namely
						\begin{subequations}
							\begin{align} \label{avac1}
								A_{t_1,\text{vac}}^{\text{smKdV}}&=A_{x}^{\text{vac}}= \Om_1\\ \label{avac3}
								A_{t_3,\text{vac}}^{\text{smKdV}}&= \Om_3+\frac{v_0^2}{2}\left(\Gamma_{1}- \Om_{1}\right)\\ \label{avac5}
								A_{t_5,\text{vac}}^{\text{smKdV}}&= \Om_{5}+\frac{v_0^2}{2} \left(\Gamma_{3}- \Om_{3}\right)+\frac{3v_0^4}{8} \left(\Om_{1}-\Gamma_{1}\right)\\
								\;\; \; \; \; \vdots \nonumber\\ \label {avac.gen.nz.p}
								A_{t_{2m+1},\text{vac}}^{\text{smKdV}}&= \Om_{2m+1}+ \sum_{i=0}^{m-1} v_0^{2(m-i)} \left( a_i \Om_{2i+1}+b_i\G_{2i+1}\right),
							\end{align}
						\end{subequations}
						where $a_i \;, b_i$ are constants coefficients. We therefore find
						\begin{tcolorbox}[enhanced,attach boxed title to top center={yshift=-3mm,yshifttext=-1mm}, colback=blue!5!white,colframe=blue!70!black,colbacktitle=red!80!black,fonttitle=\bfseries, 
							boxed title style={size=small,colframe=red!50!black}]
							For positive smKdV flows, the relation \eqref{zc.vac.mkdvp} is valid for any combination of constant vacuum  $(0,0)$, $(v_0,0)$, $(0,\bp_0)$, and $(v_0,\bp_0)$.
						\end{tcolorbox}
						%
						%
						%
						\noindent Notice that for the special case of the zero vacuum state $(0,0)$, $\G_{2m+1}$ vanishes  and hence
						\begin{equation} \label{avac.gen.z.p}
							A_{t_{2n+1},\text{vac}}^{\text{smKdV}}=\Om_{2n+1}\Big\vert_{(v_0,\bp_0)=(0,0)}=\mathcal{E}^{(2n+1)}.
						\end{equation}
						For negative smKdV flows, the situation is a bit more intricated. If $v_0 \neq 0$,  from the lowest grade equation
						\begin{equation} \label{zc.vac.even}
							\comm{v_0 M_1^{(0)}}{ D^{(-N)}_{-N,\text{vac}}}=0,
						\end{equation}
						we get that the only possible value is $N=-2m$, since no odd element commutes with $M_1$. Then, considering the case where $N=-2m$ , we get from eqs. \eqref{zc.vac} and \eqref{vacua.proj.high.b}, the following relation 
						\begin{equation} \label{zc.vac.mkdvn}
							\comm{A_{\text{$x$,vac}}^{\text{mKdV}}}{A_{\text{$t_{-N}$,vac}}^{\text{mKdV}}}=\comm{\cE+v_0 M_1^{(0)}+\bp_0 G_2^{\frac{1}{2}}}{ M_1^{(-2m)} + D^{(-2m+\frac{1}{2})}_{-2m,\text{vac}} + \cdots + D^{(-1)}_{-2m,\text{vac}}}=0,
						\end{equation}
						from which we can easily obtain the elements of $\mathcal{K}_\Om$, as long as we introduce a factor of $v_0^{-1}$. For instance, in the $N=-2$ case, we have
						\begin{equation} \label{avacm2}
							A_{t_{-2},\text{vac}}^{\text{smKdV}}=\frac{1}{v_0} \left(\Om_{-1}-\G_{-1}\right)+ \G_{-1},
						\end{equation}
						and for a general vacuum operator, we can write
						\begin{equation} \label{avac.gen.nz.n}
							A_{t_{-2m},\text{vac}}^{\text{smKdV}}=\sum_{i=1}^{m} v_0^{-2(m-i)-1} \left( c_i \Om_{-2i+1}+d_i\G_{-2i+1}\right).
						\end{equation}
						Notice that in this case we can not take the limit $v_0 \to 0$. Then, 
						
						\begin{tcolorbox}[enhanced,attach boxed title to top center={yshift=-3mm,yshifttext=-1mm}, colback=blue!5!white,colframe=blue!70!black,colbacktitle=red!80!black,fonttitle=\bfseries, boxed title style={size=small,colframe=red!50!black}]
							For negative even smKdV flows,  only two possibilities  $(v_0,0)$ and $(v_0,\bp_0)$ are  allowed.
						\end{tcolorbox}
						\noindent On the other hand, if $v_0 \equiv 0$, the lowest grade equation is now given by
						\begin{equation} \label{zc.vac.odd1}
							\comm{\bp_0 G_2^{\frac{1}{2}}}{ D^{(-N)}_{-N,\text{vac}}}=0,
						\end{equation}
						which is clearly not satisfied by a even flow ($N=-2m$). However, for a odd flow, we get the following  restriction on $ D^{(-N)}_{-N,\text{vac}}$
						\begin{equation} \label{rest.v2}
							D^{(-N)}_{-N,\text{vac}} = a_{N, \text{vac}}^{-2m+1} \left(K_1 ^{(-2m+1)} -K_2^{(-2m+1)}\right).
						\end{equation}
						A quick check on equation \eqref{vacua.proj.high.c} with $v_0=\phi_0=0$ shows that the super sinh-Gordon and other lower  flows within the hierarchy do not obey this restriction\footnote{By solving the ZCC for odd negative flows of the $N = -1$ case, we can perform some modifications of the constant parameters in such a way that we obtain a distinct temporal Lax operator, associated with a \textit{modified super sinh-Gordon} that allows us to set  both $(0,0)$ or $(0,\bp_0)$ as a vacuum configuration (restriction \eqref{rest.v2} is completely satisfied), i.e.,
							\begin{equation}
								\pa_x \pa_{t_{-1}} \phi = 2\sinh{2 \phi} - 2\bpsi \psi \cosh{\phi},
								\quad 
								\pa_{t_{-1}} \bpsi = 2 \psi \sinh{\phi},
								\quad 
								\pa_x \psi= 2 \bp \sinh\phi.
						\end{equation}}
					\begin{equation} \label{rest.ssg}
						D^{(-2m+1)}_{\text{vac}} =  \left(K_1 ^{(-2m+1)} +K_2^{(-2m+1)}\right) = \mathcal{E}^{(-2m+1)}
					\end{equation}
					and hence  $\bp_0 = 0$. Therefore,  for super sinh-Gordon and other negative odd flows, both the  bosonic  and fermionic vacuum  vanish.  In fact, the lowest equation in such case is
					\begin{equation}
						\comm{\cE}{ D^{(-N)}_{-N,\text{vac}}}=0,
					\end{equation}
					which is automatically satisfied for negative odd flows. In general, we have
					\begin{equation} \label{avac.gen.z.n}
						A_{t_{-2n+1},\text{vac}}^{\text{smKdV}}= \mathcal{E}^{(-2n+1)}
					\end{equation}
					and the vacuum structure is such that
					\begin{tcolorbox}[enhanced,attach boxed title to top center={yshift=-3mm,yshifttext=-1mm}, colback=blue!5!white,colframe=blue!70!black,colbacktitle=red!80!black,fonttitle=\bfseries, boxed title style={size=small,colframe=red!50!black}]
						For negative odd smKdV flows, only combinations with both bosonic and fermionic zero vacuum are possible, i.e. $(v_0,\bp_0)=(0,0)$.
					\end{tcolorbox}
					Finally, we state that for both positive and negative sKdV flows,  all the vacuum combinations are possible (see more details in Appendix \ref{ap.vacuumkdv}).
					
					\begin{tcolorbox}[enhanced,attach boxed title to top center={yshift=-3mm,yshifttext=-1mm}, colback=blue!5!white,colframe=blue!70!black,colbacktitle=red!80!black,fonttitle=\bfseries, 
						boxed title style={size=small,colframe=red!50!black}]
						For sKdV flows, any combination of constant vacuum is possible, $(0,0)$, $(v_0,0)$, $(0,\bp_0)$, $(v_0,\bp_0)$.
					\end{tcolorbox}
					\noindent	For such vacuum structure, it is possible to determine which flows are in involution with each other\footnote{The argument for commuting flows follows the same line of reasoning used in the bosonic case \cite{adans_negative_2023, hollowood_tau-functions_1993, aratyn_integrable_2003}.}. First, we recall that for a given integrable flow described by a general Lax operator
					\begin{equation}
						\mathcal{L}_N=\pa_{t_N}+A_{t_N},
					\end{equation}
					a sufficient condition to prove the commutation of two different flows, say $N$ and $M$,  is to show that their corresponding operators commute with each other, i.e.
					\begin{equation}
						\comm{	\mathcal{L}_N}{	\mathcal{L}_M}=0.
					\end{equation}
					As a direct consequence of the dressing operator \eqref{gauge.zcc}, this expression can be rewritten as,
					\begin{equation} \label{inv.eq}
						\comm{	\mathcal{L}_N}{	\mathcal{L}_M}=\Theta\comm{	\mathcal{L}_{N,\text{vac}}}{\mathcal{L}_{M,\text{vac}}}\Theta^{-1}=0,
					\end{equation}
					showing  that commutation of the  flows is equivalent to prove that the corresponding vacuum operators commutes with each other. For smKdV, the vacuum operators  are defined in terms of three objects $\mathcal{E}^{2m+1}$, $\Om_{2m+1}$ and $\G_{2m+1}$. As one can easily verify, two distinct abelian subalgebras can be formed with these objects. The first one is composed by
					\begin{equation} \label{inv.a}
						\comm{\mathcal{E}^{2m+1}}{\mathcal{E}^{2n+1}}=0,
					\end{equation}
					and the second is
					\begin{equation} \label{inv.b}
						\comm{\Om_{2m+1}}{\Om_{2n+1}}=0, \quad
						\comm{\G_{2m+1}}{\Om_{2n+1}}=0, \quad
						\comm{\G_{2m+1}}{\G_{2n+1}}=0.
					\end{equation}	 
					When a central extension is considered, they correspond to the so-called \textit{Heisenberg sub algebras}. This leads us to conclude that  two different hierarchies (in the sense of flows in involution) exist within smKdV. The first one is formed for flows where $\mathcal{L}_{N,\text{vac}} = \mathcal{E}^{(N)}$ as vacuum operator, namely, positive odd \eqref{avac.gen.z.p} and negative odd \eqref{avac.gen.z.n} flows. Hence, using \eqref{inv.eq} and  \eqref{inv.a}, we find that	 
					\begin{equation} \label{inv.smkdv1}
						\comm{	\mathcal{L}_{2m+1}^{\text{smKdV}}}{	\mathcal{L}_{2n+1}^{\text{smKdV}}}=\Theta_I\comm{	\mathcal{E}^{2m+1}}{\mathcal{E}^{2n+1}}\Theta_I^{-1}=0
					\end{equation}
					with $n,m \in \mathbb{Z}$. Then, we can claim that
					
					\begin{tcolorbox}[enhanced,attach boxed title to top center={yshift=-3mm,yshifttext=-1mm}, colback=blue!5!white,colframe=blue!70!black,colbacktitle=red!80!black,fonttitle=\bfseries, 
						boxed title style={size=small,colframe=red!50!black}]
						Positive and negative odd flows within smKdV commute with each other and form a hierarchy with zero vacuum orbit, \textit{smKdV - I}.
					\end{tcolorbox}
					\noindent Similarly, we obtain a second hierarchy of vacuum operators formed by the linear combination, $\mathcal{L}_{N,\text{vac}} =  \sum_{i}^{N} \alpha_i \Om_{2i+1}+\beta_i\G_{2i+1}$, namely, positive odd \eqref{avac.gen.nz.p} and negative even \eqref{avac.gen.nz.n}. Then, eq.   \eqref{inv.b} leads to
					\begin{equation} \label{inv.smkdv2}
						\comm{	\mathcal{L}_{-2m}^{\text{smKdV}}}{	\mathcal{L}_{2n+1}^{\text{smKdV}}}=
						\Theta_{II}\comm{\sum_{i}^{m} \alpha_i \Om_{-2i+1}+\beta_i\G_{-2i+1}} {\sum_{j}^{n} \g_j \Om_{2j+1}+\om_j\G_{2j+1}}\Theta_{II}^{-1}=0
					\end{equation}
					with $n,m \in \mathbb{Z}$ and $\alpha_i, \beta_i, \g_i, \om_i$ coefficients. Therefore, we find that\footnote{For simplicity, we are considering any combination with a non-zero bosonic vacuum orbit $(v_0,0)$ or $(v_0,\bp_0)$, as a non-zero vacuum. This works very well since the commutation relations holds for any of them. The same is valid for sKdV.}
					\begin{tcolorbox}[enhanced,attach boxed title to top center={yshift=-3mm,yshifttext=-1mm}, colback=blue!5!white,colframe=blue!70!black,colbacktitle=red!80!black,fonttitle=\bfseries, 
						boxed title style={size=small,colframe=red!50!black}]
						Positive odd and negative even flows within smKdV commute with each other and form a hierarchy that poses non-zero vacuum orbit, \textit{smKdV - II}.
					\end{tcolorbox}
					For the sKdV the procedure is quite similar (see appendix \ref{ap.vacuumkdv}). In this case, the Lax operators are constructed based on $\mathcal{E}^{2m+1}$, $\Lambda_{2m+1}$ and $\Pi_{2m+1}$ operators, which also form two different abelian subalgebras:
					\begin{equation} \label{inv.c}
						\comm{\mathcal{E}^{2m+1}}{\mathcal{E}^{2n+1}}=0,
					\end{equation}
					and 
					\begin{equation} \label{inv.d}
						\comm{\Lambda_{2m+1}}{\Lambda_{2n+1}}=0, \quad
						\comm{\Lambda_{2m+1}}{\Pi_{2n+1}}=0, \quad
						\comm{\Pi_{2m+1}}{\Pi_{2n+1}}=0.
					\end{equation}	 
					From them, we can conclude that:
					\begin{tcolorbox}[enhanced,attach boxed title to top center={yshift=-3mm,yshifttext=-1mm}, colback=blue!5!white,colframe=blue!70!black,colbacktitle=red!80!black,fonttitle=\bfseries, 
						boxed title style={size=small,colframe=red!50!black}]
						\begin{itemize}
							\item Positive odd and negative odd flows within sKdV commute with each other and form a hierarchy with zero vacuum orbit, \textit{sKdV - I}.
							\item Positive odd and negative even flows within sKdV commute with each other and form a hierarchy with non-zero vacuum orbit, \textit{sKdV - II}.
						\end{itemize}
					\end{tcolorbox}
					\noindent At this point, several important remarks can be made:%
					\begin{itemize}
						\item[$\diamond$] For smKdV, the hierarchy splits in two different classes, both constructed using the same $A_x$ but with different vacuum orbits. Both  type I and II share the same positive flows in this case, but distinct negative flows. 
						\item[$\diamond$] The smKdV - I and smKdV - II hierarchies are constructed using different dressing operators $\Theta_I$ and $\Theta_{II}$, since the operators $\theta_i$ depend explicitly on the vacuum chosen. This will lead to different vertex operators and soliton solutions. 
						\item[$\diamond$] Although sKdV - I and sKdV - II are constructed using different vacua and  dressing operators, they  share the same positive and negative flows.
						\item[$\diamond$] The super Miura transformation maps type I/II smKdV hierarchy into type I/II sKdV hierarchy. Although the positive flows coincide in both types for smKdV, the negative ones do not. In this way, it may appear that two different flows lead to the same sKdV equation. However, each flow actually arises from a specific vacuum orbit, either
						\begin{equation}
							\begin{tikzcd}[row sep=0.1em, column sep=3em]
								t_{N}^{\smkdv-I} \arrow[r, "\cS"]  &  t_{N}^{\skdv-I} 
							\end{tikzcd},
						\end{equation}
						where $N=2m+1$  for $m \in \mathbb{Z}$, or
						\begin{equation} 
							\begin{tikzcd}[row sep=0.1em, column sep=3em]
								t_{M}^{\smkdv-II} \arrow[r, "\cS"]  &  t_{N}^{\skdv-II} ,
							\end{tikzcd}
						\end{equation}
						for $M = 2m+1$ or $M=-2m$ with $m \in \mathbb{Z}^{+}$ and $N=2l+1$  for $l \in \mathbb{Z}$.
					\end{itemize}

								Having concluded our analysis of the vacuum structure and commutativity of  flows, in the next section we will obtain the vertex operators, and construct the one-soliton solutions for different vacuum configurations.
								
								\section{Vertex operator for smKdV and soliton solutions}
								\label{mkdv.solutions}
								
								We recall that the final step to obtain the soliton solution involves determining  vertex operators that depend upon the vacuum structure.
								The general vacuum configuration, is written in terms of the operators $\Om_{2m+1}$ and $\G_{2m+1}$ given by (\ref{o1}) and (\ref{g1}), i.e., 
								\begin{equation} \label{autv.vacuum}
									\comm{A_{\mu,\text{vac}}^{\text{smKdV}}}{ V(k_i)}= \om_\mu (k_i) V(k_i).
								\end{equation}
								Let us consider a generic vertex operator as follows
								\begin{equation}
									\begin{split}
										V(k_1)= \sum_{j=-\infty}^{\infty} k_1^{-2j} \left[
										a_0  \delta_{j,0}\, \hat{\kp}+a_1 M_1^{(2j)}+a_2 K_1^{(2j+1)}+a_3 K_2^{(2j+1)}+a_4 M_1^{(2j+1)} \right.
										\\
										\left. \!\!\! +a_5 F_1^{(2j+\frac{1}{2})} +a_6 G_2^{(2j+\frac{1}{2})} +a_7 F_2^{(2j+\frac{3}{2})} +a_8 G_1^{(2j+\frac{3}{2})}\right],
									\end{split}
								\end{equation}
								from which we can compute the following matrix elements for a one-soliton solution
								\begin{subequations}
									\begin{align}
										\tau_0 &= 1 + \langle \lambda_0|V(k_1)|\lambda_0\rangle \rho  = 1+ a_0 \rho \\[0.1cm]
										\tau_1 &= 1 + \langle \lambda_1|V(k_1)|\lambda_1\rangle \rho  = 1+ \left(a_0+a_1\right) \rho\\[0.1cm]
										\tau_2 &=\langle \lambda_0| F_1^{\left(\frac{1}{2}\right)}V(k_1)|\lambda_0\rangle \rho  =  \left(a_8-a_7\right) k_1^2 \rho\\[0.1cm]
										\tau_3 &=\langle \lambda_1| F_1^{\left(\frac{1}{2}\right)}V(k_1)|\lambda_1\rangle \rho  =  -\left(a_8+a_7\right) k_1^2 \rho
									\end{align}
								\end{subequations}
								where  the space-time dependence is given by $\rho=\rho(x,t_N) = e^{-\om_{x} x-\om_{N} t_N}$. Now, in order to simplify the computations of our solutions, let us summarize the results obtained for vertex operators  and their eigenvalues. We divide them into two categories, those that contain only pure fermionic solutions and those that include mixed solutions:
								\begin{enumerate}
									\item \emph{Pure fermionic vertex}
									\begin{enumerate}
										\item Non-zero fermionic vacuum ($\bp_0 \neq 0$, $c_1$ is a generic bosonic constant)
										\begin{equation} 
											\begin{split}
												V_{1}(\Bar{k}_1)= \sum_{j=-\infty}^{\infty} &\left(\frac{(2\Bar{k}_1-v_0)(2\Bar{k}_1+v_0)}{4\Bar{k}_1}\right)^{-2j}c_1 \bp_0\Biggl[ -\frac{v_0}{2\Bar{k}_1} F_1^{(2j+\frac{1}{2})} + G_2^{(2j+\frac{1}{2})}
												\\
												&+\frac{2v_0}{4\Bar{k}_1^2-v_0^2} F_2^{(2j+\frac{3}{2})} -\frac{4\Bar{k}_1}{4\Bar{k}_1^2-v_0^2} G_1^{(2j+\frac{3}{2})}\Biggr], 
											\end{split}
										\end{equation}
										with eigenvalues
										\begin{subequations} \label{eigen.1a}
											\begin{align}
												\left[\Omega_{2m+1},V_{1}(\Bar{k}_1)\right]&= 2 \Bar{k}_1 \left(\frac{4\Bar{k}_1^2-v_0^2}{4\Bar{k}_1}\right)^{2m} V_{1} (\Bar{k}_1),	\\[0.2cm]	
												\left[\G_{2m+1},V_{1}(\Bar{k}_1)\right]&=  \left(\frac{4\Bar{k}_1^2-v_0^2}{4\Bar{k}_1}\right)^{2m+1} V_{1} (\Bar{k}_1).
											\end{align}				
										\end{subequations}
										\item Zero fermionic vacuum ($\bp_0 = 0$, $\nu_1$ is a generic fermionic constant)
										
										\begin{equation} 
											\begin{split}
												V_{2}(\Bar{k}_1)= \sum_{j=-\infty}^{\infty} &\left(\frac{(2\Bar{k}_1-v_0)(2\Bar{k}_1+v_0)}{4\Bar{k}_1}\right)^{-2j}\nu_1 \Biggl[ -\frac{v_0}{2\Bar{k}_1} F_1^{(2j+\frac{1}{2})} + G_2^{(2j+\frac{1}{2})}
												\\
												&+\frac{2v_0}{4\Bar{k}_1^2-v_0^2} F_2^{(2j+\frac{3}{2})} -\frac{4\Bar{k}_1}{4\Bar{k}_1^2-v_0^2} G_1^{(2j+\frac{3}{2})}\Biggr] ,
											\end{split}
										\end{equation}
										with eigenvalues
										\begin{subequations}
											\begin{align}
												\left[\Omega_{2m+1},V_{2}(\Bar{k}_1)\right]&= 2 \Bar{k}_1 \left(\frac{4\Bar{k}_1^2-v_0^2}{4\Bar{k}_1}\right)^{2m} V_{2}(\Bar{k}_1),	\\[0.2cm]	
												\left[\G_{2m+1},V_{2}(\Bar{k}_1)\right]&=  \left(\frac{4\Bar{k}_1^2-v_0^2}{4\Bar{k}_1}\right)^{2m+1} V_{2}(\Bar{k}_1).
											\end{align}				
										\end{subequations}		
										In both cases, there are no problems taking the limit $v_0 \to 0$ to obtain other vacuum configurations, such as $(0,\bp_0)$.
									\end{enumerate}
									\item \emph{Bosonic-fermionic vertex}
									\begin{enumerate}
										\item Non-zero bosonic vacuum ($v_0 \neq 0$):
										\begin{equation} 
											\begin{split}
												V_{3}(\Bar{k}_1)&= \sum_{j=-\infty}^{\infty} \left(\Bar{k}_1^2-v_0^2\right)^{-j} \Biggl[ M_1^{(2j)}- \frac{\left(v_0+\Bar{k}_1\right)}{2\Bar{k}_1} \hat{\kappa} \delta_{j,0}+\frac{v_0}{v_0^2-\Bar{k}_1^2} K_1^{(2j+1)} \\
												&+  \frac{\Bar{k}_1}{v_0^2-\Bar{k}_1^2} M_2^{(2j+1)}+v_0^{-1} \bp_0 G_2^{(2j+\frac{1}{2})} - \frac{\bp_0}{v_0^2-\Bar{k}_1^2} F_2^{(2j+\frac{3}{2})} +\frac{\bp_0 \Bar{k}_1 v_0^{-1}}{v_0^2-\Bar{k}_1^2}  G_1^{(2j+\frac{3}{2})}\Biggr],
											\end{split}
										\end{equation}
										with eigenvalues
										\begin{subequations} \label{eigen.2a}
											\begin{align}
												\left[\Omega_{2m+1},V_{3}(\Bar{k}_1)\right]&= 2 \Bar{k}_1 \left(\Bar{k}_1^2-v_0^2\right)^{m} V_{3} (\Bar{k}_1),	\\[0.2cm]	
												\left[\G_{2m+1},V_{3}(\Bar{k}_1)\right]&=  0.
											\end{align}				
										\end{subequations}
										By taking the  limit of $\bp_0 \to 0$ one obtains the vacuum configurations $(v_0,0)$.
										\item Zero bosonic  and non-zero fermionic vacuum ($v_0 = 0$,$\bp_0 \neq  0$):
										\begin{equation} 
											\begin{split}
												V_{4}= \sum_{j=-\infty}^{\infty} k_1^{-2j} \biggl\{ \biggl[ M_1^{(2j)}&- \frac{1}{2} \hat{\kappa} \delta_{j,0} -  \frac{1}{k_1} M_2^{(2j+1)} + \frac{\bp_0}{2k_1} F_1^{(2j+\frac{1}{2})} + \frac{\bp_0}{2k_1^2} F_2^{(2j+\frac{3}{2})}
												\biggr] \\
												&+c_1 \biggl[\frac{ \psi_0 }{2k_1}  G_2^{(2j+\frac{1}{2})}  -\frac{ \psi_0 }{2k_1^2}  G_1^{(2j+\frac{3}{2})}\biggr] \biggr\}
											\end{split}
										\end{equation}
										with eigenvalues
										\begin{subequations}
											\begin{align}
												\left[\Omega_{2m+1},V_{4}\right]&= 2 k_1^{2m+1} V_{4} ({k}_1), \\[0.2cm]	
												\left[\G_{2m+1},V_{4}({k}_1)\right]&=  0.
											\end{align}				
										\end{subequations}
										
										\item Zero bosonic  and  zero fermionic vacuum ($v_0 = 0$,$\bp_0 =  0$):
										\begin{equation} 
											\begin{split}
												V_{5}({k}_1)= \sum_{j=-\infty}^{\infty} {k}_1^{-2j} \biggl\{ &\biggl[ M_1^{(2j)}- \frac{1}{2} \hat{\kappa} \delta_{j,0} -  \frac{1}{{k}_1} M_2^{(2j+1)} + \frac{\bp_0}{2{k}_1} F_1^{(2j+\frac{1}{2})}\biggr]\\
												&+\nu_1 \biggl[  G_2^{(2j+\frac{1}{2})}  -\frac{ 1 }{{k}_1}  G_1^{(2j+\frac{3}{2})}\biggr] \biggr\},
											\end{split}
										\end{equation}
										with eigenvalues
										\begin{subequations}
											\begin{align}
												\left[\Omega_{2m+1},V_{5}({k}_1)\right]&= 2 {k}_1^{2m+1} V_{5}({k}_1), \\[0.2cm]	
												\left[\G_{2m+1},V_{5}({k}_1)\right]&=  0.
											\end{align}				
										\end{subequations}
									\end{enumerate}
								\end{enumerate}
								Once we have constructed the vertex operators and its eigenvalues, we can finally obtain one-soliton solutions for the different hierarchies related to smKdV. Here, we will use $c_1$ to denote  a bosonic constant, and $\nu_1$ to denote a fermionic constant. It is also worth mentioning that although we are mainly focusing in one-soliton solutions\footnote{All the solutions we obtained were double checked using the \textit{Hirota Method} implemented in \textit{Mathematica} }, we can use the vertex operators to obtain in general multi-soliton solutions for the models \cite{gomes_soliton_2006}.

								\subsection{The smKdV-I (zero vacuum) solutions}
								
								For both positive and  odd negative flows, we must have a zero vacuum solution such $(v_0,\bp_0)=(0,0)$. In such case, we consider the vertex $	V_{1}$ and $V_{5}$ leading to two \mbox{different} solutions:
								\begin{equation}
									v_{2} =  \pa_x ln \left(\frac{\ta_{0}}{\ta_{1}}\right) = 0, \qquad 	\bp_{2} = \frac{\tau_3}{\tau_1} - \frac{\tau_2}{\tau_0} = 2{k}_1 \nu_1 \rho_N;
								\end{equation}
								and
								\begin{equation}
									v_{5}=\pa_x ln\left(\frac{1-\frac{1}{2} \rho_N}{1+\frac{1}{2} \rho_N}\right), \qquad \bp_{5} = \nu_1 \; \pa_x ln\left(\frac{1-\frac{1}{2} \rho_N}{1+\frac{1}{2} \rho_N}\right),
								\end{equation}
								with $\rho_N=\rho(x,t_N) = e^{-2{k}_1 x-\om_{N} t_N}$, where the subscript number on the fields corresponds to the vertex that was used. 
								For both cases, the eigenvalues are given by
								\begin{itemize}
									\item The smKdV ($N=3$)
									\begin{equation}
										\om_3= 2 {k}_1^3,
									\end{equation}
									\item 		The sshG  ($N=-1$)
									\begin{equation}
										\om_{-1}=  2 {k}_1^{-1}
									\end{equation}
									\item General Case  ($N=2m+1$)
									\begin{equation}
										\om_{2n+1}=  2 {k}_1^{2n+1}
									\end{equation}
								\end{itemize}
								These solutions are in agreement with the ones reported previously in the literature \cite{gomes_soliton_2006}.
								
								\begin{figure}[H]
									\centering
									\subfigure(a){\includegraphics[width=0.29\textwidth]{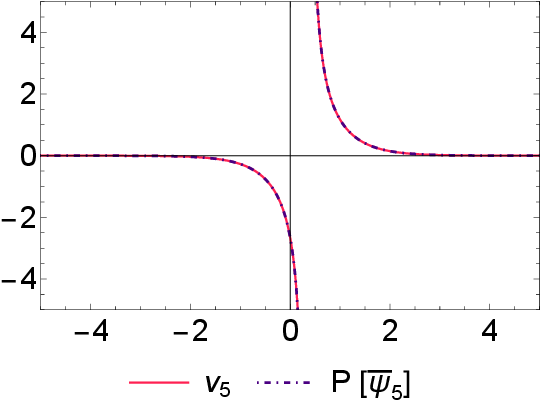}} 
									\subfigure(b){\includegraphics[width=0.29\textwidth]{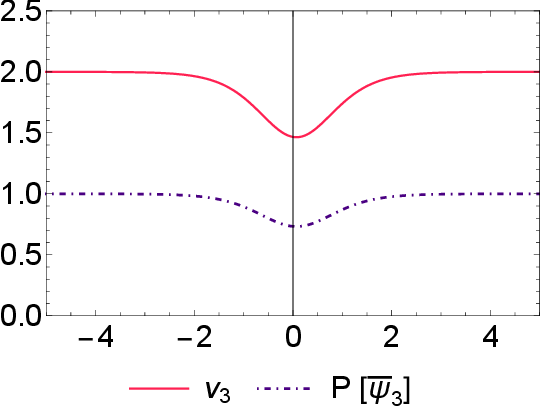}} 
									\subfigure(c){\includegraphics[width=0.29\textwidth]{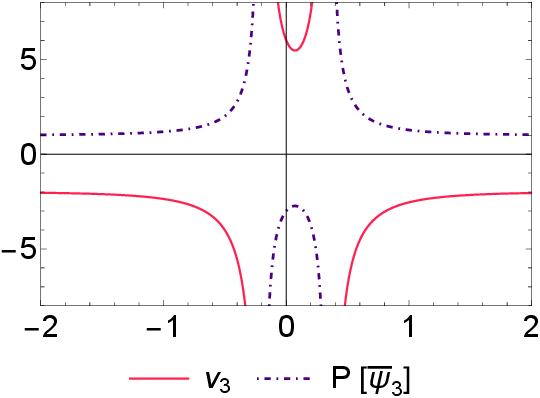}} 
									\subfigure(d){\includegraphics[width=0.29\textwidth]{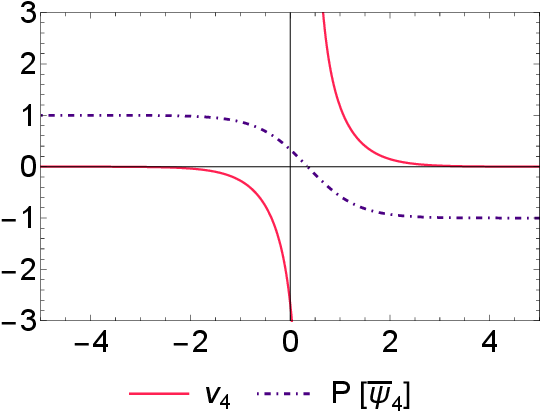}} 
									\subfigure(e){\includegraphics[width=0.29\textwidth]{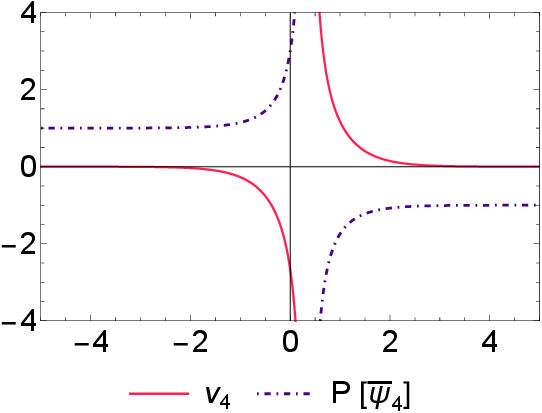}} 
									\caption{ The smKdV-I and smKdV-II solutions. In all cases, $k_1 =1$ and $t=0$. \textbf{(a)} Classical profile for the zero vacuum solution $v_5, \bp_5$. In this case, both the bosonic field and the projection of the fermionic field matches. \textbf{(b)} Setting the bosonic vacuum background to be positive $v_0=2$ for the $v_3, \bp_3$ solution, we obtain a anti-solitonic profile (\textit{dark soliton}). In this case, the projection background for the fermionic field will depend on $\bp_0$ \textbf{(c)} On the other hand, choosing a negative vacuum background $v_0=-2$ for $v_3, \bp_3$ solution leads to a discontinuous solution (\textit{peakon}). Notice that due to the inversion of the vacuum, the fermionic projection will have a overall signed compared to the bosonic solution. \textbf{(d - e)} Analysis of the $c_1$ dependent solution $v_4, \bp_4$. Both share the same bosonic profile since it does not depend on the constant but have different fermionic projection. In (d) we have chosen $c_1=-1$, leading to anti-kink profile, while in (e) we have chosen $c_1=1$, leading to a mirrored standard solution. }	
									\label{fig:smkdv}
								\end{figure}

								\subsection{The smKdV-II (non-zero vacuum) solutions}
								
								Now, for the flows possessing a wider range of vacuums solutions, we can obtain novel combinations. First of all, for both positive and negative even flows, we must have a non-zero bosonic vacuum solution $(v_0,\bp_0)=(v_0,0)$ or $(v_0,\bp_0)$. In that case, we can consider the vertex $V_1$, $V_2$ or $V_3$, leading to three different solutions:
								\begin{eqnarray}
									v_{1} &=&  v_0 , \qquad 	\bp_{1} = \bp_0 + \left(\frac{4\Bar{k}_1^2-v_0^2}{2\Bar{k}_1}\right) c_1\bp_0 \rho_N;\\
									v_{2} &=&  v_0 , \qquad 	\bp_{2} = \left(\frac{4\Bar{k}_1^2-v_0^2}{2\Bar{k}_1}\right) \nu_1 \rho_N.
								\end{eqnarray}
								where $\rho_N=\rho(x,t_N) = e^{-2{k}_1 x-\om_{N} t_N}$, with  eigenvalues given by\footnote{A general formula for the eigenvalues can be obtained by using \eqref{avac.gen.nz.n}, \eqref{avac.gen.nz.p} and  \eqref{eigen.1a}.}
								\begin{eqnarray}
									\om_3 &=& 2 \Bar{k}_1^3-\frac{3\Bar{k}_1 v_0^2}{2} , \qquad \qquad\quad  \,\,\,\,\qquad \qquad \mbox{for} \qquad N=3,\\[0.2cm]
									\om_{-2} &=& \frac{4\Bar{k}_1v_0^2\left(v_0-1\right)+16\Bar{k}_1^3\left(v_0+1\right) }{v_0 \left(4\Bar{k}_1^2-v_0^2\right)^2} , \qquad \mbox{for} \qquad N=-2,
								\end{eqnarray}
								Regarding $V_3(\Bar{k}_1)$, we have a more interesting solution that includes mixed terms
								\begin{equation}
									v_{3} = v_0  + \pa_x  \ln\left(\frac{1-\frac{\left(\Bar{k}_1+v_0\right)}{2\Bar{k}_1} \rho_N}{1+\frac{\left(\Bar{k}_1-v_0\right)}{2\Bar{k}_1} \rho_N}\right), \qquad 	\bp_{3} = \bp_0 +\frac{\bp_0}{v_0} \; \pa_x \ln\left(\frac{1-\frac{\left(\Bar{k}_1+v_0\right)}{2\Bar{k}_1} \rho_N}{1+\frac{\left(\Bar{k}_1-v_0\right)}{2\Bar{k}_1} \rho_N}\right),
								\end{equation}
								%
								with the eigenvalues
								\begin{eqnarray}
									\om_3 &=& 2\Bar{k}_1^3-3v_0^2\Bar{k}_1, \qquad \,\mbox{for} \qquad N=3,\\
									\om_{-2} &=& \frac{2\Bar{k}_1}{v_0\left(\Bar{k}_1^2-v_0^2\right)}, \qquad \mbox{for} \qquad N=-2,
								\end{eqnarray}
								which can also be generalized using \eqref{eigen.2a}.
								In addition, we can also construct solutions for positive times with a different mixed vacuum $(v_0,\bp_0)=(0,\bp_0)$, by using the vertices $V_1$ and $V_4$. We get
								\begin{equation}
									v_{1} = 0, \qquad 	\bp_{1} = \bp_0 + 2{k}_1\bp_0c_1 \rho_N,
								\end{equation}
								and
								\begin{equation}
									v_{4}=\pa_x \ln\left(\frac{1-\frac{1}{2} \rho_N}{1+\frac{1}{2} \rho_N}\right), \qquad \bp_{4} = \bp_0+ \frac{\left(2c_1 \bp_0+\bp_0 \rho_N \right)}{4 \Bar{k_1}} \pa_x \ln\left(\frac{1-\frac{1}{2} \rho_N}{1+\frac{1}{2} \rho_N}\right),
								\end{equation}
								both with eigenvalue $\om_3=2k_1^3$. 
								In this case, we notice that $v_0$ does not contribute for the eigenvalue.
								
								These results show us that introducing different parameters to the vacuum configuration leads to novel and interesting solutions that mix bosonic and fermionic parameters. The possibility of combining vacuum configurations lead us to six different solutions for  smKdV-II models. In order to visualize the difference in the behavior of the solutions, in Figure \ref{fig:smkdv} we have plotted the profile for the dynamical solutions $v_3, \bp_3$, $v_4, \bp_4$ and $v_5, \bp_5$. For the fermionic fields $\bp_i$, we have plotted only the bosonic part of the solution, factorizing the Grassmann constant, namely the \textit{bosonic projection}  $P[\bp_i]$, such that $\bp_i = \bp_0 P[\bp_i]$, or $\bp_i = \nu_1 P[\bp_i]$. We see that different kind of solutions, kinks, anti-kinks, dark-solitons and peakons, appear by choosing the parameters accordingly.
								
								In the next section, we will combine these solutions with  the Gauge-super Miura transformation to obtain new solutions for the sKdV system.

								\section{The sKdV solutions via super Miura transformation}
								\label{kdv.solutions}
								As highlighted in section \ref{sec:Miura}, one can map the smKdV solutions into the sKdV solutions by using the super Miura transformation \eqref{smiura.x.s.1}:
								\begin{equation} 
									\begin{split}
										J &= v^2 - \pa_x v + \bpsi \pa_x \bpsi ,
										\\
										\bar{\chi} &= - v \bpsi + \pa_x \bpsi.
									\end{split}
								\end{equation}
								Due the fact that the smKdV equation is invariant under the parity transformation $v \to -v$ and $\bp\to -\bp$, it is possible to obtain three other different kind of super Miura transformations, leading to four different possibilities. Here, we evaluate each one of them to obtain the corresponding sKdV solutions\footnote{Here, we use the same notation for subscript label of the vertex. The subscript $\pm$ sign denote if the solution is obtained using either $v$ or $-v$, while the upper $\pm$ sign  denote if the solution is obtained using either $\bp$ or $-\bp$. In addition $\rho_N$ and $\om_N$ are the same as defined by smKdV in the previos section}.

								\subsection{The sKdV-I (non-zero vacuum) solutions}
								
								For the sKdV-I hierarchy, we must have a zero vacuum solution such $(J_0,\bchi_0)=(0,0)$. In this case, we get six different solutions:
								\begin{equation}
									J_{2,-}^{\pm} = J_{2,+}^{\pm} =  0, \qquad 	\chi_{2,-}^{\pm} =\chi_{2,+}^{\mp} = \pm 4{k}_1^2 \nu_1 \rho_N,
								\end{equation}
								and
								\begin{equation}
									J_{5,-}^{\pm}=-2 \pa_x^2 \ln\left(1- \frac{1}{2} \rho_N\right), \qquad \bchi_{5,-}^{\pm} = \pm 2 \nu_1 \; \pa_x^2 \ln\left(1- \frac{1}{2} \rho_N\right),
								\end{equation}
								
								\begin{equation}
									J_{5,+}^{\pm}=-2 \pa_x^2 \ln\left(1+ \frac{1}{2} \rho_N\right), \qquad \bchi_{5,-}^{\pm} = \pm 2 \nu_1 \; \pa_x^2 \ln\left(1+\frac{1}{2} \rho_N\right),
								\end{equation}
								with $\rho_N=\rho(x,t_N) = e^{-2{k}_1 x-\om_{N} t_N}$.

								\begin{figure}[H]
									\centering
									\subfigure(a){\includegraphics[width=0.29\textwidth]{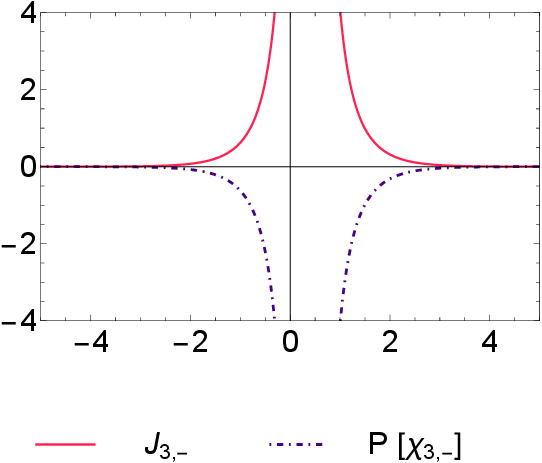}} 
									\subfigure(b){\includegraphics[width=0.29\textwidth]{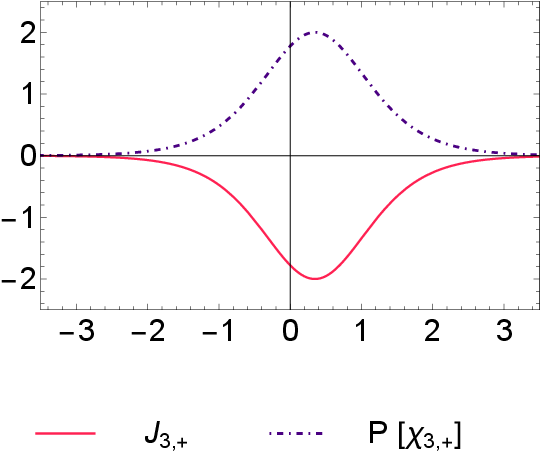}} 
									\subfigure(c){\includegraphics[width=0.29\textwidth]{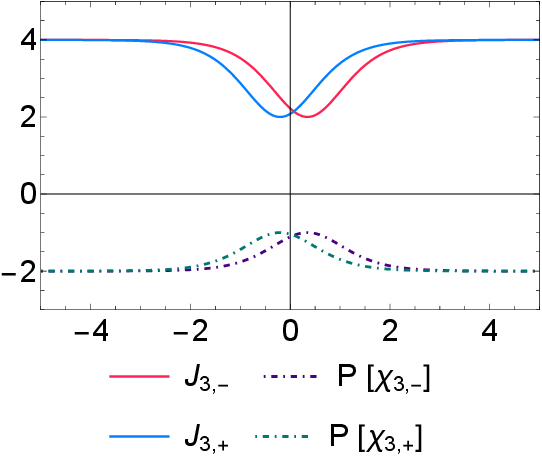}} 
									\subfigure(d){\includegraphics[width=0.29\textwidth]{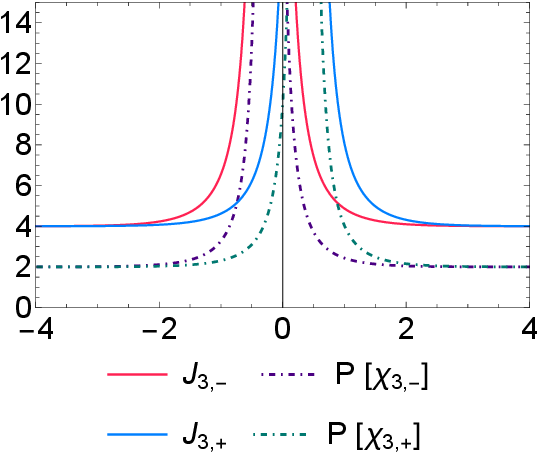}} 
									\subfigure(e){\includegraphics[width=0.29\textwidth]{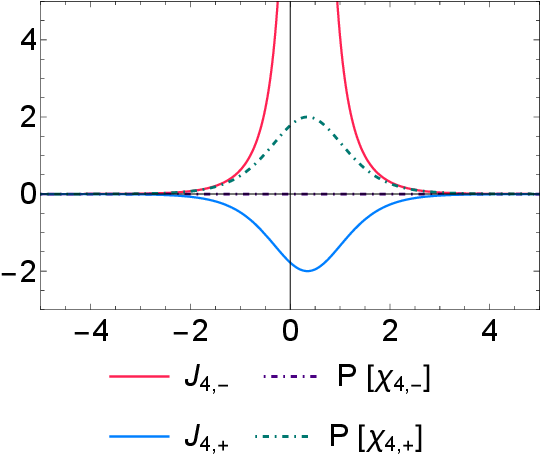}}
									\subfigure(f){\includegraphics[width=0.29\textwidth]{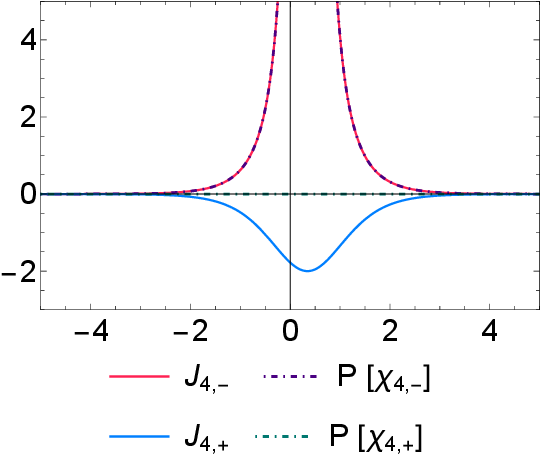}}  
									\caption{ Plots of the KdV-I and KdV-II solutions. We use both the positive and negative super Miura transformation for the bosonic field, and only the positive one for the fermionic field. \textbf{(a-b)} Using the zero vacuum solution $v_5, \bp_5$, we obtain two different solutions for sKdV. We obtain a {peakon} solution by using the first Miura, and by changing the SMiura leads to {soliton/dark-soliton profile}. \textbf{(c)} Now using the non-zero positive vacuum $v_0=2$, we obtain dark-solitons for the bosonic field and solitons for the fermionic projection for sKdV-II solutions. In such case, changing the super Miura only leads to a phase shift. \textbf{(d)} Setting a negative smKdV-II vacuum $v_0=-2$ leads to {peakons} solutions, although now the quadratic nature of the super Miura transformation leads to a positive vacuum for the sKdV-II field. \textbf{(e - f)} The $c_1$ dependent solution $v_4, \bp_4$ is quite interesting for the sKdV case. As the bosonic field does not depend on $c_1$, we have in both cases a {peakon} for the first super Miura, and {dark-soliton} for the second one. For the fermionic projection, in {\bf (e)} the choice $c_1=-1$, leads to a {null solution} for the first Miura, and a dark-soliton for the second one. In turn, in {\bf (f)} with the choice $c_1=1$, the first super Miura leads to a peakon solution, while the second one gives a solution that is always zero. }	
									\label{fig:skdv}
								\end{figure}
								These solutions agree with the well-known format for KdV system, which includes a second-order derivative acting on the logarithmic term \cite{sawada_method_1974, luis_miramontes_tau-functions_1999}. Notice that the parity transformation for the bosonic field $v \to -v$ interchanges the $\tau_0$ function with the  $\tau_1$ function. This features will be also present for a non-zero vacuum. On the other hand, the parity transformation for the fermionic field just changes a global sign.
								
								\subsection{The sKdV-II (non-zero vacuum) solutions}
								
								Now, for the sKdV-II flows, we find again several solutions associated to the different vacuum configurations. For  $(J_0,\bchi_0)=(J_0,0)$ or $(J_0,\bchi_0)$, we find twelve different solutions. The first set contains eight solutions where only the fermionic field is dynamical, namely
								\begin{equation}
									J_{1,-}^{\pm} =  v_0^2, \qquad 	\bchi_{1,-}^{\pm}  = \mp v_0 \bp_0 \mp \frac{\left(4\Bar{k}_1^2-v_0^2\right)\left(v_0+2\Bar{k}_1\right)}{2\Bar{k}_1} c_1\bp_0 \rho_N;
								\end{equation}
								\begin{equation}
									J_{1,+}^{\pm}  =  v_0^2, \qquad 	\bchi_{1,+}^{\pm}  = \mp v_0 \bp_0 \mp \frac{\left(4\Bar{k}_1^2-v_0^2\right)\left(v_0-2\Bar{k}_1\right)}{2\Bar{k}_1} c_1\bp_0 \rho_N;
								\end{equation}
								\begin{equation}
									J_{2,-}^{\pm}  =  v_0^2, \qquad 	\bchi_{2,-}^{\pm}  =  \mp \frac{\left(4\Bar{k}_1^2-v_0^2\right)\left(v_0+2\Bar{k}_1\right)}{2\Bar{k}_1} \nu_1 \rho_N;
								\end{equation}
								\begin{equation}
									J_{2,+}^{\pm}  =  v_0^2, \qquad 	\bchi_{2,+}^{\pm}  =  \mp \frac{\left(4\Bar{k}_1^2-v_0^2\right)\left(v_0-2\Bar{k}_1\right)}{2\Bar{k}_1} \nu_1 \rho_N.
								\end{equation}
								Notice that the sKdV and smKdV vacuum is related by $J_0=v_0^2$, no matter if we are using the positive or negative sign for the bosonic field. Then, for the sKdV systems, the background for the bosonic vacuum is {always positive}. On the other hand, for the fermionic vacuum is $\chi_0=\pm v_0 \bp_0$, allowing both positive and negative configurations.
								
								Finally, the remaining solutions contain both dynamical fields, but only a positive bosonic vacuum is possible. They have the following form,
								\begin{equation}
									J_{3,-}^{\pm} = v_0^2-2 \pa_x^2 \ln\Bigg(1-\frac{\left(v_0+\Bar{k}_1\right)}{2\Bar{k}_1} \rho_N\Bigg), \qquad \bchi_{3,-}^{\pm}  = \mp v_0 \bp_0 \; \pm \frac{2\bp_0}{v_0} \pa_x^2 \ln\Bigg(1-\frac{\left(v_0+\Bar{k}_1\right)}{2\Bar{k}_1} \rho_N\Bigg)
								\end{equation}
								\vspace{-0.1cm}
								\begin{equation}
									J_{3,+}^{\pm} = v_0^2-2 \pa_x^2 \ln\Bigg(1+\frac{\left(\Bar{k}_1-v_0\right)}{2\Bar{k}_1} \rho_N\Bigg), \qquad \bchi_{3,+}^{\pm}  = \mp v_0 \bp_0 \; \pm \frac{2\bp_0}{v_0} \pa_x^2 \ln\left(1+\frac{\left(\Bar{k}_1-v_0\right)}{2\Bar{k}_1} \rho_N\right)
								\end{equation}
								In addition, for the mixed vacuum configuration $(v_0,\bp_0)=(0,\bp_0)$, we have
								\begin{equation}
									J_{1,-}^{\pm}  = J_{1,+}^{\pm}  =  0, \qquad 	\bchi_{1,-}^{\pm}  =\bchi_{1,+}^{\pm}  =\pm 4{k}_1^2 c_1 \bp_0 \rho_N,
								\end{equation}
								and
								\begin{equation}
									J_{4,-}^{\pm} =-2 \pa_x^2 \ln\left(1- \frac{1}{2} \rho_N\right), \qquad \bchi_{4,-}^{\pm}  = \pm  \frac{\left(c_1+1\right) \bp_0}{k_1} \; \pa_x^2 \ln\left(1- \frac{1}{2} \rho_N\right);
								\end{equation}
								\begin{equation}
									J_{4,+}^{\pm} =-2 \pa_x^2 \ln\left(1+ \frac{1}{2} \rho_N\right), \qquad \bchi_{4,-}^{\pm}  = \pm \frac{\left(c_1-1\right) \bp_0}{k_1}\; \pa_x^2 \ln\left(1+\frac{1}{2} \rho_N\right).
								\end{equation}
								An interesting feature about this solution is the possibility of having a vanishing fermionic solution for $\bchi_{4,-}$ if $c_1=-1$, or for $\bchi_{4,+}$ if $c_1=1$. In Figure \ref{fig:skdv}, we have plotted the obtained solutions for skdV, where the upper index is always considered to be positive. Again, several interesting solutions appears, such as  solitons, dark-solitons and peakons.

								\section{Discussion and further developments}
								\label{discussion}
								
								In this paper we have extended   the algebraic construction of positive flows to  integrate the negative grade sector  for both smKdV and sKdV hierarchies. These,   follows  the spirit  developed  for the pure bosonic case   \cite{adans_negative_2023} with the inclusion  of     fermi  fields.

								The vacuum orbit plays a crucial role in classifying the hierarchies, splitting both smKdV and sKdV into two different types: the type-I hierarchy, which allows only a zero bosonic and fermionic vacuum, and a type-II hierarchy, which only have a non-zero bosonic vacuum. For the smKdV, the type I and II share the same positive flows but has different negative flows, while for the sKdV they share the same equations of motion for both the positive and negative part.

								Soliton solutions for the smKdV were constructed by applying the dressing method. This approach generates different vacuum operators that form an abelian subalgebra, which is essential for the involution of the flows: $\left\{\mathcal{E}^{(2n+1)}\right\}$ for the smKdV-I, and $\left\{\Om_{2m+1}, \G_{2m+1} \right\}$ for smKdV-II. Five different vertex operators were obtained in such context, leading to two different types of one-soliton solution for smKdV-I, and five solutions for smKdV-II. 

								Concerning the super Miura map, we have extend this transformation to all  integer flows, whether they are positive or negative. For the negative flows, it was necessary to introduce an additional condition on the time derivatives of the sKdV field, the so-called temporal super Miura. These temporal relations are extremely important to  show consistency of  the gauge transformation for both super Miura and the dressing transformation, but also to prove the supersymmetry of $sKdV(-1)$.
								Another interesting result is the coalescence of two subsequent negative flows of smKdV hierarchy into a single flow of the sKdV. 
								The soliton solutions  for the sKdV hierarchy  was obtained by using the super Miura transformation. Due to the existence of four possible super Miura transformation, we find that each smKdV solution lead to multiples solutions for sKdV.

								

								
								On the other hand, regarding the vacuum structure, note that $\mathcal{E}^{(2n+1)}$ has degree $2n+1$ according the grading operator $\widehat{Q} \equiv 2 \widehat{d} + \tfrac{1}{2} M_1^{(0)}$, whether $\Om_{2m+1}$ and $\G_{2m+1}$ would have different grades in each part of its components\footnote{ The grades would  be $2n$, $2n+\frac{1}{2}$, $2n+1$ and  $2n+1$, $2n+\frac{3}{2}$ respectively. }. This issue can be solved if we associate degree $1$ to $\v_0$ and $\frac{1}{2}$ to $\bp_0$, by redefining the grading operator as follows,
								\begin{equation}
									\widetilde{Q} \equiv \widehat{Q} + \widetilde{d} \quad \text{where} \quad \widetilde{d} \equiv \v_0 \tfrac{\pa}{\pa \v_0} + \frac{1}{2}\bp_0 \tfrac{\pa}{\pa \bp_0},
								\end{equation}
								in such a way that both $\Om_{2m+1}$ and $\G_{2m+1}$ have degree  $2n+1$, and the vacuums $\v_0,\; \bp_0$ could be interpreted as the new spectral parameters. Thus, each term in $\mathcal{L}_{2n+1,\text{vac}}^{\smkdv \text{-II}}$ and $\mathcal{L}_{-2n,\text{vac}}^{\smkdv\text{-II}}$ has now grade $2n+1$ and $2n$, as would be expected.  This idea had already been discussed previously in \cite{adans_negative_2023}, but only $\v_0$ was assumed to be a new spectral parameter, being possible to define a two-loop algebra \cite{aratyn_new_1992} associated to it. In the supersymmetric case is necessary to introduce two new spectral parameters, making unclear whether this concept can be extended to a three-loop algebra. We intend to investigate this further.
								
								In addition, it would be also interesting to explore different solutions with non-zero vacuum, such as multi-soliton solutions and quasi-periodic solutions in future investigations. The quasi-periodic solutions have already been obtained for supersymmetric systems by using different methods \cite{gao_bosonization_2012, hon_super_2011}, but we would like to extend the dressing method to obtain it more systematically. We also would like to implement the Bäcklund transformations \cite{gao_bosonization_2013} via a regular ansatz \cite{de_carvalho_ferreira_generalized_2021}. This will allow us to obtain the defect matrix for systems such as sKdV($-1$) and smKdV($-2$), and to use it for performing the scattering of the solutions. Finally, it would be also interesting to explore what would happen to the ZCC if we use a semi-integer ansatz for the Lax Pair, i.e. $N \in \mathbb{Z}+ \frac{1}{2}$. Some of these issues are currently under investigations, and we hope to report it in the future.
									
								\section*{Acknowledgments}
								
								JFG and AHZ thank CNPq and FAPESP for support. YFA thanks FAPESP for financial support under grant \#2022/13584-0. ARA thanks CAPES for financial support. GVL thanks Jogean Carvalho for valuable discussions.

								\newpage
								
								\section*{Appendix}
								
								\appendix
								\section{The $sl(2,1)$ superalgebra}
								
								\label{algebra.sl(2,1)}
								
								In this section, we present the algebraic structure used to construct the smKdV and sKdV integrable hierarchies. Let us consider super Kac-Moody algebra $\lie = sl(2,1)$ generated by
								\begin{eqnarray}
									L_0 &=& \left\{ h_1^{(m)} = \la^{m} h_1, \; h_2^{(m)} = \la^{m} h_2, \; E_{\pm \alpha_1}^{(m)} = \la^{m} E_{\pm \alpha_1}, \hat{\kp} \right\},
									\nonumber
									\\
									L_1 &=& \left\{ E_{\pm \alpha_2}^{(m)} = \la^m E_{\pm \alpha_2}, \; E_{\pm (\alpha_1+\alpha_2)}^{(m)} = \la^m E_{\pm (\alpha_1 + \alpha_2)} \right\},
								\end{eqnarray}
								where $\la \in \mathbb{C}$, $m \in \mathbb{N}$, $\hat{\kp}$ is central extension term and 
								\begin{eqnarray}
									h_1 &=& \left( \begin{matrix} 1 & 0 & 0 \\ 0 &-1 & 0 \\ 0 & 0 & 0 \\ \end{matrix}\right),
									\quad
									h_2 = \left( \begin{matrix} 0 & 0 & 0 \\ 0 & 1 & 0 \\ 0 & 0 & 1 \\ \end{matrix}\right),
									\quad 
									E_{\alpha_1} = \left( \begin{matrix} 0 & 1 & 0 \\ 0 & 0 & 0 \\ 0 & 0 & 0 \\ \end{matrix}\right), \quad E_{-\alpha_1} = E_{\alpha_1}^{\dagger} \nonumber\\[0.2cm]
									E_{\alpha_2} &=& \left( \begin{matrix} 0 & 0 & 0 \\ 0 & 0 & 1 \\ 0 & 0 & 0 \\ \end{matrix}\right),
									\quad
									E_{\alpha_1+\alpha_2} = \left( \begin{matrix} 0 & 0 & 1 \\ 0 & 0 & 0 \\ 0 & 0 & 0 \\ \end{matrix}\right), \quad E_{-\alpha_2} = E_{\alpha_2}^{\dagger}, \quad E_{-(\alpha_1+\alpha_2)} = E_{(\alpha_1+\alpha_2)}^{\dagger}. \quad \mbox{}
								\end{eqnarray}
								The $L_0$ and $L_1$ are called the bosonic and fermionic parts of algebra, respectively, satisfying the following relations
								\begin{equation}
									\comm{L_0}{L_0} \subset L_0,
									\qquad
									\comm{L_0}{L_1} \subset L_1,
									\qquad
									\comm{L_1}{L_1} \subset L_0.
									\nonumber
								\end{equation}
								The \textit{principal grading operator} is defined by
								\begin{equation}
									Q = 2 \hat{d} + \frac{1}{2} h_1,
									\label{op.grad_super}
								\end{equation}
								where $\hat{d}$ is a derivation operator,
								\begin{equation}
									\comm{\hat{d}}{T_a^{(m)}} = m \; T_a^{(m)},
									\quad
									\quad
									\quad
									T_a^{(m)} \in \lie \,.
									\nonumber
								\end{equation}
								The \textit{principal grading operator} decomposes the algebra in graded subspaces, $\lie  = \bigoplus_a \lie_a$, where
								\begin{equation}
									\left[Q, \mathcal{G}_a\right] = a \; \mathcal{G}_a, \qquad \left[\mathcal{G}_a, \mathcal{G}_b\right] \in \mathcal{G}_{a+b},
									\nonumber   
								\end{equation}
								for $a, b \in  \mathbb{Z}$. For our purposes, the subspaces to consider are:
								\begin{equation}
									\begin{split}
										\lie_{2m} &= \left\{ h_1^{(m)} \right\},
										\\
										\lie_{2m+\frac{1}{2}} &= \left\{ E_{\alpha_2}^{(m+\frac{1}{2})}, \; E_{\alpha_1 + \alpha_2}^{(m)}, \; E_{-\alpha_2}^{(m)}, \; E_{-\alpha_1 - \alpha_2}^{(m+\frac{1}{2})} \right\},
										\\
										\lie_{2m+1} &= \left\{ E_{\alpha_1}^{(m)}, \; E_{-\alpha_1}^{(m+1)}, \; h_2^{(m+\frac{1}{2})} \right\},
										\\
										\lie_{2m+\frac{3}{2}} &= \left\{ E_{\alpha_2}^{(m+1)}, \; E_{\alpha_1 + \alpha_2}^{(m+\frac{1}{2})}, \; E_{-\alpha_2}^{(m+\frac{1}{2})}, \; E_{-\alpha_1 - \alpha_2}^{(m+1)} \right\}.
									\end{split}
									\label{subspace}
								\end{equation}
								Another key ingredient in constructing our models is the grade one constant element, defined as
								\begin{equation}
									\label{super_E1}
									\mathcal{E}^{(1)} = E_{\alpha_1}^{(0)} + E_{-\alpha_1}^{(1)} + h_1^{(\frac{1}{2})} + 2 h_2^{(\frac{1}{2})}
									= \mqty( \sqrt{\la} & 1 & 0 \\
									\la & \sqrt{\la} & 0 \\
									0 & 0 & 2 \sqrt{\la}),
								\end{equation}
								which decomposes the algebra into $\lie = \mathcal{K} \bigoplus \mathcal{M}$, where $\mathcal{K}_{\mathcal{E}} =  \left\{ x \in \lie \, | \, \comm{\mathcal{E}}{x} = 0 \right\}$ is the kernel of $\mathcal{E}$, and $\mathcal{M}_{\mathcal{E}}$ its complement,
								\begin{equation}
									\begin{split}
										\mathcal{K}_\text{Bose} = \mathcal{K}_{\mathcal{E}} \cap L_0 &= \left\{ K_1^{(2m+1}, \; K_2^{(2m+1)} \right\},
										\\
										\mathcal{K}_\text{Fermi} = \mathcal{K}_{\mathcal{E}} \cap L_1 &= \left\{ F_1^{(2m+\frac{1}{2})}, \; F_2^{(2m+3/2)}  \right\},
									\end{split}
									\quad
									\begin{split}
										\mathcal{M}_\text{Bose} = \mathcal{M}_{\mathcal{E}} \cap L_0 &= \left\{ M_1^{(2m)} \right\},
										\\
										\mathcal{M}_\text{Fermi} = \mathcal{M}_{\mathcal{E}} \cap L_1 &= \left\{ G_1^{(2m+3/2)}, \; G_2^{(2m+\frac{1}{2})} \right\}.
									\end{split}
									\label{kernel.super.E1}
								\end{equation}
								The bosonic generators are be defined as
								\begin{equation}
									\begin{split}
										K_1^{(2m+1)} &= E_{\alpha_1}^{(m)} + E_{-\alpha_1}^{(m+1)},\\
										K_2^{(2m+1)} &= h_1^{(m+\frac{1}{2})} + 2 h_2^{(m+\frac{1}{2})},
									\end{split}
									\qquad
									\begin{split}
										M_1^{(2m)} &=  h_1^{(m)},\\
										M_2^{(2m+1)} &= E_{\alpha_1}^{(m)} - E_{-\alpha_1}^{(m+1)},   
									\end{split}
									\nonumber
								\end{equation}
								and the fermionic generators are
								\begin{subequations}
									\begin{align*}
										F_{1}^{(2 m+\frac{1}{2})}&=\left(E_{\alpha_2}^{(m+\frac{1}{2})}+E_{\alpha_1+\alpha_2}^{(m)}\right)+\left(E_{-\alpha_2}^{(m)}+E_{-\left(\alpha_1+\alpha_2\right)}^{(m+\frac{1}{2})}\right),
										\\
										F_{2}^{(2 m+\frac{3}{2})}&=\left(E_{\alpha_2}^{(m+1)}+E_{\alpha_1+\alpha_2}^{(m+\frac{1}{2})}\right)-\left(E_{-\alpha_2}^{(m+\frac{1}{2})}+E_{-\left(\alpha_1+\alpha_2\right)}^{(m+1)}\right),
										\\
										G_{1}^{(2 m+\frac{3}{2})}&=\left(E_{\alpha_2}^{(m+1)}-E_{\alpha_1+\alpha_2}^{(m+\frac{1}{2})}\right)+\left(E_{-\alpha_2}^{(m+\frac{1}{2})}-E_{-\left(\alpha_1+\alpha_2\right)}^{(m+1)}\right),
										\\
										G_{2}^{(2 m+\frac{1}{2})}&=\left(E_{\alpha_2}^{(m+\frac{1}{2})}-E_{\alpha_1+\alpha_2}^{(m)}\right)-\left(E_{-\alpha_2}^{(m)}-E_{-\left(\alpha_1+\alpha_2\right)}^{(m+\frac{1}{2})}\right).
									\end{align*}
								\end{subequations}
								From \eqref{op.grad_super} and \eqref{super_E1}, we can reorganize the graded subspaces \eqref{subspace} in terms of the kernel-image decomposition in the following way,
								\begin{equation}
									\begin{split}
										\lie_{2m} &= \left\{ M_1^{(2m)} \right\},
										\\
										\lie_{2m+\frac{1}{2}} &= \left\{F_1^{(2m+\frac{1}{2})}, \; G_2^{(2m+\frac{1}{2})} \right\},
										\\
										\lie_{2m+1} &= \left\{ K_1^{(2m+1)}, \; K_2^{(2m+1)}, \; M_2^{(2m+1)} \right\},
										\\
										\lie_{2m+\frac{3}{2}} &= \left\{ F_2^{(2m+\frac{3}{2})}, \; G_1^{(2m+\frac{3}{2})}\right\}.
									\end{split}
									\label{subspace-k+i}
								\end{equation}
								The commutation relations  of the algebra are then given by
								\begin{equation*}
									\begin{split}
										&\comm{K_1^{(2m+1)}}{K_1^{(2n+1)}} = \hat{\kappa}(m-n)\delta_{n+m+1,0}, \\
										&\comm{K_1^{(2m+1)}}{K_2^{(2n+1)}} = 0, \\
										&\comm{K_1^{(2m+1)}}{M_1^{(2n)}} = -2 M_2^{(2m+2n+1)}, \\
										&\comm{K_1^{(2m+1)}}{M_2^{(2n+1)}} = -2 M_1^{(2(m+n+1))} -\hat{\kappa} (m+n)\delta_{n+m+1,0}, \\
										&\comm{K_2^{(2m+1)}}{K_2^{(2n+1)}} = -\hat{\kappa} (2m+1) \delta_{n+m+1,0}, \\
										&\comm{K_2^{(2m+1)}}{M_1^{(2n)}} = 0, \\
										&\comm{K_2^{(2m+1)}}{M_2^{(2n+1)}} = 0, \\
										&\comm{M_1^{(2m)}}{M_1^{(2n)}} = 2 \hat{\kappa} m \delta_{m+n,0}, \\
										&\comm{M_1^{(2m)}}{M_2^{(2n+1)}} = 2 K_1^{(2m+2n+1)}, \\
										&\comm{M_2^{(2m+1)}}{M_2^{(2n+1)}} = - \hat{\kappa} (m-n) \delta_{n+m+1,0}, \\									
										&\comm{K_1^{(2m+1)}}{F_1^{(2n+1/2)}} = F_2^{(2(m+n)+3/2)}, \\
										&\comm{K_1^{(2m+1)}}{F_2^{(2n+3/2)}} = F_1^{(2(m+n+1)+1/2)}, \\
										&\comm{K_1^{(2m+1)}}{G_1^{(2n+3/2)}} = - G_2^{(2(m+n+1)+1/2)},
									\end{split}
									\begin{split}									
										&\comm{K_1^{(2m+1)}}{G_2^{(2n+1/2)}} = - G_1^{(2(m+n)+3/2)}, \\
										&\comm{K_2^{(2m+1)}}{F_1^{(2n+1/2)}} = - F_2^{(2(m+n)+3/2)}, \\
										&\comm{K_2^{(2m+1)}}{F_2^{(2n+3/2)}} = - F_1^{(2(m+n+1)+1/2)}, \\
										&\comm{K_2^{(2m+1)}}{G_1^{(2n+3/2)}} = - G_2^{(2(m+n+1)+1/2)}, \\
										&\comm{K_2^{(2m+1)}}{G_2^{(2n+1/2)}} = - G_1^{(2(m+n)+3/2)}, \\		
										&\comm{M_1^{(2m)}}{F_1^{(2n+1/2)}} = - G_2^{(2(m+2)+1/2)}, \\
										&\comm{M_1^{(2m)}}{F_2^{(2n+3/2)}} = - G_1^{(2(m+2)+3/2)}, \\
										&\comm{M_1^{(2m)}}{G_1^{(2n+3/2)}} = - F_2^{(2(m+2)+3/2)}, \\
										&\comm{M_1^{(2m+1)}}{G_2^{(2n+1/2)}} = - F_1^{(2(m+2)+1/2)}, \\
										&\comm{M_2^{(2m+1)}}{F_1^{(2n+1/2)}} = - G_1^{(2(m+n)+3/2)}, \\
										&\comm{M_2^{(2m+1)}}{F_2^{(2n+3/2)}} = - G_2^{(2(m+n+1)+1/2)}, \\
										&\comm{M_2^{(2m+1)}}{G_1^{(2n+3/2)}} = F_1^{(2(m+n+1)+1/2)}, \\
										&\comm{M_2^{(2m+1)}}{G_2^{(2n+1/2)}} = F_2^{(2(m+n)+3/2)}.
									\end{split}
								\end{equation*}
								and the anti-commutations relations, given by 
								\begin{equation}
									\begin{split}
										\acomm{F_1^{(2m+1/2)}}{F_1^{(2n+1/2)}} &= 2 (K_1^{(2m+2n+1)} + K_2^{(2m+2n+1)}) - \hat{\kappa} (4m+1) \delta_{n+m+1,0}, \\
										\acomm{F_1^{(2m+1/2)}}{F_2^{(2n+3/2)}} &= 0, \\
										\acomm{F_1^{(2m+1/2)}}{G_1^{(2n+3/2)}} &=  -2 M_1^{(2(m+n+1))} + \hat{\kappa} \delta_{n+m+1,0}, \\
										\acomm{F_1^{(2m+1/2)}}{G_2^{(2n+1/2)}} &= -2 M_2^{(2m+2n+1)}, \\ 
										\acomm{F_2^{(2m+3/2)}}{F_2^{(2n+3/2)}} &= - (K_1^{(2(m+n+1)+1)} + K_2^{(2(m+n+1)+1)}), \\
										\acomm{F_2^{(2m+3/2)}}{G_1^{(2n+3/2)}} &= 2 M_2^{(2(m+n+1)+1)}, \\
										\acomm{F_2^{(2m+3/2)}}{G_2^{(2n+1/2)}} &= 2 M_1^{(2(m+n+1))} - \hat{\kappa} \delta_{n+m+1,0}, \\ 
										\acomm{G_1^{(2m+3/2)}}{G_1^{(2n+3/2)}} &= -2 (K_1^{(2(m+n+1)+1)} - K_2^{(2(m+n+1)+1)}), \\
										\acomm{G_1^{(2m+3/2)}}{G_2^{(2n+1/2)}} &= \hat{\kappa} \left(2m-2n-1\right)\delta_{n+m+1,0}, \\ 
										\acomm{G_2^{(2m+1/2)}}{G_2^{(2n+1/2)}} &= 2 (K_1^{(2m+2n+1)} + K_2^{(2m+2n+1)}).
									\end{split}
									\nonumber
								\end{equation}
								
								\section{Grade decomposition of zero curvature condition}
								\label{ap.gradezcc}
								In this appendix, we present  explicitly the grade by grade decomposition proposed for each ansatz in section \ref{sec.flows}.
								\begin{itemize}
									\item The smKdV positive flows
									\begin{equation}
										\comm{\pa_x + A_x}{\; \pa_{t_N} +  D^{(N)}_N+ D^{(N-\frac{1}{2})}_N +\cdots + D^{(0)}_N} = 0,
										\label{zcc.1}
									\end{equation}
									decomposes as follows
									\begin{subequations}
										\begin{align}
											\comm{\cE}{ D^{(N)}_N} &= 0, \label{highgrade.smkdv}\\
											\comm{\cE}{D^{(N-\frac{1}{2})}_N} + \comm{A_{\frac{1}{2}}}{D^{(N)}_N} &= 0, \\
											\comm{\cE}{D^{(N-1)}_N} + \comm{A_{\frac{1}{2}}}{D^{(N-\frac{1}{2})}_N} + \comm{A_0}{D^{(N)}_N} + \pa_x D^{(N)}_N &= 0, \\
											\nonumber & \; \; \vdots \\
											\comm{A_{\frac{1}{2}}}{D^{(0)}_N} + \comm{A_0}{D^{(\frac{1}{2})}_N} + \pa_x D^{(\frac{1}{2})}_N - \pa_{t_{N}} A_{\frac{1}{2}} & = 0, \\
											\comm{A_0}{D^{(0)}_N} + \pa_x D^{(0)}_{N} - \pa_{t_{N}}A_0 &= 0
										\end{align}
									\end{subequations}
									\item For sKdV positive flows, we have
									\begin{equation}
										\comm{\pa_x + A_x}{\; \pa_{t_N} +  \cD^{(N)}_N+ \cD^{(N-\frac{1}{2})}_N +\cdots + \cD^{(-1)}_N} = 0,
										\label{zcc.2}
									\end{equation}
									\begin{subequations}
										\begin{align}
											\comm{\cE}{ \cD^{(N)}_N} &= 0, \label{highgrade.skdv}\\
											\comm{\cE}{ \cD^{(N-\frac{1}{2})}_N} &= 0, \\
											\pa_x \cD^{(N)}_N + \comm{\cE}{\cD^{(N-1)}_N} &= 0, \\
											\nonumber & \; \; \vdots \\
											\pa_x \cD^{(-\frac{1}{2})}_N + \comm{A_{-1}}{\cD^{(\frac{1}{2})}_N} + \comm{A_{-\frac{1}{2}}}{\cD^{(0)}_N} - \pa_{t_{N}} A_{-\frac{1}{2}} &=0, \\
											\pa_x \cD^{(-1)}_N + \comm{A_{-1}}{\cD^{(0)}_N} + \comm{A_{-\frac{1}{2}}}{\cD^{(-\frac{1}{2})}_N} - \pa_{t_{N}} A_{-1} &=0.
										\end{align}
									\end{subequations}
									\item The smKdV negative flows
									\begin{equation}
										\comm{\pa_x + A_x}{\; \pa_{t_{-N}} + D^{(-\frac{1}{2})}_{-N} + D^{(-1)}_N+\cdots + D^{(-N)}_{-N}} = 0,
										\label{zcc.3}
									\end{equation}
									decomposes as follows
									\begin{subequations}
										\begin{align}
											\label{lowgrade.smkdv} \comm{A_0}{ D^{(-N)}_{-N}} + \pa_x  D^{(-N)}_{-N} &= 0, \\
											\comm{A_{\frac{1}{2}}}{D^{(-N)}_{-N}} + \comm{A_0}{D^{(-N+\frac{1}{2})}_{-N}} + \pa_x D^{(-N+\frac{1}{2})}_{-N} &= 0, \\
											\nonumber \; \; \vdots & \\
											\comm{\cE}{D^{(-1)}_{-N}} + \comm{A_0}{D^{(-\frac{1}{2})}_{-N}} - \pa_{t_{-N}} A_0 &= 0, \label{2c} \\
											\comm{\cE}{D^{(-\frac{1}{2})}_{-N}} - \pa_{t_{-N}} A_{\frac{1}{2}} &= 0. \label{2d}
										\end{align}
									\end{subequations}
									\item The sKdV negative flows
									\begin{equation}
										\comm{\pa_x + A_x}{\; \pa_{t_{-N}} + \cD^{(-\frac{1}{2})}_{-N} + \cD^{(-1)}_N+\cdots + \cD^{(-N-2)}_{-N}} = 0,
										\label{zcc.4}
									\end{equation}
									decomposes in the following way
									\begin{subequations}
										\begin{align}
											\comm{A_{-1}}{\cD^{(-N-2)}_{-N}} &= 0, \label{lowest.grade.skdv.neg} \\
											\comm{A_{-\frac{1}{2}}}{\cD^{(-N-2)}_{-N}} + \comm{A_{-1}}{\cD^{(-N-\frac{3}{2})}_{-N}} &= 0, \label{lowest.grade.skdv.neg2}\\
											\pa_x \cD^{(-N-2)}_{-N} + \comm{A_{-1}}{\cD^{(-N-1)}_{-N}} + \comm{A_{-\frac{1}{2}}}{\cD^{(-N-\frac{3}{2})}_{-N}} &= 0, \\
											\nonumber & \; \; \vdots  \\
											\pa_x \cD^{(-1)}_{-N} + \comm{A_{-\frac{1}{2}}}{\cD^{(-\frac{1}{2})}_{-N}} + \comm{\cE}{\cD^{(-2)}_{-N}} - \pa_{t_{-N}} A_{-1} &= 0, \label{10d}\\
											\pa_x \cD^{(-\frac{1}{2})}_{-N} + \comm{\cE}{\cD^{(-\frac{3}{2})}_{-N}} - \pa_{t_{-N}} A_{-\frac{1}{2}} &= 0, \label{10e}\\
											\comm{\cE}{\cD^{(-1)}_{-N}} &= 0, \label{10f} \\
											\comm{\cE}{\cD^{(-\frac{1}{2})}_{-N}} &= 0 \label{10g}.
										\end{align}
									\end{subequations}
									
								\end{itemize}
								
								\section{Lax Pairs for smKdV and sKdV}
								\label{ap.laxpairs}
								Here we present the explicit form of the Lax pairs for some specific flows of both sKdV and smKdV hierarchies.
								\subsection{The smKdV hierarchy}
								
								\begin{itemize}
									\item $N=3$
									\begin{equation}
										\begin{split} \label{smkdv.t3.lax}
											A_{t_{3}}^{\smkdv} = & \;\; K_1^{(3)} + K_2^{(3)}+\bar{\psi} \; G_2^{(\frac{5}{2})}+v \; M_1^{(2)}+ \frac{1}{2} v \; \bar{\psi} \; F_2^{(\frac{3}{2})} + \frac{1}{2} \pa_x \bar{\psi} \; G_1^{(\frac{3}{2})}\\
											&-\frac{1}{2} \left\{ v^2 + \bpsi \pa_x \bpsi \right\} K_1^{(1)} + \frac{1}{2} \bpsi \pa_x \bpsi \; K_2^{(1)} + \frac{1}{2} \pa_x v \; M_2^{(1)}\\
											&+\frac{1}{4} \left\{ v\pa_x \bpsi - \pa_x v \bpsi \right\} \; F_1^{(\frac{1}{2})} + \left\{ \frac{1}{4} \pa_x^2 \bpsi - \frac{1}{2} v^2 \bpsi \right\} \; G_2^{\frac{1}{2}}\\
											&+\left\{ \frac{1}{4} \pa_x^2 v - \frac{1}{2} v^3 - \frac{3}{4 }v \bpsi \pa_x \bpsi \right\} \; M_1^{(0)}
										\end{split}
									\end{equation}
									\item $N=5$
									\begin{equation}
										A^{\smkdv}_{t_{5}} = D^{(5)}_{5} + D^{(\frac{9}{2})}_{5} + \cdots + D^{(0)}_{5}
										\label{smkdv.tm5.lax}
									\end{equation}
									with
									\begin{subequations}
										\begin{align*}
											D^{(5)}_{5}  &= K_1^{(5)}+ K_2^{(5)} = \mathcal{E}^{(5)}, \\[3mm]
											D^{(\frac{9}{2})}_{5} &= \bar{\psi} \; G_2^{(9/2)}, \\[3mm]
											D^{(4)}_{5} &= v \; M_1^{(4)},\\[3mm]
											D^{(\frac{7}{2})}_{5}&= \frac{1}{2} v \bar{\psi} \; F_2^{(7/2)} + \frac{1}{2} \pa_x \bar{\psi} \; G_1^{(7/2)},\\[3mm]
											D^{(3)}_{5} &= - \frac{1}{2} \left( \bar{\psi} \pa_x \bar{\psi} + v^2 \right) \; K_1^{(3)} + \frac{1}{2} \bar{\psi} \pa_x \bar{\psi} \; K_2^{(3)} + \frac{1}{2} \pa_x v \; M_2^{(3)},  \\[3mm]
											D^{(\frac{5}{2})}_{5}&=\frac{1}{4}\left(v \pa_x \bar - \pa_x v \bar{\psi} \right) F_1^{(5/2)} + \frac{1}{4}\left(-2v^2 \bar{\psi} + \pa_x^2 \bar{\psi} \right) \; G_2^{(5/2)},\\[3mm]
											D^{(2)}_{5} &= \frac{1}{4} \left(-3 v \bar{\psi}\pa_x \bar{\psi} - 2 v^3 + \pa_x^2 v \right) \; M_1^{(2)}, 	\\[3mm]
											D^{(\frac{3}{2})}_{5} &= \frac{1}{8} \left(v \pa_x^2 \bar{\psi} + \bar{\psi} \pa_x^2 v -v^3 \bar{\psi} - \pa_x v \pa_x \bar{\psi} \right) \; F_2^{(3/2)} + \frac{1}{8} \left( \pa_x^3 \bar{\psi} - 3 v^2 \pa_x \bar{\psi} - 3 v \pa_x v \bar{\psi} \right) \; G_1^{(3/2)},\\[3mm]
											D^{(1)}_{5}&= \frac{1}{8} \left( \bar{\psi} \pa_x^3 \bar{\psi} - \pa_x \bar{\psi} \pa_x^2 \bar{\psi} - 8 v^2 \bar{\psi} \pa_x \bar{\psi} - (\pa_x v)^2 + 2 v \pa_x^2 v - 3 v^4 \right) \; K_1^{(1)} \\
											&+ \frac{1}{8} \left(\bar{\psi} \pa_x^3 \bar{\psi} - \pa_x \bar{\psi} \pa_x^2 \bar{\psi} - 4 v^2 \bar{\psi} \pa_x \bar{\psi} \right) \; K_2^{(1)} + \frac{1}{8} \left( \pa_x^3 v - 6 v^2 \pa_x v - 4 \pa_x v \bar{\psi} \pa_x \bar{\psi} - 2 v \bar{\psi} \pa_x^2 \bar{\psi} \right) \; M_2^{(1)}, \\[3mm]
											D^{(\frac{1}{2})}_{5}&= \frac{1}{16} \left( 4 v^2 \pa_x v \bar{\psi} - 4 v^3 \pa_x\bar{\psi} - \pa_x v \pa_x^2 \bar{\psi} + \pa_x^2 v \pa_x \bar{\psi} + v \pa_x^3 \bar{\psi} - \pa_x^3v \bar{\psi} \right) \; F_1^{(1/2)} \\
											&+ \frac{1}{16} \left(\pa_x^4 \bar{\psi} - 4 v^2 \pa_x^2 \bar{\psi} - 8 v \pa_x v \pa_x \bar{\psi} -2 (\pa_x v)^2 \bar{\psi} - 6 v \pa_x^2 v \bar{\psi} + 6 v^4 \bar{\psi}  \right) \; G_2^{(1/2)},\\[3mm]	
											D^{(0)}_{5}&=\frac{1}{16} \left( \pa_x^4 v - 10 v^2 \pa_x^2 v - 10 v (\pa_x v)^2 + 6 v^5 + 20 v^3 \bar{\psi} \pa_x \bar{\psi} - 5 \pa_x ( \pa_x v \bar{\psi} \pa_x \bar{\psi} ) - 5 v \bar{\psi} \pa_x^3 \bar{\psi} \right) M_1^{(2)}.							
										\end{align*}
									\end{subequations}

									\item $N=\frac{1}{2}$
									\begin{equation}
										\label{smkdv.t12.lax} 
										A_{t_{\frac{1}{2}}}^{\smkdv} = \xi F_{1}^{(\frac{1}{2})} + \xi \bp M_1^{(0)}
									\end{equation}
									\item $N=-1$
									\begin{equation}
										\label{smkdv.tm1.lax} 
										A_{t_{-1}}^{\smkdv} = \cosh{2\phi} \; K_1^{(-1)} + K_2^{(-1)} -  \sinh{2 \phi} \; M_2^{(-1)}  -  \psi\sinh{\phi}  \; F_{2}^{(-\frac{1}{2})} - \psi  \cosh{\phi} \; G_1^{(-\frac{1}{2})} 
									\end{equation}
									where  $v(x,t_{-1}) = \partial_x \phi(x,t_{-1})$.
									\item $N=-2$
									\begin{align}
										\label{smkdv.tm2.lax}
										A_{t_{-2}}^{\text{smKdV}} &= M_1^{(-2)}-\frac{e^{\phi} \psi_{-}}{2} \left(F_1^{(-\frac{3}{2})}+G_2^{(-\frac{3}{2})}\right) -\frac{e^{-\phi} \psi_{+}}{2} \left(F_1^{(-\frac{3}{2})}-G_2^{(-\frac{3}{2})}\right) \nonumber \\
										&+a_{-}\left(K_1^{(-1)}+M_2^{(-1)}\right)-a_{+}\left(K_1^{(-1)}-M_2^{(-1)}\right)+
										\left(1 + \psi_{-} \psi_{+} \right)K_2^{(-1)}  \\
										&+ \Om_{+} \left(F_2^{(-\frac{1}{2})}+G_1^{(-\frac{1}{2})}\right)+ \Om_{-} \left(F_2^{(-\frac{1}{2})}-G_1^{(-\frac{1}{2})}\right) \nonumber
									\end{align}
									where
									\begin{align*}
										v= \pa_x \phi \quad \quad \psi_{\pm} = \pa_x^{-1}\left( e^{\pm \phi} \bp\right),
										\quad
										\quad
										a_{\pm} = e^{\pm 2\phi} \pa_{x}^{-1} \left[e^{\mp 2\phi} \left( 1+\psi_{\mp} \pa_x \psi_{\pm} \right)\right],
									\end{align*}
									and
									\begin{equation*}
										\Om_{\pm} = \frac{e^{\pm \phi}}{2} \pa_x^{-1} \left[e^{\mp 2\phi} \psi_{\pm}-\psi_{\mp} \mp \pa_x \psi_{\mp} \left(1+\psi_{-}\psi_{+}\mp 2a_{\pm}\right)\right].
									\end{equation*}
								\end{itemize}
								
								\subsection{The sKdV hierarchy}
								
								\begin{itemize}
									\item $N=3$
									\begin{equation}
										\begin{split} \label{skdv.t3.lax}
											A_{t_{3}}^{\skdv} = & \;\; K_1^{(3)} + K_2^{(3)}+\frac{\bar{\chi}}{2} G_1^{\left(\frac{3}{2}\right)}-\frac{J}{2} M_2^{(1)}+\frac{\partial_x \bar{\chi}}{4} G_2^{\left(\frac{1}{2}\right)}\\
											&-\frac{\partial_x J}{4}M_1^{(0)}+\left(\frac{\partial_x^2 \chi}{8}-\frac{J \bar{\chi}}{4}\right)\left(G_1^{\left(\frac{2}{2}\right)}+F_2^{\left(-\frac{1}{2}\right)}\right)\\
											&+\left(\frac{\partial_x^2 J}{8}-\frac{J^2}{4}-\frac{\bar{\chi} \partial_x \bar{\chi}}{8} \right)\left(K_1^{(-1)}-M_2^{(-1)}\right)
										\end{split}
									\end{equation}
									\item $N=5$
									\begin{equation}
										A^{\skdv}_{t_{5}} = \cD^{(5)}_{5} + \cD^{(\frac{9}{2})}_{5} + \cdots + \cD^{(-1)}_{5}
										\label{skdv.tm5.lax}
									\end{equation}
									with
									\begin{subequations}
										\begin{align*}
											\cD^{(5)}_{5}  &= K_1^{(5)}+ K_2^{(5)}= \mathcal{E}^{(5)}, \\[3mm]
											\cD^{(\frac{9}{2})}_{5} &=0, \\[3mm]
											\cD^{(4)}_{5} &=0,\\[3mm]
											\cD^{(\frac{7}{2})}_{5}&=\frac{\bchi}{2}G_1^{(\frac{7}{2})}\\[3mm]
											\cD^{(3)}_{5} &=-\frac{J}{2}M_2^{(3)},\\[3mm]
											\cD^{(\frac{5}{2})}_{5}&=\frac{\pa_x\bchi}{4}G_2^{(\frac{5}{2})}\\[3mm]
											\cD^{(2)}_{5} &=-\frac{\pa_x J}{4}M_1^{(2)},	\\[3mm]
											\cD^{(\frac{3}{2})}_{5}&=\frac{J\bchi}{8}F_2^{(\frac{3}{2})}+\frac{1}{8}\left(\pa_x^2\bchi-2J\bchi\right)G_1^{(\frac{3}{2})},\\[3mm]
											\cD^{(1)}_{5}&=\frac{1}{8}\left(J^2+\bchi\pa_x \bchi\right)K_1^{(1)}-\frac{1}{8}\bchi\pa_x \bchi K_2^{(1)}+\frac{1}{8}\left(-\pa_x^2 J+2J^2+\bchi\pa_x \bchi\right)M_2^{(1)},\\[3mm]
											\cD^{(\frac{1}{2})}_{5}&=\frac{1}{16}\left(-\pa_xJ \bchi+J\pa_x\bchi\right)F_1^{(\frac{1}{2})}+\frac{1}{16}\left(-\pa_x^3 \bchi -3 \pa_x \left(J \bchi\right)\right)G_2^{(\frac{1}{2})},\\[3mm]	
											\cD^{(0)}_{5}&=\frac{1}{16}\left(-\pa_x^3J+3\pa_xJ^2+2\bchi\pa_x^2\bchi\right)M_1^{(0)},\\[3mm]
											\cD^{(-\frac{1}{2})}_{5}&=\frac{1}{32}\left(\pa_x^4 \bchi -3\pa_x^2(J\bchi)+6J^2\bchi-J\pa_x^2\bchi-\pa_x^2J\bchi\right)\left(F_2^{(-\frac{1}{2})}+G_1^{(-\frac{1}{2})}\right),\\[3mm]
											\cD^{(-1)}_{5}&=\frac{1}{32}\left(\pa_x^4 J -6(\pa_xJ)^2-8J\pa_x^2J+6J^3-4\bchi\pa_x^3\bchi+8J\bchi\pa_x\bchi\right)\left(K_1^{(-1)}-M_2^{(-1)}\right).
										\end{align*}
									\end{subequations}						
									
									\item $N=\frac{1}{2}$
									\begin{equation}
										\label{skdv.t12.lax} 
										A_{t_{\frac{1}{2}}}^{\skdv} = \xi F_{1}^{(\frac{1}{2})} - \xi \bchi \left(K_1^{(-1)}-M_2^{(-1)}\right)
									\end{equation}
									\item $N=-1$
									\begin{equation}
										A^{\skdv}_{t_{-1}} = \cD^{(-3)}_{-1} + \cD^{(-\frac{5}{2})}_{-1} + \cdots + \cD^{(-\frac{1}{2})}_{-1}
										\label{skdv.tm1.lax}
									\end{equation}
									with
									\begin{subequations}
										\begin{align*}
											\cD^{{(-3)}}_{-1} &= - \frac{1}{8} \left\{ \pa_x (\pa_{t_{-1}} \pa_x \eta + \pa_x \gamma)  - 2 \pa_x \eta \left( \pa_{t_{-1}}\eta + \gamma \right) \right. 
											\nonumber \\[1mm]
											& \qquad \qquad \qquad \left. + \pa_x \bar{\eta} \left( \bnu_{-} - \bnu_{+} + 2 \bg - \pa_{t_{-1}} \pa_x \bg \right) \right\} \left(K_1^{(-3)} - M_2^{(-3)}\right), \\[3mm]
											\cD^{(-\frac{5}{2})}_{-1} &= \frac{1}{8}\left( \pa_x (\bnu_{+} + \bnu_{-}) - \pa_{t_{-1}} \pa_x^2 \bg + 2 \pa_x \bg \;\g + 2 \pa_x \bg \right) \left( F_2^{(-\frac{5}{2})} + G_1^{(-\frac{5}{2})}\right), \\[3mm]
											\cD^{(-2)}_{-1} &= \frac{1}{4} \left( \pa_{t_{-1}} \pa_x \eta + \pa_x \g \right) \; M_1^{(-2)}, \\[3mm]
											\cD^{(-\frac{3}{2})}_{-1} &= \frac{1}{4} \left( \bnu_{+} - \bnu_{-} - 2 \bg \right) \; F_1^{(-\frac{3}{2})} - \frac{1}{4} \pa_{t_{-1}}\pa_x \bg \; G_2^{(-\frac{3}{2})},
											\\[3mm]
											\cD^{(-1)}_{-1} &= \frac{1}{2} \g \left(K_1^{(-1)} + K_2^{(-1)}\right) + \frac{1}{2} \pa_{t_{-1}} \eta \; K_1^{(-1)} + K_2^{(-1)}, \\[3mm]
											\cD^{(-\frac{1}{2})}_{-1} &= \frac{1}{2} \pa_{t_{-1}} \bg \; F_2^{(-\frac{1}{2})}.
										\end{align*}
									\end{subequations}
								\end{itemize}
								
								\section{Supersymmetry transformation for smKdV$(-2)$} \label{ap.susy.tm2}
								Let us consider the field equations for $smKdV(-2)$,
								\begin{subequations}
									\begin{align}
										\pa_{t_{-2}}\pa_x \phi &= -2\left(a_{-}+a_{+}\right)+2 \bp\left(\Om_{-}+\Om_{+}\right),
										\\[0.2cm]
										\pa_{t_{-2}} \bp &= -2 \left(\Omega_{+}-\Omega_{-}\right),
									\end{align}
								\end{subequations}
								with
								\begin{subequations}
									\begin{align}
										\psi_{\pm} &= \pa_x^{-1}\left( e^{\pm \phi} \bp\right),
										\\[0.2cm]
										a_{\pm} &= e^{\pm 2\phi} \pa_{x}^{-1} \left[e^{\mp 2\phi} \left( 1 + \psi_{\mp} \pa_x \psi_{\pm} \right)\right],	
									\end{align}
								\end{subequations}
								and
								\begin{equation}
									\Om_{\pm} = \frac{e^{\pm \phi}}{2} \pa_x^{-1} \left[e^{\mp 2\phi} \psi_{\pm}-\psi_{\mp} \mp \left( \pa_x \psi_{\mp} \right) \left(1+\psi_{-}\psi_{+}\mp 2a_{\pm} \right)\right],
								\end{equation}
								where the anti-derivative operator is defined by $\pa_x^{-1} f = \int^x f(y) \dd{y}$. By applying the supersymmetry transformation
								\begin{subequations}
									\begin{align}
										\delta v &= \xi \; \pa_x\bp \quad \to \quad \delta \phi = \xi \bp,
										\\[0.2cm]
										\delta \bp &= -\xi \; v = -\xi \pa_x \phi,
									\end{align}	
								\end{subequations}
								on the fields $\psi_{\pm}$, $a_{\pm}$ and $\Om_{\pm}$, we have:
								\begin{equation}
									\pa_{x} \de\psi_{+} = \de\left(e^{\phi}\bp\right)= -\xi \pa_{x} \left(e^{\phi}\right)
								\end{equation}
								and then
								\begin{equation} \label{vp.p}
									\de\psi_{+} = -\xi e^{\phi}.
								\end{equation}
								Similarly, for $\psi_{-}$ we get
								\begin{equation} \label{vp.n}
									\de\psi_{-} = \xi e^{-\phi}.
								\end{equation}
								Now consider for $a_{+}$,
								\begin{equation}
									\begin{split}
										\de a_{+} &= \de \left\{ e^{2\phi} \pa_{x}^{-1} \left[e^{- 2\phi} \left( 1 + \psi_{-} \pa_x \psi_{+} \right)\right]\right\}\\
										&= 2\de \phi a_{+}+e^{2\phi} \pa_x^{-1} \left\{\de\left[e^{- 2\phi} \left( 1 + \psi_{-} \pa_x \psi_{+} \right)\right]\right\}.
									\end{split}
								\end{equation}
								Now, let us evaluate the second term
								\begin{equation*}
									\begin{split}
										\de\left[e^{- 2\phi} \left( 1 + \psi_{-} \pa_x \psi_{+} \right)\right] &= -2 \de \phi e^{- 2\phi} \left( 1 + \psi_{-} \pa_x \psi_{+} \right) + e^{-2\phi} \de \left(\psi_{-} \pa_x \psi_{+} \right) \\
										&= -2\xi \bp e^{-2\phi }+ e^{-2\phi} \xi e^{-\phi} e^{\phi} \bp - e^{-2\phi} \psi_{-} \xi \pa_x{\phi} e^{\phi}\\
										&=-\xi \pa_x \left(\psi_{-}  e^{-\phi}\right) 
									\end{split}
								\end{equation*}
								leading to
								\begin{equation}  \label{va.p}
									\begin{split}
										\de a_{+} &= 2\de \phi a_{+}+e^{2\phi} \pa_x^{-1} \left\{\de\left[e^{- 2\phi} \left( 1 + \psi_{-} \pa_x \psi_{+} \right)\right]\right\}\\
										&= 2 \xi \bp a_{+}-\xi e^{2\phi} \pa_x^{-1} \left\{\pa_x \left(\psi_{-}  e^{-\phi}\right) \right\}\\
										&=  2 \xi \bp a_{+}-\xi e^{\phi} \psi_{-}.
									\end{split}
								\end{equation} 
								For $a_{-}$, the calculation is rather similar and give us the following result
								\begin{equation}  \label{va.n}
									\de a_{-} =- 2 \xi \bp a_{-}+\xi e^{-\phi} \psi_{+}.
								\end{equation}
								Finally for $\Om_{+}$, we have
								\begin{equation}
									\begin{split}
										\de \Om_{+} = \de \phi \Om_{+} + \frac{e^{\phi}}{2} \pa_x^{-1} \left\{ \de\left[e^{- 2\phi} \psi_{+}-\psi_{-} -\left( \pa_x \psi_{-} \right) \left(- 2a_{+}+\psi_{-}\psi_{+} + 1\right)\right]\right\} 	.
									\end{split}
								\end{equation}
								Calculating the second term, we find
								\begin{equation}
									\begin{split}
										\de&\left[e^{- 2\phi} \psi_{+}-\psi_{-} -\left( \pa_x \psi_{-} \right) \left(- 2a_{+}+\psi_{-}\psi_{+} + 1\right)\right]=\\
										\xi e^{-2\phi}&\psi_{+}\bp-2\xi  e^{-\phi}  +\pa_x\phi e^{-\phi}\left(- 2a_{+}+\psi_{-}\psi_{+} + 1\right)+3\xi \bp \psi_{-}. \\
									\end{split}
								\end{equation}
								By noting that the following relation holds,
								\begin{equation}
									\begin{split}
										\pa_x&\left(e^{-\phi} \left(-2a_{+}-1-\psi_{-}\psi_{+}\right)\right)= \\
										\pa_x& \phi e^{-\phi}\left(-2a_{+}+1+\psi_{-}\psi_{+}\right)-2\xi  e^{-\phi} +\xi e^{-2\phi}\psi_{+}\bp +3\xi \bp \psi_{-}
									\end{split}
								\end{equation}
								we can write
								\begin{equation}
									\de\left[e^{- 2\phi} \psi_{+}-\psi_{-} -\left( \pa_x \psi_{-} \right) \left(- 2a_{+}+\psi_{-}\psi_{+} + 1\right)\right]= \xi \pa_x\left(e^{-\phi} \left(-2a_{+}-1-\psi_{-}\psi_{+}\right)\right)
								\end{equation}
								and finally, we get
								\begin{equation} \label{vom.p}
									\begin{split}
										\de \Om_{+} &= \xi \bp \Om_{+} +\frac{\xi e^{\phi}}{2} \pa_x^{-1} \left\{ \pa_x\left(e^{-\phi} \left(-2a_{+}-1-\psi_{-}\psi_{+}\right)\right)\right\}\\ 
										\de \Om_{+} &= \xi \bp \Om_{+}- \xi a_{+}- \xi \frac{\psi_{-}\psi_{+}+1}{2}.	
									\end{split}
								\end{equation}
								Now, one can calculate for $\Om_{-}$ and obtain
								\begin{equation} \label{vom.n}
									\begin{split}
										\de \Om_{-} &= -\xi \bp \Om_{-}+\xi a_{-}-\xi \frac{\psi_{-}\psi_{+}+1}{2}.	
									\end{split}
								\end{equation}
								By using \eqref{vp.p},\eqref{vp.n}, \eqref{va.p}, \eqref{vom.p} and \eqref{vom.n}, we can finally evaluate the supersymmetry of the field equation, namely
								\begin{equation}
									\begin{aligned}
										\delta \pa_{t_{-2}} \bp &= -2 \left(\de \Omega_{+}-\de \Omega_{-}\right)\\
										&= -2 \left(\xi \bp \Om_{+}- \xi a_{+}- \xi \frac{\psi_{-}\psi_{+}+1}{2}+\xi \bp \Om_{-}- \xi a_{-}+ \xi \frac{\psi_{-}\psi_{+}+1}{2}\right)\\
										&= -\xi \left[2\bp \left( \Om_{+}+ \Om_{-}\right)- 2 \left( a_{+}+ a_{-}\right)\right] = -\xi\pa_{t_{-2}}\pa_x \phi
									\end{aligned}
								\end{equation}
								implying that
								\begin{equation}
									\begin{aligned}
										\delta \pa_{t_{-2}} \bp = -\xi\pa_{t_{-2}}\pa_x \phi.
									\end{aligned}
								\end{equation}
								If we check the other way around, we have
								\begin{equation}
									\begin{aligned}
										\delta \pa_x \pa_{t_{-2}} \phi &= -2\left(\de a_{-}+\de a_{+}\right)+2 \de \bp\left(\Om_{-}+\Om_{+}\right)+2  \bp\left(\de \Om_{-}+\de \Om_{+}\right)\\
										&=\xi \left[2 \bp \left(a_{-} - a_{+}+1+\psi_{-}\psi_{+}\right)+2 \psi_{-}e^{\phi}- 2 \psi_{+} e^{-\phi} - 2 \pa_x \phi \left(\Om_{+}+\Om_{-}\right)\right]
									\end{aligned}
								\end{equation}
								and then we can verify that the fermionic equation satisfies
								\begin{equation}
									\begin{split}
										\pa_{t_{-2}} \pa_x \bp &= - 2 \left(\pa_x \Om_{+}- \pa_x \Om_{-} \right)\\
										&= 2\bp \left(a_{-} - a_{+}+1+\psi_{-}\psi_{+}\right)+2 \psi_{-}e^{\phi}- 2 \psi_{+} e^{-\phi} - 2 \pa_x \phi \left(\Om_{+}+\Om_{-}\right),
									\end{split}
								\end{equation}
								from which we conclude that this pair is supersymmetric, namely
								\begin{equation}
									\begin{aligned}
										\delta \pa_x \pa_{t_{-2}} \phi &=  \xi	\pa_{t_{-2}} \pa_x \bp.
									\end{aligned}
								\end{equation}

								\section{Determining $\theta_{-\frac{1}{2}}$}
								\label{dif.dress}
								In order to completely determine the element $\theta_{-\frac{1}{2}}$ in (\ref{eq4.28}), we need to solve the following ordinary differential equation 
								\begin{equation} \label{eqx.1}
									(\pa_x\phi) \bp \,=\pa_x \Xi - \bp_0 v_0 +\frac{3\Xi}{2}\bp_0 \bp.
								\end{equation}
								Therefore, let us analyze it for different vacuum configurations:

								\begin{enumerate}
									\item For $(v_0,\bp_0) = (0,0)$ and $(v_0,\bp_0)= (v_0,0)$, the 
									equation \eqref{eqx.1} simplifies as follows
									\begin{equation} 
										(\pa_x\phi) \bp \,=\pa_x \Xi,
									\end{equation}
									and then we get the following solution
									\begin{equation}
										\Xi = \pa_{x}^{-1} \left(\phi_x \bp \right).
									\end{equation}

									\item For $(v_0,\bp_0)= (v_0,\bp_0)$, the equation \eqref{eqx.1} simplifies as follows
									\begin{equation} 
										\left(\pa_x\phi+v_0\right) \bp \,=\pa_x \Xi +\frac{3\Xi}{2}\bp_0 \bp.
									\end{equation}
									In this case, the solution is slightly more complex, and can be written as follows
									\begin{equation}
										\Xi = e^{ -\pa_{x}^{-1} \left(\frac{3\bp_0 \bp}{2} \right)} \pa_{x}^{-1} \left(e^{\pa_{x}^{-1} \left(\frac{3\bp_0 \bp}{2} \right)} \left(\phi_x \bp+v_0\bp_0 \right)\right).
									\end{equation}
									where is possible to take the limit $\bp_0 \to 0$.
								\end{enumerate}
								\section{Vacuum projection and Heisenberg subalgebra for sKdV hierarchy}
								\label{ap.vacuumkdv}
								
								In section \ref{kdv.solutions},  we have obtained the solutions for sKdV hierarchy from the smKdV solutions by applying the super Miura transformation. However, we have not exhibited explicitly the sKdV  vacuum operators. They are presented in this appendix.
								
								Let us start from the vacuum projection of the spatial sKdV Lax
								\begin{equation}
									A_{x,\text{vac}}^{\text{sKdV}}=\Lambda=	\cE+\frac{J_0}{2} \left(K_1^{(-1)}-M_2^{(-1)}\right)+\frac{\chi_0}{2} \left(F_2^{-\frac{1}{2}}+G_1^{-\frac{1}{2}}\right).
								\end{equation}
								The Kernel of $\Lambda$ is given by
								\begin{equation}
									\mathcal{K}_\Lambda = \left\{\Lambda_{2m+1}, \Pi_{2m+1}\right\} \quad \text{such} \quad \comm{X}{\Lambda}=0 \; , \quad X \in \mathcal{K}_\Lambda,
								\end{equation}
								where 
								\begin{equation}
									\label{aux.vacuum.kdv}
									\begin{split}
										\Lambda_{(2m+1)}&=\mathcal{E}^{(2m+1)}+\frac{J_0}{2}\left(K_1^{(2m-1)}-M_2^{(2m-1)}\right)+\frac{\bchi_0}{2}\left(G_1^{(2m-\frac{1}{2})}+F_2^{(2m-\frac{1}{2})}\right),
										\\
										\Pi_{(2m+1)}&=K_2^{(2m+1)}-\frac{\bchi_0}{J_0}F_2^{(2m+\frac{3}{2})}.
									\end{split}
								\end{equation}
								We notice that any linear combination of these operators will satisfy the ZCC in the vacuum. 
								
								\noindent Firstly, in the case of positive flows, we set $(J,\bchi)= (J_0, \chi_0)$, and obtain the following vacuum projection
								\begin{equation}
									\label{vacuum.kdv.proj}
									\begin{split}
										A_{t_1,\text{vac}}^{\text{sKdV}}&=A_{x,\text{vac}}^{\text{sKdV}}= \Lambda_1\\
										A_{t_3,\text{vac}}^{\text{sKdV}}&= \Lambda_3+ \frac{J_0}{2}\left(\Pi_1-\Lambda_1\right)\\
										A_{t_{5},\text{vac}}^{\text{sKdV}}&=\Lambda_{5}+\frac{J_0}{2} \left(\Pi_{3}- \Lambda_{3}\right)+\frac{3J_0^2}{8} \left(\Lambda_{1}-\Pi_{1}\right)\\
										A_{t_{2m+1},\text{vac}}^{\text{sKdV}}&= \Lambda_{2m+1}+ \sum_{i=0}^{m-1} v_0^{2(m-i)} \left( a_i \Lambda_{2i+1}+b_i\Pi_{2i+1}\right).
									\end{split}
								\end{equation}
								Clearly, the zero vacuum limit is given by
								\begin{equation}
									A_{t_{2m+1},\text{vac}}^{\text{sKdV}}=	\mathcal{E}^{(2m+1)}.
								\end{equation}
								Then, the zero curvature condition projected on the vacuum, i.e.
								\begin{equation} \label{zc.vac.kdvp}
									\comm{A_{\text{$x$,vac}}^{\text{KdV}}}{A_{\text{$t_{N}$,vac}}^{\text{KdV}}}=0,
								\end{equation}
								is satisfied as long as we can express the vacuum potential in terms of $\Lambda_{(2m+1)}$ and $\Pi_{(2m+1)}$.
								
								For the negative sKdV flows, the vacuum projection is a little more subtle. We notice that the Lax pair given by eqs. \eqref{skdv.tm1} and \eqref{skdv.tm1.aux}, is highly dependent of $\pa_{t_{-1}} \eta$  and $\pa_{t_{-1}} \bg$. Therefore, besides specifying the vacuum,   $(J,\bchi)= (\pa_{x} \eta,\pa_{x}\bg)= (J_0, \chi_0)$, we must also specify the time derivatives. To do that, we must use the temporal and spatial Miura relations for a specified vacuum. For a zero vacuum solution, we can project \eqref{smiura.tm1.b} and  \eqref{smiura.tm1.f} to obtain:
								\begin{equation}
									\eta_x = \bg_x = \bg_t=0 \quad \text{and} \quad \eta_t=2,
								\end{equation}
								leading to the following vacuum operator
								\begin{equation}
									A_{t_{-1},\text{vac}}^{\text{sKdV}}=K_1^{(-1)}+K_2^{(-1)}=\mathcal{E}^{(-1)},
								\end{equation}
								which only commutes with other zero vacuum flows.
								However, we can also use the relations \eqref{smiura.tm2.b} and  \eqref{smiura.tm2.f}, which also holds for non-zero vacuum flows, and obtain 
								\begin{equation}
									\eta_x = v_0^2, \quad \bg_x = -v_0 \bp_0, \quad \bg_t=\frac{\bp_0(v_0-1)}{v_0^2} \quad \text{and} \quad \eta_t=\frac{2}{v_0},
								\end{equation}
								and then the vacuum operator is given by
								\begin{equation}
									A_{t_{-1},\text{vac}}^{\text{sKdV}}=\frac{1}{v_0}\left(\Lambda_{-1}- \Pi_{-1}\right)+\Pi_{-1}= J_0^{-\frac{1}{2}}\left(\Lambda_{-1}- \Pi_{-1}\right)+\Pi_{-1},
								\end{equation}
								which only commutes with non-zero vacuum operators. For lower flows, the procedure follows the same approach: to identify the negative vacuum operator, we must first establish the vacuum configuration. This will then lead to:
								\begin{equation} \label{avac.kgen.z.n}
									A_{t_{-2m+1},\text{vac}}^{\text{sKdV}}=\mathcal{E}^{(-2m+1)}
								\end{equation}
								for a zero vacuum, or to	
								\begin{equation} \label{avac.kgen.nz.n}
									A_{t_{-2m+1},\text{vac}}^{\text{sKdV}}=\sum_{i=1}^{m} J_0^{-(m-i)-\frac{1}{2}} \left( c_i \Lambda_{-2i+1}+d_i\Pi_{-2i+1}\right)
								\end{equation}
								for a non-zero vacuum.	\\
								\mbox{}
								\hspace{1cm}


								
								\bibliography{bibliography.bib}

\begin{thebibliography}{10}

\bibitem{miura_korteweg-vries_1968}
Robert~M. Miura.
\newblock Korteweg-de {Vries} {Equation} and {Generalizations}. {I}. {A}
  {Remarkable} {Explicit} {Nonlinear} {Transformation}.
\newblock {\em Journal of Mathematical Physics}, 9(8):1202--1204, August 1968.

\bibitem{gardner_method_1967}
Clifford~S. Gardner, John~M. Greene, Martin~D. Kruskal, and Robert~M. Miura.
\newblock Method for {Solving} the {Korteweg}-{deVries} {Equation}.
\newblock {\em Physical Review Letters}, 19(19):1095--1097, November 1967.

\bibitem{gardner_kortewegdevries_1974}
Clifford~S. Gardner, John~M. Greene, Martin~D. Kruskal, and Robert~M. Miura.
\newblock Korteweg‐devries equation and generalizations. {VI}. methods for
  exact solution.
\newblock {\em Communications on Pure and Applied Mathematics}, 27(1):97--133,
  January 1974.

\bibitem{gomes_negative_2009}
J~F Gomes, G~Starvaggi França, G~R De~Melo, and A~H Zimerman.
\newblock Negative even grade {mKdV} hierarchy and its soliton solutions.
\newblock {\em Journal of Physics A: Mathematical and Theoretical},
  42(44):445204, November 2009.

\bibitem{aratyn_integrable_2003}
H.~Aratyn, J.F. Gomes, and A.H. Zimerman.
\newblock Integrable hierarchy for multidimensional {Toda} equations and
  topological–anti-topological fusion.
\newblock {\em Journal of Geometry and Physics}, 46(1):21--47, April 2003.

\bibitem{zamolodchikov_infinite_1985}
A.~B. Zamolodchikov.
\newblock Infinite additional symmetries in two-dimensional conformal quantum
  field theory.
\newblock {\em Theoretical and Mathematical Physics}, 65(3):1205--1213,
  December 1985.

\bibitem{sasaki_virasoro_nodate}
Ryu Sasaki and Itaru Yamanaka.
\newblock Virasoro {Algebra}, {Vertex} {Operators}, {Quantum} {Sine}-{Gordon}
  and {Solvable} {Quantum} {Field} {Theories}.
\newblock pages 271--296, Research Institute for Mathematical Sciences, Kyoto,
  Japan.

\bibitem{eguchi_deformations_1989}
Tohru Eguchi and Sung-Kil Yang.
\newblock Deformations of conformal field theories and soliton equations.
\newblock {\em Physics Letters B}, 224(4):373--378, July 1989.

\bibitem{bazhanov_integrable_1996}
Vladimir~V. Bazhanov, Sergei~L. Lukyanov, and Alexander~B. Zamolodchikov.
\newblock Integrable structure of conformal field theory, quantum {KdV} theory
  and {Thermodynamic} {Bethe} {Ansatz}.
\newblock {\em Communications in Mathematical Physics}, 177(2):381--398, April
  1996.

\bibitem{douglas_strings_1990}
Michael~R. Douglas.
\newblock Strings in less than one dimension and the generalized {KdV}
  hierarchies.
\newblock {\em Physics Letters B}, 238(2-4):176--180, April 1990.

\bibitem{frohlich_intersection_1992}
Robbert Dijkgraaf.
\newblock Intersection {Theory}, {Integrable} {Hierarchies} and {Topological}
  {Field} {Theory}.
\newblock In J.~Fröhlich, G.~’T~Hooft, A.~Jaffe, G.~Mack, P.~K. Mitter, and
  R.~Stora, editors, {\em New {Symmetry} {Principles} in {Quantum} {Field}
  {Theory}}, volume 295, pages 95--158. Springer US, Boston, MA, 1992.
\newblock Series Title: NATO ASI Series.

\bibitem{lax_integrals_1968}
Peter~D. Lax.
\newblock Integrals of nonlinear equations of evolution and solitary waves.
\newblock {\em Communications on Pure and Applied Mathematics}, 21(5):467--490,
  September 1968.

\bibitem{gelfand_asymptotic_1975}
I.~M. Gel'fand and L.~A. Dikii.
\newblock {Asymptotic} {behaviour} {of} {the} {resolvent} {of}
  {Sturm}-{Liouville} {Equations} {and} {the} {algebra} {of} {the}
  {Korteweg}-{De} {Vries} {Equations}.
\newblock {\em Russian Mathematical Surveys}, 30(5):77--113, October 1975.

\bibitem{miwa_solitons_2000}
T.~Tetsuji Miwa, M.~Masaki Jinbo, Etsurō Date, and Miles Reid.
\newblock {\em Solitons: differential equations, symmetries and infinite
  dimensional algebras}.
\newblock Number 135 in Cambridge tracts in mathematics. Cambridge University
  Press, Cambridge, 2000.

\bibitem{drinfeld_lie_1985}
V.~G. Drinfel'd and V.~V. Sokolov.
\newblock Lie algebras and equations of {Korteweg}-de {Vries} type.
\newblock {\em Journal of Soviet Mathematics}, 30(2):1975--2036, July 1985.

\bibitem{leznov_two-dimensional_1983}
A.~N. Leznov and M.~V. Saveliev.
\newblock Two-dimensional exactly and completely integrable dynamical systems:
  {Monopoles}, instantons, dual models, relativistic strings, lund-regge model,
  generalized toda lattice, etc.
\newblock {\em Communications in Mathematical Physics}, 89(1):59--75, March
  1983.

\bibitem{olive_local_1985}
D.~Olive and N.~Turok.
\newblock Local conserved densities and zero-curvature conditions for {Toda}
  lattice field theories.
\newblock {\em Nuclear Physics B}, 257:277--301, January 1985.

\bibitem{olive_affine_1993}
David~I. Olive, Neil Turok, and Jonathan~W.R. Underwood.
\newblock Affine {Toda} solitons and vertex operators.
\newblock {\em Nuclear Physics B}, 409(3):509--546, December 1993.

\bibitem{olive_solitons_1993}
D.I. Olive, N.~Turok, and J.W.R. Underwood.
\newblock Solitons and the energy-momentum tensor for affine {Toda} theory.
\newblock {\em Nuclear Physics B}, 401(3):663--697, July 1993.

\bibitem{babelon_dressing_1992}
Olivier Babelon and Denis Bernard.
\newblock Dressing symmetries.
\newblock {\em Communications in Mathematical Physics}, 149(2):279--306,
  October 1992.

\bibitem{babelon_affine_1993}
Olivier Babelon and Denis Bernard.
\newblock {Affine} {solitons}: {A} {relation} {between} {tau} {functions},
  {dressing} {and} {Bäcklund} {transformations}.
\newblock {\em International Journal of Modern Physics A}, 08(03):507--543,
  January 1993.

\bibitem{babelon_introduction_2003}
O.~Babelon, D.~Bernard, and M.~Talon.
\newblock {\em Introduction to {Classical} {Integrable} {Systems}}.
\newblock Cambridge {Monographs} on {Mathematical} {Physics}. Cambridge
  University Press, 2003.

\bibitem{aratyn_kac-moody_1991}
H.~Aratyn, L.A. Ferreira, J.F. Gomes, and A.H. Zimerman.
\newblock Kac-{Moody} construction of {Toda} type field theories.
\newblock {\em Physics Letters B}, 254(3-4):372--380, January 1991.

\bibitem{de_groot_generalized_1992}
Mark~F. De~Groot, Timothy~J. Hollowood, and J.~Luis Miramontes.
\newblock Generalized {Drinfel}'d-{Sokolov} hierarchies.
\newblock {\em Communications in Mathematical Physics}, 145(1):57--84, March
  1992.

\bibitem{hollowood_tau-functions_1993}
Timothy Hollowood and J.~Luis Miramontes.
\newblock Tau-functions and generalized intergrable hierarchies.
\newblock {\em Communications in Mathematical Physics}, 157(1):99--117, October
  1993.

\bibitem{luis_miramontes_tau-functions_1999}
J.~Luis~Miramontes.
\newblock Tau-functions generating the conservation laws for generalized
  integrable hierarchies of {KdV} and affine toda type.
\newblock {\em Nuclear Physics B}, 547(3):623--663, May 1999.

\bibitem{ferreira_tau-functions_1997}
Luiz~A. Ferreira, J.~Luis Miramontes, and Joaquín~Sánchez Guillén.
\newblock Tau-functions and dressing transformations for zero-curvature affine
  integrable equations.
\newblock {\em Journal of Mathematical Physics}, 38(2):882--901, February 1997.

\bibitem{de_carvalho_ferreira_generalized_2021}
J~M De~Carvalho~Ferreira, J~F Gomes, G~V Lobo, and A~H Zimerman.
\newblock Generalized {Bäcklund} transformations for affine {Toda}
  hierarchies.
\newblock {\em Journal of Physics A: Mathematical and Theoretical},
  54(6):065202, February 2021.

\bibitem{de_carvalho_ferreira_gauge_2021}
J~M De~Carvalho~Ferreira, J~F Gomes, G~V Lobo, and A~H Zimerman.
\newblock Gauge {Miura} and {Bäcklund} transformations for generalized
  $\text{A}_n$-{KdV} hierarchies.
\newblock {\em Journal of Physics A: Mathematical and Theoretical},
  54(43):435201, October 2021.

\bibitem{mathieu_supersymmetric_1988}
Pierre Mathieu.
\newblock Supersymmetric extension of the {Korteweg}–de {Vries} equation.
\newblock {\em Journal of Mathematical Physics}, 29(11):2499--2506, November
  1988.

\bibitem{kersten_higher_1988}
Paul~H.M. Kersten.
\newblock Higher order supersymmetries and fermionic conservation laws of the
  supersymmetric extension of the {KdV} equation.
\newblock {\em Physics Letters A}, 134(1):25--30, December 1988.

\bibitem{delduc_supersymmetric_1998}
F.~Delduc and L.~Gallot.
\newblock Supersymmetric {Drinfeld}–{Sokolov} reduction.
\newblock {\em Journal of Mathematical Physics}, 39(9):4729--4745, September
  1998.

\bibitem{inami_lie_1991}
Takeo Inami and Hiroaki Kanno.
\newblock Lie superalgebraic approach to super {Toda} lattice and generalized
  super {KdV} equations.
\newblock {\em Communications in Mathematical Physics}, 136(3):519--542, March
  1991.

\bibitem{inami_n_1991}
T.~Inami and H.~Kanno.
\newblock N = 2 super {KdV} and super sine-{Gordon} equations based on {Lie}
  super algebra $\text{A}(1,1)^(1)$.
\newblock {\em Nuclear Physics B}, 359(1):201--217, July 1991.

\bibitem{madsen_non-local_2001}
Jens~Ole Madsen and J.~Luis Miramontes.
\newblock Non-local conservation laws and flow equations for supersymmetric
  integrable hierarchies.
\newblock {\em Communications in Mathematical Physics}, 217(2):249--284, March
  2001.

\bibitem{aratyn_supersymmetry_2004}
H.~Aratyn, J.F. Gomes, and A.H. Zimerman.
\newblock Supersymmetry and the {KdV} equations for integrable hierarchies with
  a half-integer gradation.
\newblock {\em Nuclear Physics B}, 676(3):537--571, January 2004.

\bibitem{yamanaka_super_1988}
I.~Yamanaka and R.~Sasaki.
\newblock Super {Virasoro} {Algebra} and {Solvable} {Supersymmetric} {Quantum}
  {Field} {Theories}.
\newblock {\em Progress of Theoretical Physics}, 79(5):1167--1184, May 1988.

\bibitem{di_vecchia_classical_1977}
P.~Di~Vecchia and S.~Ferrara.
\newblock Classical solutions in two-dimensional supersymmetric field theories.
\newblock {\em Nuclear Physics B}, 130(1):93--104, November 1977.

\bibitem{chaichian_method_1978}
M.~Chaichian and P.P. Kulish.
\newblock On the method of inverse scattering problem and {Bäcklund}
  transformations for supersymmetric equations.
\newblock {\em Physics Letters B}, 78(4):413--416, October 1978.

\bibitem{chodos_simple_1980}
Alan Chodos.
\newblock Simple connection between conservation laws in the {Korteweg}-de
  {Vries} and sine-{Gordon} systems.
\newblock {\em Physical Review D}, 21(10):2818--2822, May 1980.

\bibitem{fioravanti_hidden_2000}
D.~Fioravanti and M.~Stanishkov.
\newblock Hidden {Virasoro} symmetry of (soliton solutions of) the
  sine-{Gordon} theory.
\newblock {\em Nuclear Physics B}, 591(3):685--700, December 2000.

\bibitem{gomes_soliton_2006}
J.F. Gomes, L.H. Ymai, and A.H. Zimerman.
\newblock Soliton solutions for the super {mKdV} and sinh-{Gordon} hierarchy.
\newblock {\em Physics Letters A}, 359(6):630--637, December 2006.

\bibitem{aguirre_n1_2015}
A.~R. Aguirre, J.~F. Gomes, N.~I. Spano, and A.~H. Zimerman.
\newblock N=1 super sinh-{Gordon} model with defects revisited.
\newblock {\em Journal of High Energy Physics}, 2015(2):175, February 2015.

\bibitem{aguirre_type-ii_2015}
A.~R. Aguirre, J.~F. Gomes, N.~I. Spano, and A.~H. Zimerman.
\newblock Type-{II} super-{Bäcklund} transformation and integrable defects for
  the {N} = 1 super sinh-{Gordon} model.
\newblock {\em Journal of High Energy Physics}, 2015(6):125, June 2015.

\bibitem{aguirre_defects_2018}
A.R. Aguirre, A.L. Retore, J.F. Gomes, N.I. Spano, and A.H. Zimerman.
\newblock Defects in the supersymmetric {mKdV} hierarchy via {Bäcklund}
  transformations.
\newblock {\em Journal of High Energy Physics}, 2018(1):18, January 2018.

\bibitem{xue_backlund-darboux_2014}
Ling-Ling Xue.
\newblock Bäcklund-{Darboux} {Transformations} and {Discretizations} of
  {Super} {KdV} {Equation}.
\newblock {\em Symmetry, Integrability and Geometry: Methods and Applications},
  April 2014.

\bibitem{zhou_darboux_2014}
Ruguang Zhou.
\newblock A {Darboux} transformation of the super {KdV} hierarchy and a super
  lattice potential {KdV} equation.
\newblock {\em Physics Letters A}, 378(26-27):1816--1819, May 2014.

\bibitem{gomes_nonvanishing_2012}
J.~F. Gomes, Guilherme~S. França, and A.~H. Zimerman.
\newblock Nonvanishing boundary condition for the {mKdV} hierarchy and the
  {Gardner} equation.
\newblock {\em Journal of Physics A: Mathematical and Theoretical},
  45(1):015207, January 2012.

\bibitem{adans_negative_2023}
Ysla~F. Adans, Guilherme França, José~F. Gomes, Gabriel~V. Lobo, and
  Abraham~H. Zimerman.
\newblock Negative flows of generalized {KdV} and {mKdV} hierarchies and their
  gauge-{Miura} transformations.
\newblock {\em Journal of High Energy Physics}, 2023(8):160, August 2023.

\bibitem{adans_skdv_2024}
Y.~F. Adans, A.~R. Aguirre, J.~F. Gomes, G.~V. Lobo, and A.~H. Zimerman.
\newblock {SKdV}, {SmKdV} flows and their supersymmetric gauge-{Miura}
  transformations.
\newblock {\em Open Communications in Nonlinear Mathematical Physics},
  Proceedings: OCNMP...:13294, April 2024.

\bibitem{fordy_factorization_1980}
Allan~P. Fordy and John Gibbons.
\newblock Factorization of operators {I}. {Miura} transformations.
\newblock {\em Journal of Mathematical Physics}, 21(10):2508--2510, October
  1980.

\bibitem{guil_homogeneous_1991}
F.~Guil and M.~Mañas.
\newblock Homogeneous manifolds and modified {KdV} equations.
\newblock {\em Journal of Mathematical Physics}, 32(7):1744--1749, July 1991.

\bibitem{gomes_miura_2016}
J.~F. Gomes, A.~L. Retore, and A.~H. Zimerman.
\newblock Miura and generalized {Bäcklund} transformation for {KdV} hierarchy.
\newblock {\em Journal of Physics A: Mathematical and Theoretical},
  49(50):504003, December 2016.

\bibitem{sawada_method_1974}
K.~Sawada and T.~Kotera.
\newblock A {Method} for {Finding} {N}-{Soliton} {Solutions} of the {K}.d.{V}.
  {Equation} and {K}.d.{V}.-{Like} {Equation}.
\newblock {\em Progress of Theoretical Physics}, 51(5):1355--1367, May 1974.

\bibitem{aratyn_new_1992}
H.~Aratyn, L.A. Ferreira, J.F. Gomes, and A.H. Zimerman.
\newblock A new deformation of {W}-infinity and applications to the two-loop
  {WZNW} and conformal affine {Toda} models.
\newblock {\em Physics Letters B}, 293(1-2):67--71, October 1992.

\bibitem{gao_bosonization_2012}
Xiao~Nan Gao and S.Y. Lou.
\newblock Bosonization of supersymmetric {KdV} equation.
\newblock {\em Physics Letters B}, 707(1):209--215, January 2012.

\bibitem{hon_super_2011}
Y.~C. Hon and Engui Fan.
\newblock Super quasiperiodic wave solutions and asymptotic analysis for
  $\mathcal{N} = 1$ supersymmetric {KdV}-type equations.
\newblock {\em Theoretical and Mathematical Physics}, 166(3):317--336, March
  2011.

\bibitem{gao_bosonization_2013}
Xiao~Nan Gao, S.~Y. Lou, and Xiao~Yan Tang.
\newblock Bosonization, singularity analysis, nonlocal symmetry reductions and
  exact solutions of supersymmetric {KdV} equation.
\newblock {\em Journal of High Energy Physics}, 2013(5):29, May 2013.

\end{thebibliography}
								\bibliographystyle{unsrt}

							\end{document}